\documentclass[prx,reprint,notitlepage,superscriptaddress,shortbibliography,noeprint]{revtex4-1}

\usepackage[english]{babel}

\usepackage{amsmath,amssymb,bbm,mathrsfs,mathtools,braket,color,%
  graphicx,comment,textcomp,amsfonts,dsfont,lettrine,units,comment}
\usepackage[colorlinks,citecolor=blue,urlcolor=blue]{hyperref}

\AtBeginDocument{%
  \newwrite\bibnotes
  \def\bibnotesext{Notes.bib}
  \immediate\openout\bibnotes=\jobname\bibnotesext
  \immediate\write\bibnotes{@CONTROL{REVTEX41Control}}
  \immediate\write\bibnotes{@CONTROL{%
      apsrev41Control,author="08",editor="1",pages="1",title="0",year="1",eprint=""}}
  \if@filesw
  \immediate\write\@auxout{\string\citation{apsrev41Control}}%
  \fi
}%

\newcommand{\hc}{\mathrm{H.c.}}
\newcommand{\id}{\mathbbm{1}}

\newcommand{\N}{\mathbb{N}}
\newcommand{\Z}{\mathbb{Z}}

\newcommand{\imag}{\mathrm{i}}
\newcommand{\e}{\mathrm{e}}

\makeatletter
\newcommand*{\transpose}{%
  {\mathpalette\@transpose{}}%
}
\newcommand*{\@transpose}[2]{%
  \raisebox{\depth}{$\m@th#1\intercal$}%
}
\makeatother

\newcommand{\abs}[1]{\left\lvert #1 \right\rvert}

\newcommand{\tr}{\mathop{\mathrm{tr}}}
\newcommand{\Li}{\mathop{\mathrm{Li}_3}}

\renewcommand{\Re}{\mathop{\mathrm{Re}}}

\newcommand{\iqoqi}{\affiliation{Institute for Quantum Optics and Quantum Information of the Austrian Academy of Sciences, Innsbruck, Austria}}
\newcommand{\qci}{\affiliation{Center for Quantum Physics, University of Innsbruck, Innsbruck, Austria}}
\newcommand{\exphys}{\affiliation{Institut f\"ur Experimentalphysik, Universit\"at Innsbruck, Innsbruck, Austria}}

\begin{document}

\title{Scalable and Parallel Tweezer Gates for Quantum Computing with Long Ion Strings}





\author{Tobias Olsacher} \qci \iqoqi
\author{Lukas Postler} \exphys
\author{Philipp Schindler} \exphys
\author{Thomas Monz} \exphys
\author{Peter Zoller} \qci \iqoqi
\author{Lukas M. Sieberer} \qci \iqoqi
\email{lukas.sieberer@uibk.ac.at}

\date{\today}

\begin{abstract}
  Trapped-ion quantum computers have demonstrated high-performance gate operations in registers of about ten qubits. However, scaling up and parallelizing quantum computations with long one-dimensional (1D) ion strings is an outstanding challenge due to the global nature of the motional modes of the ions which mediate qubit-qubit couplings. Here, we devise methods to implement scalable and parallel entangling gates by using engineered localized phonon modes. We propose to tailor such localized modes by tuning the local potential of individual ions with programmable optical tweezers. Localized modes of small subsets of qubits form the basis to perform entangling gates on these subsets in parallel.
  We demonstrate the inherent scalability of this approach by
  presenting analytical and numerical results for long 1D ion chains and even for infinite
  chains of uniformly spaced ions. Furthermore, we show that combining our methods with optimal coherent control techniques allows to realize maximally dense universal parallelized quantum circuits.
\end{abstract}

\maketitle


\section{Introduction}
\label{sec:introduction}

\begin{figure*}
  \centering  
  \includegraphics[width=\linewidth]{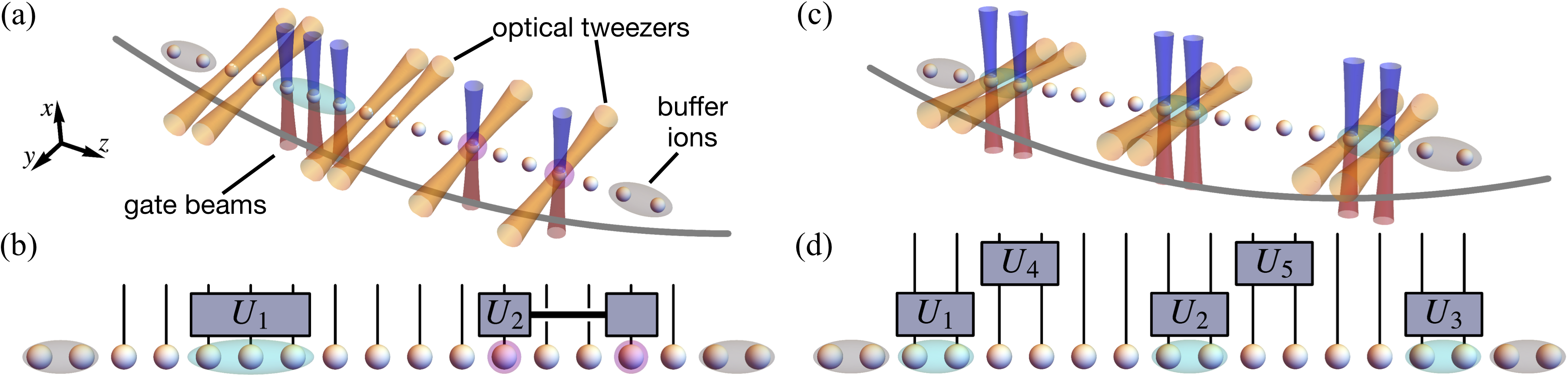}
  \caption{Schematic setup and circuits. (a) Specific ions in a long 1D chain are pinned by optical tweezers. This leads to the formation of localized phonon modes, e.g., for the subregister formed by the three ions which are marked by cyan shading, and for the pair of distant qubits which are marked by pink shading and which are pinned by equally strong optical potentials. (b)~The localized phonon modes can be utilized to implement multi-qubit entangling gates. Buffer ions at the ends of the chain are not used as qubits. Addressed lasers which are required to perform entangling gates are indicated schematically in (a) as blue and red beams. (c) In this work, we focus on implementing parallel two-qubit entangling gates which use the localized phonon modes of pairs of neighboring pinned ions. (d) After a first layer of the circuit which corresponds to the product of gates $U_1 U_2 U_3$ that act on distinct qubits, the tweezers are focused on different ions to perform the second layer corresponding to the gate operation $U_4 U_5$.}
  \label{fig:schematic}
\end{figure*}

Trapped atomic ions are a leading platform for quantum information
processing~\cite{Cirac1995,gardiner2015quantum,Schindler2013,Bruzewicz2019,Ballance2016,Shapira2018,Negnevitsky2018,Harty2014,Gaebler2016}. High-fidelity gate operations, qubit initialization and readout, as
well as long coherence times have been demonstrated in trapped-ion quantum
computers which consist of tens of individually addressable qubits~\cite{Wright2019,Erhard2020,Lu2019,Zhu2019,Wan2019}. However,
while state-of-the-art ion traps can sustain 1D chains of more than a hundred
ions, scaling up and, in particular, parallelizing gate operations is
challenging due to the increasing complexity of the spectrum of phonon modes,
which serve as data buses for the transmission of quantum information~\cite{Cirac1995,James1998Quantum}. 


The spectrum of phonon modes becomes dense for long ion crystals~\cite{Schiffer1993, Zhu2006}, and, consequently, gate
operations which rely on using a single phonon mode become slow due to the
necessity to spectrally resolve this mode. In contrast, implementing gates which
use multiple modes requires solving optimal control problems~\cite{Zhu2006a, Roos2008, Choi2014, Debnath2016, Wu2018, Landsman2019a, Figgatt2019, Lu2019a,Leung2018} which become
exceedingly complex, in particular, when gates should be performed on many
qubits in parallel~\cite{Figgatt2019, Landsman2019a}. This additional
complication to implement parallel gates is due to the collective and thus
nonlocal nature of phonon modes, which stands in contrast to the goal of
effecting two-qubit entangling operations that locally address individual pairs
of ions.

An alternative approach is provided by segmented traps, in which small ion crystals are shuttled between different zones that are dedicated to qubit storage and manipulation, respectively~\cite{Kielpinski2002}. Shuttling operations are comparatively time-consuming, and form a bottleneck for fast quantum computations. This disadvantage is mitigated by the fact that in traps with several interaction zones, gates can be performed in parallel. Although the fabrication and operation of segmented traps is technologically challenging, quantum information processing which relies heavily on ion shuttling has been demonstrated experimentally~\cite{Kaufmann2017, blakestad-prl-102-153002, Pino2020}.

In this paper, we address the challenge of scaling up and parallelizing 
quantum computations with long 1D ion strings by combining well-developed conventional linear Paul traps~\cite{Leibfried2003a} with programmable optical tweezer arrays, which are commonly utilized as a powerful tool in quantum simulation with neutral Rydberg atoms~\cite{Omran570,Barredo2018}.
Here, instead, we consider using programmable optical tweezer arrays
to pin specific ions in a linear trap~\cite{Shen2019}, and thereby engineer localized
phonon modes~\cite{Loye2020,Ivanov2009,Abdelrahman2017}. These designer phonon modes form the basis to implement scalable and parallel entangling
gates. The required capabilities to optically address individual ions are available in current experiments~\cite{Barredo2020,Schneider2010}. By dynamically reconfiguring the tweezer array, the designer phonon modes can be
adjusted during running quantum computations, and thus offer great flexibility to achieve an effective
``optical segmentation'' of the ion chain.

Possible quantum circuits provided by optical segmentation are illustrated schematically in Figs.~\ref{fig:schematic}(a) and~(b): In the first example, a subregister formed by three consecutive qubits which are marked by cyan shading is decoupled from the other qubits by ``optical tweezer walls''~\cite{Shen2019} that consist of 
pairs of optically pinned ions. By using the phonon modes which are localized in between the tweezer walls, a multi-qubit gate can be performed on the subregister. The second example relies on the long-range connectivity in chains of trapped ions to implement an entangling gate between the two qubits marked by pink shading and which are both pinned by optical tweezers.

In this work, we focus on the implementation of quantum circuits which are composed of parallel gates between pairs of nearest-neighboring qubits. Pinning pairs of neighboring ions with optical tweezers as
illustrated in Fig.~\ref{fig:schematic}(c) gives rise to localized phonon modes
which correspond to center-of-mass (COM) and stretch oscillations of the pinned
pairs. In micro traps with segmented electrodes, an analogous local mode structure arises for two ions which are shuttled to an interaction zone of the trap.
The optical segmentation enables performing entangling gates on all pairs of pinned
ions in parallel, which corresponds to the first layer of the quantum circuit
shown in Fig.~\ref{fig:schematic}(d). 

Due to the emergence of a local phonon mode structure with only two relevant
localized phonon modes per pinned pair, high-fidelity scalable and parallel
gates can be performed in long ion chains without resorting to optimal coherent
control techniques~\cite{Zhu2006a, Roos2008, Choi2014, Debnath2016, Wu2018, Landsman2019a, Figgatt2019, Lu2019a,Leung2018}. Further, the gate duration is determined by the splitting of the localized phonon modes, and is thus independent of the total number of ions in the chain. Indeed, as we show analytically and numerically, tweezer gates can be performed even
in infinitely long chains~\cite{Landsman2019a}.

Finally, we discuss different ways to minimize
crosstalk between parallel two-qubit entangling gates based on optimal control of time-modulated laser-pulse amplitudes as developed by Duan et al.~\cite{Zhu2006}. This enables the implementation of
dense ``brick wall circuits.'' Such dense circuits have a wide range of applications, for example, in
digital quantum simulation~\cite{Martinez2016, Barreiro2011, Lanyon57} of spin models with local interactions, or to realize
random circuits~\cite{Arute2019}, which are models for strongly chaotic quantum dynamics~\cite{Nahum2017, Nahum2018, VonKeyserlingk2018, Rakovszky2018, Khemani2018}.

The paper is structured as follows: We start in Sec.~\ref{sec:review} with a short review of some aspects of quantum gates with trapped ions. In Sec.~\ref{sec:optical-design-of-modes}, we discuss how to design localized phonon modes for quantum computing. Subsequently, in Secs.~\ref{sec:entangl-gates-minimal-control} and \ref{sec:optimized-gates}, we show how these localized modes can be used to implement scalable and parallel tweezer gates with and without optimal control, respectively. We give an outlook in Sec.~\ref{sec:outlook}.

\section{Entangling quantum gates with trapped ions}
\label{sec:review}

For reference, we find it convenient
to summarize some fundamentals of quantum computing with trapped ions. Based on this, we state the decomposition of quantum circuits into single-qubit and nearest-neighbor entangling gates and discuss how one can quantify the performance of a two-qubit quantum gate.


\subsection{Quantum computing with trapped ions}
\label{sec:nn-quantum-circuits}

We consider an implementation of quantum logic gates in 1D chains of laser-cooled
trapped ions that relies on the laser-induced coupling between long-lived
internal states of the ions, which encode the qubits, and the phonon modes of the
ion chain, which serve as quantum data buses. Entanglement between qubits is established through the exchange of real~\cite{Cirac1995} or virtual~\cite{Milburn1999, Molmer1999, Sorensen1999, Sorensen2000} phonons. The latter approach is realized by geometric phase gates, which offer the advantage of being insensitive to finite temperatures of the phonon modes. We employ a particular type of geometric phase gate, that is known as the M\o lmer-S\o rensen gate~\cite{Sorensen1999}. In a suitable rotating frame, the corresponding qubit-phonon coupling
Hamiltonian for $N$ ions with $3 N$ motional modes reads
\begin{multline}
  \label{eq:H-qubit-ph}
  H = \sum_{\alpha  \in \{ x, y, z \}} \sum_{i, n = 1}^N \eta_{\alpha, i}^n \hbar \Omega_i(t)
  \sin(\mu_i t) \\ \times \left( a_{\alpha, n}^{\dagger} \e^{\imag
      \omega_{\alpha, n} t} + \hc \right)
  \sigma^x_i,
\end{multline}
where the first sum is over the spatial directions $\alpha$ of the 3D motion of the ions. As illustrated in Fig.~\ref{fig:schematic}(a), we choose the coordinate system such that $x$ and $y$ directions are transverse to the weak trap axis, which is along the $z$ direction. In the form given above, the Hamiltonian is valid in the Lamb-Dicke limit, in which the amplitudes of oscillations of the ions around their equilibrium positions are small in comparison to optical wavelengths. This justifies an expansion to first order in the Lamb-Dicke parameter matrix, which we define as 
\begin{equation}
  \label{eq:Lamb-Dicke}
  \eta_{\alpha, i}^n = k_{\mathrm{L}, \alpha} \sqrt{\frac{\hbar}{2 m \omega_{\alpha, n}}} M_{\alpha, i}^n,
\end{equation}
with the effective laser wave vectors with components $k_{\mathrm{L}, \alpha}$ that are assumed to be equal for all ions~\cite{Wu2018}. Here $m$ is the mass of a single
ion, $\omega_{\alpha, n}$ is the frequency of the phonon mode $n \in \{ 1, \dotsc, N \}$ in spatial direction $\alpha$, and the element $M_{\alpha, i}^n$ of the phonon mode matrix is given by the
amplitude of the phonon mode vector $n$ on ion $i \in \{ 1, \dotsc, N \}$. The phonon mode matrix is derived in Appendix~\ref{sec:ion-chain-tweezers}.

The Hamiltonian~\eqref{eq:H-qubit-ph} results from addressing the ions with pairs of laser
beams which are detuned by $\pm \mu_i$ from the qubit
transition~\cite{Milburn1999, Molmer1999, Sorensen1999, Sorensen2000}. We assume that the
detuning $\mu_i$ and the laser pulse shape as described by a time-modulated Rabi
frequency $\Omega_i(t)$ can be chosen for each ion $i$
individually. The annihilation and creation operators of phonons with frequency $\omega_{\alpha, n}$ are denoted by $a_{\alpha, n}$ and $a_{\alpha, n}^{\dagger}$, respectively, and $\sigma^x_i$ is the Pauli matrix which
acts in the Hilbert space of the qubit which is encoded in ion $i$. We assume in the following that the lasers dominantly excite transverse $x$
modes, which is justified if the detuning from the excited transversal $x$ modes is much smaller than
the frequency difference between the transverse $x$ modes and the
transverse $y$ and longitudinal $z$ modes,
respectively~\cite{Zhu2006}.



The time evolution that is generated by the time-dependent qubit-phonon Hamiltonian~\eqref{eq:H-qubit-ph} is described by the unitary~\cite{Zhu2003, Zhu2006a, Zhu2006, Garcia-Ripoll2005, Wu2018, Monroe2019}
\begin{equation}
  \label{eq:U-gate}
  U = \exp \! \left( \sum_{i = 1}^N \phi_i \sigma^x_i + \imag \sum_{i < i' =
      1}^N \chi_{i, i'} \sigma^x_i \sigma^x_{i'} \right).
\end{equation}
The first term in the exponent with the operator
\begin{equation}
  \label{eq:phi}
  \phi_i = \sum_{n = 1}^N \left( \alpha^n_i a_n^{\dagger} + \hc \right),
\end{equation}
induces qubit-state-dependent displacements of the phonon modes where
\begin{equation}
  \label{eq:alpha}
  \alpha^n_i = - \imag \eta^n_i \int_0^{\tau} \mathrm{d} t \, \Omega_i(t)
  \sin(\mu_i t) \e^{\imag \omega_n t}.
\end{equation}
If $\phi_i \neq 0$ for a given qubit $i$, unwanted entanglement is created between that qubit and the phonon modes. This can be prevented by carefully choosing the detuning and gate duration depending on the phonon spectrum. The desired qubit-qubit coupling term is given by
\begin{multline}
  \label{eq:chi}
  \chi_{i, i'} = \sum_{n = 1}^N \eta^n_i \eta^n_{i'} \int_0^{\tau} \mathrm{d} t
  \int_0^t \mathrm{d} t' \sin(\omega_n \left( t - t' \right)) \\ \times \left(
    \Omega_i(t) \Omega_{i'}(t') \sin(\mu_i t) \sin(\mu_{i'} t') \right. \\
  \left. + \Omega_i(t') \Omega_{i'}(t) \sin(\mu_i t') \sin(\mu_{i'} t) \right).
\end{multline}

We focus in the following on the implementation, based on Eq.~\eqref{eq:U-gate}, of quantum circuits with entangling operations between only neighboring qubits as illustrated in Fig.~\ref{fig:schematic}(d). Such circuits can be arranged in consecutive layers where gates within each layer are executed in parallel. A single layer can be written as
\begin{equation}
  \label{eq:U-0}
  U_0 = \prod_{\left( i, i' \right) \in I} U_{i, i'},
\end{equation}
where $I$ is the set of all pairs of neighboring qubits on which gates are to be performed in the layer under consideration. Since any entangling two-qubit gate can be decomposed into single-qubit rotations and entangling operations which are generated by
$\sigma^x_i \sigma^x_{i'}$~\cite{Kraus2001}, we can assume without loss of generality that $U_{i, i'}$ is an $xx$ gate,
\begin{equation}
  \label{eq:U-xx}
  U_{i, i'} = \e^{\imag \chi^0_{i, i'} \sigma^x_i \sigma^x_{i'}}
\end{equation}
with couplings $\chi^0_{i, i'}$ for different pairs $\left( i, i' \right)$ all having the same positive or negative sign and $\abs{\chi^0_{i, i'}} \in [0, \pi/4]$~\cite{Kraus2001}. Throughout this work we consider maximally entangling gates with $\chi^0_{i, i'} = \pm \pi/4$ as a benchmark case. This benchmark is particularly relevant since the unitary~\eqref{eq:U-xx} with $\chi^0_{i, i'} = \pm \pi/4$ is up to single-qubit rotations equivalent to the CNOT gate~\cite{Debnath2016} which forms a universal gate set together with single-qubit rotations~\cite{Nielsen2011}.

\subsection{Gate imperfections}
\label{sec:gate-imperfections}

Any experimental implementation of quantum logic gates with trapped ions is affected by various sources of imperfections which are not captured by the model Hamiltonian~\eqref{eq:H-qubit-ph}~\cite{Wu2018}. However, even within this model, the unitary $U$~\eqref{eq:U-gate} deviates in general from the target gate unitary $U_0$~\eqref{eq:U-0}. We distinguish three types of such intrinsic gate imperfections: Residual qubit-phonon entanglement, over- and underrotation errors, and crosstalk. To formalize this distinction, we factorize the gate unitary as $U = U_{\alpha} U_{\chi}$, where
\begin{equation}
    U_{\alpha} = \e^{\sum_{i = 1}^N \phi_i \sigma^x_i}, \qquad U_{\chi} = \e^{\imag \sum_{i < i' = 1}^N \chi_{i, i'} \sigma^x_i \sigma^x_{i'}}.
\end{equation}
Deviations of $U_{\alpha}$ from the identity, $U_{\alpha} \neq \id$, imply that qubits and phonons are entangled at the end of the gate operation. Over- and underrotation errors as well as crosstalk occur for $U_{\chi} \neq U_0$.

The adverse effect of residual qubit-phonon entanglement can be quantified in terms of the average fidelity~\cite{Nielsen2002} per gate, which we define as
\begin{equation}
  \label{eq:avg-fidelity}
  F = \left( \int \mathrm{d} \Psi \bra{\Psi} U_{\chi}^{\dagger}
  {\tr}_{\mathrm{ph}} \! \left( U \ket{\Psi} \bra{\Psi} \otimes \rho_{\mathrm{th}}
    U^{\dagger} \right) U_{\chi} \ket{\Psi} \right)^{\frac{1}{G}},
\end{equation}
where the trace is over the motional degrees of freedom, $\rho_{\mathrm{th}}$ denotes a thermal state of phonons, and the integration is over the Fubini-Study measure~\cite{bengtsson_zyczkowski_2006}. We note that usually the fidelity is defined with respect to the ideal target gate, i.e., with $U_{\chi}$ replaced by $U_0$, to explicitly include over- and underrotation errors as well as crosstalk. In contrast, we quantify these types of errors separately as detailed below. To account for the exponential
dependence of the total fidelity $F_{\mathrm{tot}}$ on the number $G$ of gates
which are being performed in parallel---which in the cases of interest to us
scales with the number of qubits $G \sim N$---we include an exponent $1/G$ in
the definition of the fidelity per gate. 

The expression for the average fidelity per gate in Eq.~\eqref{eq:avg-fidelity} can be made more explicit in two steps: The first step is to perform the trace over phonon modes of the thermal density matrix $\rho_{\mathrm{th}}$, multiplied with qubit-state-dependent displacement operators which are contained in $U$ and $U^{\dagger}$ and correspond to the first term in the exponent in Eq.~\eqref{eq:U-gate}~\cite{Wu2018}. Second, the integral over random initial qubit states $\ket{\Psi}$ can be evaluated explicitly by replacing it by a sum over a discrete basis of unitary operators~\cite{Nielsen2002}. This procedure yields an expression for the infidelity per gate $\delta F = 1 - F$ which, in the limit of high fidelity, takes the form
\begin{equation}
  \label{eq:infidelity}
  \delta F = \frac{4}{5 G} \sum_{i, n = 1}^N \abs{\alpha^n_i}^2 \left( 2
    n_{\mathrm{th}}(\omega_n) + 1 \right),
\end{equation}
where $\alpha^n_i$ is the qubit-phonon coupling defined in Eq.~\eqref{eq:alpha}, and $n_{\mathrm{th}}(\omega_n) = 1/(\e^{\hbar \omega_n/T} - 1)$ is the average thermal occupation of phonons in the mode with frequency $\omega_n$ at temperature $T$. For simplicity, we assume in the following that $n_{\mathrm{th}}(\omega_n) = 0.5$ for all phonon modes. 

Over- and underrotation errors of the gate $U$ corresponds to
deviations of the qubit-qubit coupling $\chi_{i, i'}$ from to the desired value
$\chi^0_{i, i'}$ for pairs of qubits $\left( i, i' \right)$ which are contained
in $I$ in Eq.~\eqref{eq:U-0}, i.e., which are affected by the ideal gate
$U_0$. In contrast, crosstalk is due to unwanted nonzero qubit-qubit couplings for pairs
of qubits which are not contained in $I$. To separate these types of errors, we
factorize the qubit-qubit coupling in Eq.~\eqref{eq:U-gate} as
$U_{\chi} = U_1 U_C$ where
\begin{equation}
  \label{eq:U-1-C}
  U_1 = \prod_{\left( i, i' \right) \in I} \e^{\imag \chi_{i, i'} \sigma^x_i
    \sigma^x_{i'}}, \quad U_C = \prod_{\left( i, i' \right) \in I'} \e^{\imag
    \chi_{i, i'} \sigma^x_i \sigma^x_{i'}},
\end{equation}
and where $I'$ contains all ordered pairs of qubits which are not included in
$I$. We quantify these gate errors in terms of the diamond norm of the error superoperators
$\mathcal{E}_1 = \mathcal{U}_1 - \mathcal{U}_0$ and $\mathcal{E}_C = \mathcal{U}_C - \mathcal{I}$, where $\mathcal{U}_b(\rho) = U_b \rho U_b^{\dagger}$ for $b = 0, 1, C$, and
$\mathcal{I}(\rho) = \rho$~\cite{Landsman2019a, Sanders2015, Kueng2016}. Bounds
on the errors per gate are given by~\cite{Aharonov1998}
\begin{align}
  \label{eq:inaccuracy}
  \frac{1}{G} \lVert \mathcal{E}_1 \rVert_{\diamond} & \leq \delta \chi = \frac{2}{G} \sum_{\left( i, i' \right) \in I} \abs{\chi_{i, i'} - \chi^0_{i, i'}},
  \\ \label{eq:crosstalk} \frac{1}{G} \lVert \mathcal{E}_C
  \rVert_{\diamond} & \leq C = \frac{2}{G} \sum_{\left( i, i' \right) \in I'} \abs{\chi_{i, i'}},  
\end{align}
where we defined the over-/underrotation error $\delta \chi$ and the crosstalk $C$. 

\section{Optical design of localized phonon modes}
\label{sec:optical-design-of-modes}

Our goal is to design localized phonon modes in long laser-cooled ion strings, which, as we go on to show in Secs.~\ref{sec:entangl-gates-minimal-control} and~\ref{sec:optimized-gates}, enable the implementation of scalable and parallel entangling quantum gates. In particular, we engineer specific types of mode matrices $M_{\alpha,i}^n$ by using optical tweezers that are focused on the equilibrium positions of specific ions and thus pin these ions as we describe in Appendix~\ref{sec:ion-chain-tweezers}. The tweezers are realized by Gaussian laser beams along the $y$ direction as illustrated in Fig.~\ref{fig:schematic}(a). In the vicinity of the focuses of the tweezers, the optical potential can be approximated as harmonic. The optical trapping frequency along the beam axis is negligible. In contrast, the optical trapping frequency $\omega_{0,i}$ along the transverse $x$ and the longitudinal $z$ directions is determined by the beam intensity and waist at the position of ion $i$, and we assume that $\omega_{0, i}$ does not depend on the internal state
of the ions and can take on values up to $\omega_{0, i} \lesssim 0.4 \omega_x$ for
typical transverse trapping frequencies
$\omega_x = 2 \pi \times 3 \, \mathrm{MHz}$. In Appendix~\ref{sec:feasibility_study}, we discuss experimental requirements to realize such strong qubit-state-independent optical potentials for different ionic species. 

In the following, we first consider localized phonon modes of a single pair of
pinned, neighboring ions in a long chain. This allows us to delimit the regime of strong pinning in
which the phonon modes of the pinned ions decouple from the modes of the
spectator ions. We then illustrate these ideas with concrete examples of finite
and infinite chains.

\subsection{Pinning a single ion pair in a long chain}
\label{sec:local-phon-modes}

To engineer transverse $x$
phonon modes which are localized on a pair of neighboring ions $i$ and $i + 1$
and which can thus be used to perform an entangling gate on this pair, we
consider a situation in which the pinning on the ions forming the pair is the same,
$\omega_{0, i} = \omega_{0, i + 1}$, whereas the
remaining spectator ions are not pinned.
The residual Coulomb interaction (see Appendix~\ref{sec:ion-chain-tweezers} for details) has two
effects: First, the interaction between the pinned ions leads to the formation
of localized COM and stretch modes given by
\begin{equation}
  \label{eq:M-COM-stretch}
  \begin{split}
    M^{\mathrm{COM}}_{x, i'} & \approx (\delta_{i', i} + \delta_{i', i + 1})/\sqrt{2}, \\
    M^{\mathrm{stretch}}_{x, i'} & \approx (\delta_{i', i} - \delta_{i', i +
      1})/\sqrt{2}.
  \end{split}
\end{equation}
These are the desired modes to implement two-qubit
entangling gates. The frequency splitting of these modes is determined by the Coulomb interaction,
$\omega_{\mathrm{COM}} - \omega_{\mathrm{stretch}} \approx e^2/(4 \pi \epsilon_0
d_{i, i + 1}^3 m \omega_x)$,
where $d_{i, i'} = \abs{z_{i, 0} - z_{i', 0}}$ is the distance between the
equilibrium positions $z_{i, 0}$ of the ions along the trap $z$ axis, $e$ is the elementary charge and $\epsilon_0$ is the vacuum permittivity. Second, the interaction between pinned and spectator ions slightly admixes
oscillations of the spectator ions to the localized modes, i.e., the mode
vectors Eq.~\eqref{eq:M-COM-stretch} acquire nonzero amplitudes on ions
$i' \notin \{ i, i + 1 \}$. This unwanted effect is strongly suppressed if the
difference between the squares of the local oscillation frequencies of the
pinned and spectator ions, as given in Eq.~\eqref{eq:omega-alpha-i}, is large in comparison to their residual Coulomb
interaction. 

More generally, for transverse $x$ phonon modes of a chain of ions with mean
spacing $d$, the regime of strong pinning in which localized phonon modes emerge
can be conveniently characterized in terms of two dimensionless parameters:
\begin{equation}
  \label{eq:dimless-params}
  \epsilon = \sqrt{\frac{e^2}{4 \pi \epsilon_0 d^3 m \omega_x^2}}, \qquad \nu_0 =
  \frac{\omega_0}{\omega_x},
\end{equation}
where for simplicity we assume that all ions are pinned with the same optical
trapping frequency $\omega_0$. As detailed in Appendix~\ref{sec:ion-chain-tweezers}, localized COM and stretch modes of a pair of pinned ions decouple from the motion of spectator ions if $\nu_0^2/\epsilon^2 \gg 1$. While we focus here on pairs of neighboring ions, we note that the same criterion applies for pairs of pinned ions at a larger distance.

In experiments, $\epsilon$ is typically a small parameter. For example, we obtain $\epsilon \approx 0.07$ for
a spacing of $d = 10 \, \mu \mathrm{m}$ in a chain of $^{24}\mathrm{Mg}^+$ ions with a transverse trapping frequency of $\omega_x = 2 \pi \times 5.5 \, \mathrm{MHz}$, or $^{40}\mathrm{Ca}^+$ ions with $\omega_x = 2 \pi \times 4.2 \, \mathrm{MHz}$. The ratio $\nu_0^2/\epsilon^2$ can be increased by either increasing the tweezer trapping frequency $\omega_0$ or by increasing the spacing of the ions $d$.

\subsection{Localized phonon modes in a finite chain}
\label{sec:finite-chain}

\begin{figure}
  \centering  
  \includegraphics[width=\linewidth]{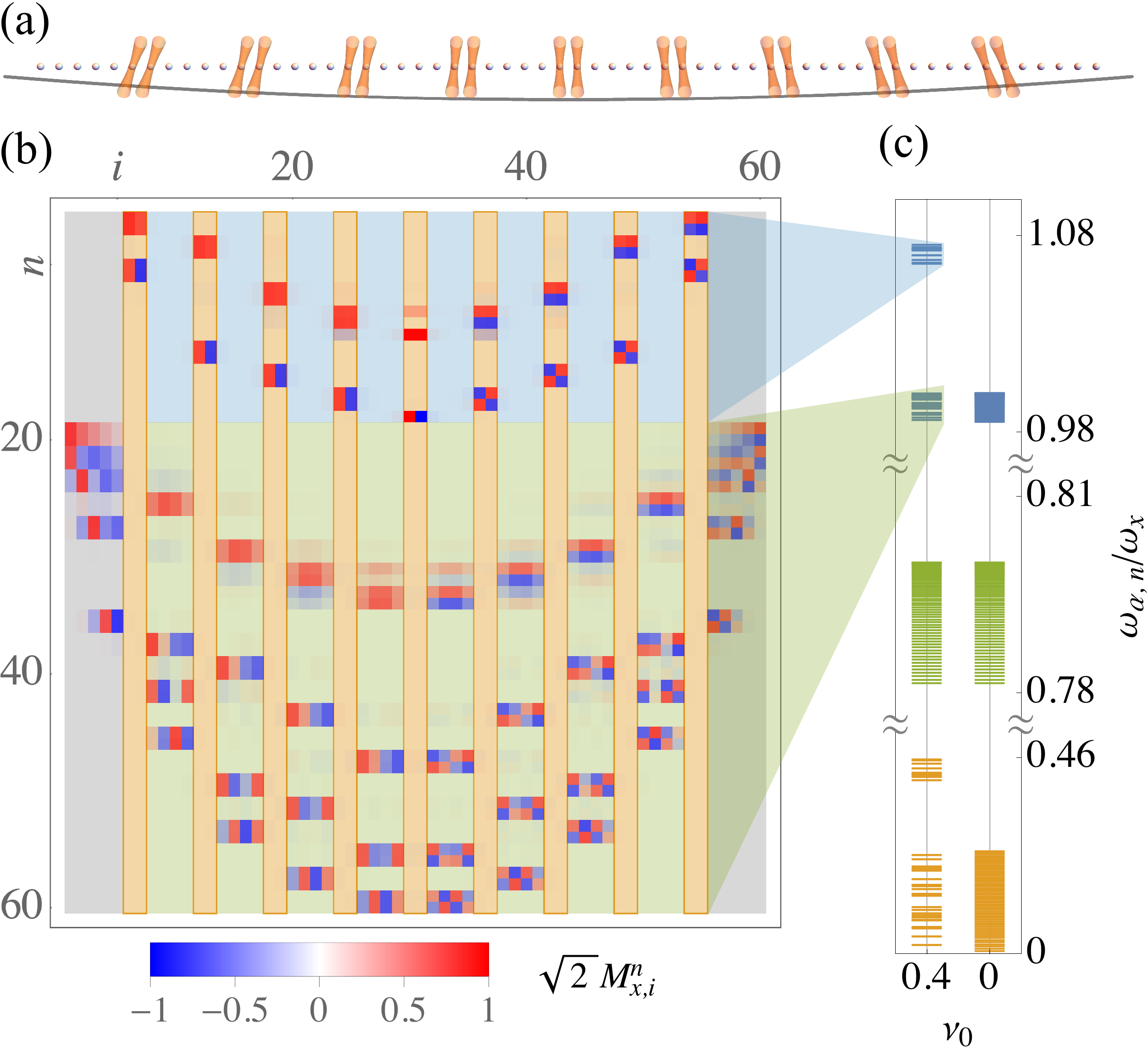}
  \caption{(a) $N = 60$ ions in an array of optical tweezers which subdivides
    the chain into groups of $p = 6$ six ions. The first and last five buffer
    ions are not used as qubits. (b) Mode matrix $M_{x, i}^n$ for transverse $x$
    modes. Orange and gray shading indicate pinned and buffer ions,
    respectively. Blue shading marks modes which are superpositions of localized
    COM and stretch modes of pairs of pinned ions. (c) Phonon mode spectrum for
    $\nu_0 = 0.4$ and, for comparison, without tweezers, i.e., for $\nu_0 = 0$. Transverse $x$ and $y$ modes and longitudinal $z$ modes are shown, respectively, in blue, green,
    and orange. We assume that the ratio of transverse trapping frequencies is given by $\omega_y/\omega_x = 0.8$.}
  \label{fig:phonon_modes_spectra_finite}
\end{figure}

As an example, we consider a chain of $N = 60$ ions as depicted in
Fig.~\ref{fig:phonon_modes_spectra_finite}(a). We assume harmonic confinement along the trap $z$ axis with trapping frequency $\omega_z$. Starting from the sixth ion, optical
tweezers are arranged to divide the chain into groups of $p = 6$ ions. The resulting mode matrix $M_{x, i}^n$ for $i, n \in \{ 1, \dotsc, N \}$ for
oscillations in the transverse $x$ direction is shown in
Fig.~\ref{fig:phonon_modes_spectra_finite}(b). Orange boxes mark pinned ions,
and gray boxes indicate buffer ions at the ends of the chain. The spacing of
these ions deviates strongly from the approximately uniform spacing in the
center of the chain, and we exclude them from gate operations. As explained
above, the residual Coulomb coupling leads to the formation of localized COM and
stretch modes of the pinned pairs. In the figure, COM and stretch modes are
distinguished by the same color of $M_{x, i}^n$ and $M_{x, i + 1}^n$ for two
neighboring ions which oscillate in phase and different colors for oscillations
with opposite phase. As an additional effect which is due to the long-range
character of the residual Coulomb interaction, the localized modes of distinct
pinned pairs hybridize, where the number of ions in pinned pairs, which is $18$
for this example, determines the number of hybridized modes. These modes are
highlighted by blue shading in the figure.

To perform entangling gates on pairs of pinned ions, the hybridization of
localized phonon modes is in principle not desirable. However, as long as the
frequency splitting of the localized modes due to the hybridization is so small
that it cannot be resolved on the time scale of the gate operation, the
hybridization has only a small effect on the gate performance. This picture
generalizes the concept of local oscillation modes from single
ions~\cite{Duan2004} to pairs of ions.

The order of mode indices $n$ in Fig.~\ref{fig:phonon_modes_spectra_finite}(b)
reflects the mode frequency, with $n = 1$ corresponding to the mode with the
highest frequency. For transverse oscillations, this is a COM-like mode, which
is given here by the in-phase oscillation of local COM modes of the outermost
pairs. The second-highest frequency mode with $n = 2$, in turn, corresponds to a
superposition of the local COM modes of the outermost pairs with opposite
phase. For this example, due to the reflection symmetry with regard to the
center of the trap, these hybridized modes are equal superpositions of
oscillations of pairs to the left and right of the trap center. Superpositions
of local COM modes are followed at lower frequencies by superpositions of local
stretch modes. 

The phonon mode spectrum of the chain, both with and without tweezers, is shown
in Fig.~\ref{fig:phonon_modes_spectra_finite}(c). The spectra of oscillations in
the longitudinal $z$ and transverse $x$ direction, which are shown in orange and
blue, respectively, are strongly modified in the presence of tweezers with
strength $\nu_0 = \omega_0/\omega_x = 0.4$: The spectrum of oscillations along
the trap $z$ axis is gapped~\cite{Shen2019}, and the almost dense set of
frequencies splits up into several subsets. Most prominently, both for the $z$
and $x$ modes, a subset of modes, which correspond to hybridized COM and
stretch modes of the pinned pairs, appear shifted above the remaining mode
frequencies. For the transverse $x$ direction, the assignment between mode
frequencies and mode vectors is indicated with blue and green shading. The
spectrum of transverse $y$ modes, which is shown in green in
Fig.~\ref{fig:phonon_modes_spectra_finite}(c), is not affected by the tweezers because we neglect the trapping along the direction of the tweezer beam.

\subsection{Phonon band structure for infinite chains}
\label{sec:phon-band-struct-infinite}

\begin{figure}
  \centering
  \includegraphics[width=.9\linewidth]{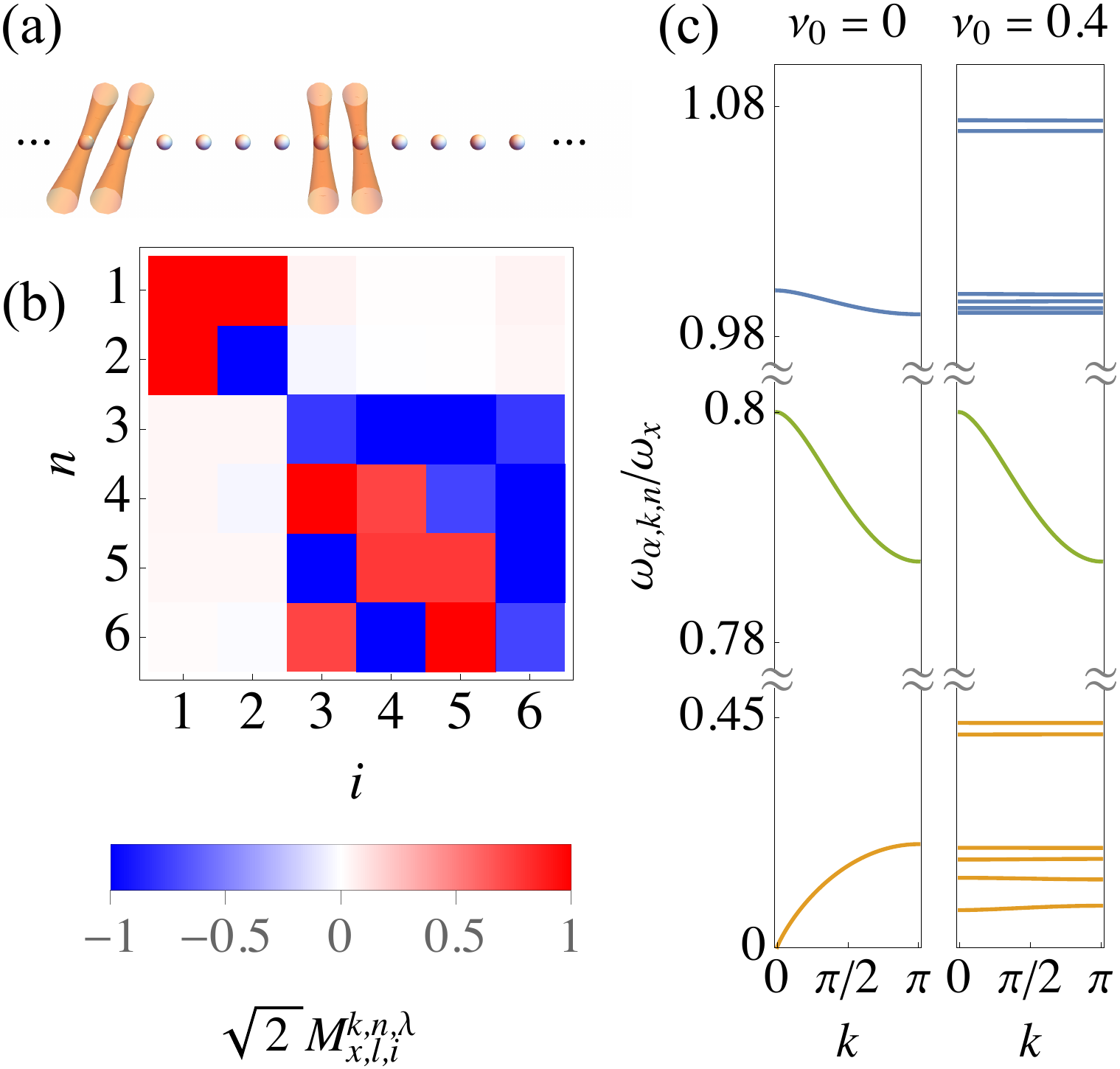}
  \caption{(a) Infinite ion chain with optical tweezers forming a periodic array
    with unit cell size $p = 6$. (b) Mode matrix $M_{x, l, i}^{k, n, \lambda}$
    for $l = k = 0$ and $\lambda = 1$. COM and stretch modes of the pinned ions
    are clearly decoupled from the ions which are not pinned. (c) Phonon mode
    spectra for an infinite chain without and with optical tweezers, for $\nu_0 = 0$ and $\nu_0 = 0.4$, respectively. Transverse $x$ and $y$ modes and
    longitudinal $z$ modes are shown, respectively, in blue, green, and
    orange. The
    spatial periodicity imposed by tweezers causes the dense bands of the $x$
    and $z$ modes in the infinite system to split up into $p = 6$ bands, where
    the highest two bands correspond to COM and stretch modes of pinned pairs.}
  \label{fig:phonon_modes_spectra_infinite}
\end{figure}

To demonstrate the inherent scalability of our approach, we also present
theoretical results for tweezer gates in infinite ion chains as illustrated in
Fig.~\ref{fig:phonon_modes_spectra_infinite}(a). In the simultaneous limit
$N \to \infty$ and $\omega_z \to 0$, the system acquires discrete translational
invariance under the transformation $i \mapsto i + p$, where $p$ is the size of
the unit cell of the spatially periodic arrangement of tweezers. We note that
discrete translational symmetry is also realized in ring traps~\cite{Li2017}. To account for
this translational invariance, it is convenient to label the positions of ions
as $\left( l, i \right)$ with unit cell index $l \in \Z$ and position
$i \in \{ 1, \dotsc, p \}$ within a unit cell. The ions at positions
$i \in \{ 1, 2 \}$ are pinned whereas the remaining ions at
$i \in \{ 3, \dotsc, p \}$ are not pinned.

The normal modes of a periodic ion crystal can be found by using concepts from
band theory of electrons in solids as detailed in
Appendix~\ref{sec:phon-band-struct}. In particular, translational invariance with
respect to the unit cell index $l$ can be accounted for by representing the
phonon modes as plane waves $\sim \e^{\imag k l}$ with quasimomentum 
$k$. Because the phonon mode vectors are real, the quasimomentum can be
restricted to the interval $k \in [0, \pi]$.  We note that the plane wave
representation implies that phonon modes are stretched out over the entire
chain. In other words, the localized COM and stretch modes of pinned pairs in
different unit cells hybridize uniformly.

The phonon mode structure for infinite systems is illustrated in terms of the
mode matrix $M^{k, n ,\lambda}_{\alpha, l, i}$ (see Appendix~\ref{sec:phon-band-struct}) in
Fig.~\ref{fig:phonon_modes_spectra_infinite}(b) for $\alpha = x$, $l = 0$,
$k = 0$, and $\lambda = 1$. Within a unit cell, there is a clear separation
between modes $n = 1$ and $n = 2$ which correspond, respectively, to COM and
stretch oscillations of the pinned ions, and the modes with
$n \in \{ 3, \dotsc, p \}$ which involve the ions which are not pinned.

As shown in Fig.~\ref{fig:phonon_modes_spectra_infinite}(c), modifications of
the mode frequencies due to tweezers are particularly clear in infinite systems:
The bands which are formed by $z$ and $x$ modes split up into $p = 6$ bands in
the presence of tweezers. In particular, COM and stretch modes of the pinned
pairs hybridize between unit cells to form bands. The widths of these bands are
vanishingly small on the scale of the figure. Moreover, close inspection reveals
that the width of the stretch band is smaller than the width of the COM band by
one order of magnitude. These features of the COM and stretch bands can be
understood in terms of perturbation theory in the small parameter
$\epsilon^2/\nu_0^2$ as we show in
Appendix~\ref{sec:phon-band-struct}. The perturbative
treatment shows that the widths of the COM and stretch bands are
$\sim \epsilon^2 \omega_x/p^3$ and $\sim \epsilon^2 \omega_x/p^5$,
respectively. That is, the widths are suppressed with the size of the unit cell
$p$, with an even stronger suppression for the stretch band.


\section{Entangling tweezer gates}
\label{sec:entangl-gates-minimal-control}

\begin{figure}
  \centering
  \includegraphics[width=.8\linewidth]{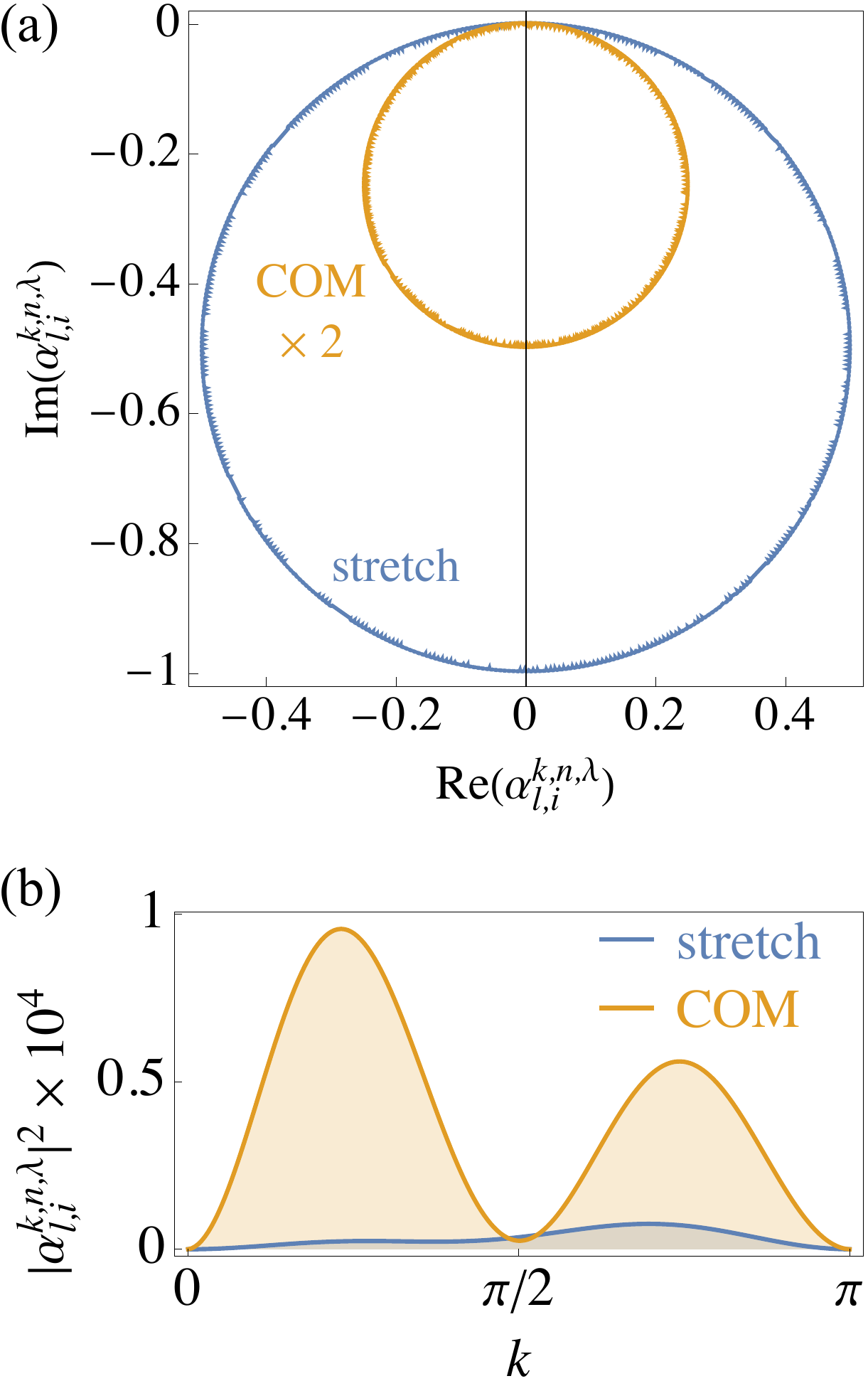}
  \caption{Phase space trajectories for parallel entangling gates in an infinite
    chain. (a) Contribution to the displacement of the phonon mode $k, n, \lambda$ due to its
    coupling~\eqref{eq:alpha} to the qubit at position $l = i = 1$, which is assumed to be in the $+ 1$ eigenstate of $\sigma^x_{l, i}$. For the chosen detuning close to
    $\mu_{\mathrm{stretch}}$ in Eq.~\eqref{eq:mu}, the stretch mode with
    $n = \lambda = 2$ and $k \approx 1.55$ such that $\omega_{k, 2} = \omega_2$
    equals the mean frequency of the stretch band experiences the strongest
    displacement. The phase space trajectory of the COM mode with the same
    values of $k$ and $\lambda$ forms a smaller circle which is traversed
    twice. Trajectories of modes with $n \geq 3$ are not visible on the scale of
    the figure. (b) Closer inspection reveals that the phase space curves for
    different values of $k$ are not perfectly
    closed. The area under the squared modulus of the qubit-phonon coupling at the
    end of the gate operation, shown here for $n = \lambda = 2$, determines the
    infidelity~\eqref{eq:infidelity}. Parameters of the ion chain are $p = 6$,
    $\nu_0 = 0.4$, and $\epsilon = 0.07$.}
  \label{fig:gates_not_optimized_infinite_phase_space}
\end{figure}

The optical design of phonon modes as described above forms the basis to implement parallel two-qubit entangling gates. In the following, we discuss the requirements for and performance of tweezer gates in infinite as well as finite ion chains.


\subsection{Infinite chains}
\label{sec:gates-not-optimized-infinite}

We consider a periodic array of pinned pairs of ions which are separated by
$p - 2$ spectator ions, and we aim at performing two-qubit entangling gates on
all pairs of pinned ions in parallel. As illustrated in
Fig.~\ref{fig:phonon_modes_spectra_infinite}(b) the spectrum of transverse $x$ phonon modes comprises $p$ bands with mean frequencies $\omega_n$ and bandwidths $\Delta_n$.

We aim at implementing gates using dominantly the COM or stretch bands with mean frequencies $\omega_1 = \omega_{\mathrm{COM}}$ and $\omega_2 = \omega_{\mathrm{stretch}}$ respectively.
This can be achieved if the resolved-sideband condition
\begin{align}
  \label{eq:resolved-sideband}
  \abs{\mu - \omega_n} \ll \mu, \omega_n,
\end{align}
is met where $\mu > \omega_1$ for the COM band with $n = 1$, and $\mu < \omega_2$ for
the stretch band with $n = 2$. The choice of tuning above the COM or below the
stretch band ensures that the detuning $\mu$ is as far away as possible from the
respective other band which should not be excited.

Note that contributions
from higher bands with $n \in \{ 3, \dotsc, p \}$ to the
qubit-phonon coupling~\eqref{eq:alpha} and qubit-qubit coupling~\eqref{eq:chi} are
strongly suppressed: First, these bands are far detuned with
$\mu - \omega_n \gtrsim \omega_0$ with the frequency of the optical potential
$\omega_0$; Second, the contribution of these bands to the
couplings~\eqref{eq:alpha} and~\eqref{eq:chi} is
proportional to $\lvert \eta^{k, n, \lambda}_{l, i} \Omega_i \rvert/\omega_0$,
which is suppressed in the limit of strong pinning by a factor of
$\eta^{k, n, \lambda}_{l, i} \sim \epsilon^2/\nu_0^2$ for pinned ions at
positions $\left( l, i \right)$ as follows from the perturbative treatment presented in
Appendix~\ref{sec:phon-band-struct}. Moreover, as detailed
below we find that typically $\Omega_i$ is smaller than $\omega_0$ by an order of
magnitude. We note that due to the proportionality to $\Omega_i$, by setting
$\Omega_i = 0$ for the ions which are not pinned, these do not contribute to the
infidelity and the crosstalk.

The precise values of $\mu$ and, in particular, $\tau$, follow from the condition of minimal infidelity. For an isolated pair of ions with COM and stretch mode frequencies $\omega_1$ and $\omega_2$, respectively, this can be achieved for~\cite{Sorensen2000,
  Zhu2006a}
\begin{equation}
  \label{eq:mu-tau}
  \left( \mu - \omega_n \right) \tau = 2 \pi l_n,
\end{equation}
where $l_n$ is an integer. Particular choices for gates which use dominantly the
COM and stretch modes are given by, respectively, $l_1 = 1$ and $l_2 = 2$, and
$l_1 = - 2$ and $l_2 = - 1$. The resulting detunings are
\begin{equation}
  \label{eq:mu} 
  \mu_{\mathrm{COM}} = 2 \omega_1 - \omega_2, \qquad \mu_{\mathrm{stretch}} = 2
  \omega_2 - \omega_1,
\end{equation}
and in both cases the gate duration $\tau = 2 \pi/(\omega_1 - \omega_2)$
is set by the mode splitting. These choices of detunings and gate duration ensure that the displacement of the phonon modes due to the qubit-phonon coupling~\eqref{eq:alpha}, when considered as a function of the upper limit of integration
$\tau' \in [0, \tau]$, performs a closed loop in phase space~\cite{Milburn1999, Molmer1999, Sorensen1999, Sorensen2000}.

In the present case of an infinite chain, condition Eq.~\eqref{eq:mu-tau} remains
valid in the case of two narrow bands, if the bandwidths $\Delta_n$ are sufficiently small
in the sense that $\Delta_n \tau \ll 1$. However, clearly this condition cannot
be met for all frequencies $\omega_{k, n}$ which form a band and correspond to
different quasimomenta $k$. For the results we show below, we first fix $\tau$
according to Eq.~\eqref{eq:mu-tau} for either the COM or the stretch band by
setting either $l_1 = 1$ or $l_2 = -1$ and by choosing $\omega_n$ as the mean
frequency of the respective band, and we then determine numerically the value of
$\mu$ which yields the lowest infidelity, which typically deviates only slightly
from the values given in Eq.~\eqref{eq:mu}. As above, the Rabi frequency
$\Omega_{l, i} = \Omega_0$ on the pinned ions is chosen according to the
condition $\chi^{l, l'}_{i, i'} = \pm \pi/4$ if $\left( l, i \right)$ and
$\left( l', i' \right)$ are neighboring pinned ions, i.e., $l' = l$ and
$i' = i + 1$. For the remaining ions which are not pinned, we set
$\Omega_{l, i} = 0$.

We first consider the implementation of parallel gates which use the stretch
band in an ion chain with $p = 6$, $\epsilon = 0.07$ and $\nu_0 = 0.4$. To minimize the infidelity we choose  $\mu/\omega_x \approx 1.065$ and $\omega_x \tau \approx 1.37 \times 10^3$ as described above. The required Rabi frequency to perform a maximally entangling gate is given by
$\eta_0 \Omega_0/\omega_x \approx 4.75 \times 10^{-3}$, where the dimensionless factor
\begin{equation}
    \label{eq:eta-0}
    \eta_0 = k_{\mathrm{L}, x} \sqrt{\hbar/(2 m \omega_x)}
\end{equation}
contains all parameters in the definition of the Lamb-Dicke parameter matrix~\eqref{eq:Lamb-Dicke} which are specific to different ionic species.

In panel Fig.~\ref{fig:gates_not_optimized_infinite_phase_space}(a), we show the
qubit-phonon coupling~\eqref{eq:alpha} as a function of the upper limit of integration
$\tau' \in [0, \tau]$ and for $l = i = 1$, $n = 1, 2$, and $k \approx 1.55$ which corresponds
to the mean frequency $\omega_2$ of the stretch band. According to Eq.~\eqref{eq:U-gate}, the actual displacement depends on the state of the qubit at position $\left( l, i \right)$, and is opposite for the states $\lvert \pm \rangle_i$ with
$\sigma^x_i \lvert \pm \rangle_i = \pm \lvert \pm \rangle_i$. The values of $n$ and
$k$ shown in Fig.~\ref{fig:gates_not_optimized_infinite_phase_space}(a) yield the largest values of the qubit-phonon coupling, i.e., the corresponding modes
contribute most to the gate. On the scale of the figure, the qubit-phonon coupling for
bands with $n \geq 3$ is not visible. The physical picture described below
Eq.~\eqref{eq:mu} is clearly reflected in the figure: The phase space
trajectories for both the stretch and the COM modes shown in the figure are
closed, where the trajectory of the COM mode is traversed twice. However, closer
inspection of the vicinity of the origin reveals that the trajectories of modes
which belong to the COM band and have different values of the quasimomentum $k$
do not close perfectly. In panel Fig.~\ref{fig:gates_not_optimized_infinite_phase_space}(b) we show the qubit-phonon coupling at the end of the gate operation as a function of $k$ for both
the stretch and the COM bands. According to Eq.~\eqref{eq:infidelity}, the area
under these curves determines the infidelity, and the figure shows that the
dominant contribution to the infidelity is indeed due to the COM modes. We find
$\delta F = 5.7 \times 10^{-4}$.

The infidelity is higher for gates which use predominantly the COM band. This is
because the width of the COM band is much larger than the width of the stretch
band and, therefore, the phase space trajectories for different $k$ deviate more
strongly from perfect closure. In particular, for $\mu/\omega_x \approx 1.079$
slightly below the COM band, $\omega_x \tau \approx 1.37 \times 10^3$, and
$\eta_0 \Omega_0/\omega_x \approx 4.77 \times 10^{-3}$, we obtain
$\delta F = 2.1 \times 10^{-3}$ which is higher than the value we obtain for the
stretch band by an order of magnitude.

We next analyze over-/underrotation errors and crosstalk as defined in Eqs.~\eqref{eq:inaccuracy} and~\eqref{eq:crosstalk},
respectively. For the implementation of gates we discuss in this section, the
Rabi frequency is chosen such that $\chi_{i, i'} = \chi_{i, i'}^0$ for
$\left( i, i' \right) \in I$, i.e., the over-/underrotation error vanishes
exactly, $\delta \chi = 0$.

However there is unwanted crosstalk corresponding to nonzero qubit-qubit couplings between ions
which belong to distinct pairs. The dominant contributions to crosstalk are due
to (i) unwanted excitation of the stretch or COM band, and (ii) the finite width
of these bands. Concerning (i), we note that for a given detuning, say
$\mu > \omega_1$, the COM and stretch bands induce ferromagnetic and
antiferromagnetic couplings. In other words, they yield contributions to
$\chi^{l, l'}_{i, i'}$ with opposite sign, which thus partially cancel each
other. This partial cancellation has to be compensated by increasing the Rabi
frequency $\Omega_0$ to achieve $\chi^{l, l'}_{i, i'} = \pm \pi/4$ on the target
ions, which then, however, also increases unwanted couplings which contribute to
the crosstalk~\eqref{eq:crosstalk}. This effect is suppressed by tuning close to
a sideband Eq.~\eqref{eq:resolved-sideband}, i.e., by dominantly exciting either
the COM or the stretch band.

With regard to (ii), the fact that a finite bandwidth leads to crosstalk can be
understood intuitively from the fact that the effective coupling between local
(within single unit cells) COM or stretch modes determines the
bandwidth. Crosstalk is negligible if the effective coupling and thus the
bandwidth is much smaller than all other relevant scales. In particular, the
finite bandwidth can be neglected and $\omega_{k, n}$ for the COM and stretch
bands can be replaced by the respective central frequencies $\omega_n$ in
Eq.~\eqref{eq:chi} if
\begin{equation}
  \abs{\mu - \omega_n} \gg \Delta_n.
\end{equation}
We note that by Eq.~\eqref{eq:resolved-sideband} this also implies that
$\mu, \omega_n \gg \Delta_n$. If we combine this condition with
Eq.~\eqref{eq:mu-tau} for the gate duration, we find the intuitive criterion that
to minimize crosstalk the gate should be performed fast on the timescale which
is set by the effective coupling between pinned pairs. This picture generalizes the concept of local oscillation modes from single ions~\cite{Duan2004} to pairs of ions. We note that the mentioned
condition can always be met efficiently by increasing the size of the unit cell
$p$. 

For the gate shown in
Fig.~\ref{fig:gates_not_optimized_infinite_phase_space}, we find
$C = 4.1 \times 10^{-2}$. The crosstalk for a gate using the COM band and with
$p = 6$ is $C = 1.7 \times 10^{-1}$, i.e., again one order of magnitude higher
than for the stretch band. The crosstalk can be reduced by increasing the unit
cell size $p$. In particular, for gates which use the stretch band and for
$p \geq 9$, we obtain $C < 10^{-2}$ such that $C/\chi_{i, i'} < 1 \, \%$ for
$\left( i, i' \right) \in I$.

Finally, we note that while the long-range character of the residual Coulomb
interaction has the adverse effect of leading to crosstalk between distant
qubits, it can also be utilized to implement gates between qubits which are not
nearest neighbors. This can be achieved, for example, by pinning the ions at
positions $1$ and $q$ where $2 < q \ll p$ within each unit cell. This leads to a
reduction of the splitting between the COM and stretch bands by a factor of
$\sim 1/(q - 1)^3$, and the gate duration increases correspondingly according to
Eq.~\eqref{eq:mu-tau}.

\subsection{Gate performance}
\label{sec:gate-performance}

\begin{figure}
  \centering
  \includegraphics[width=\linewidth]{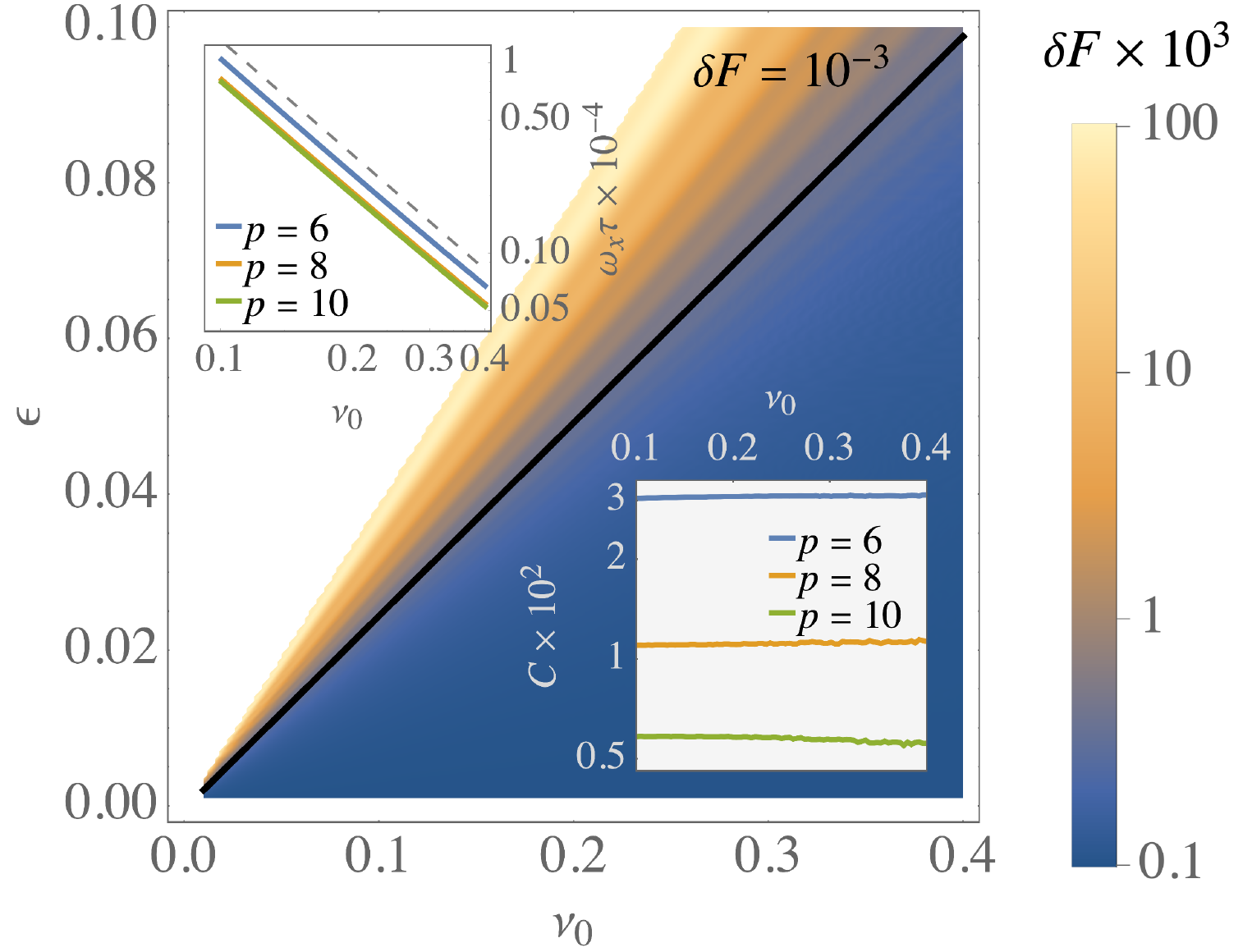}
  \caption{Gate performance for an infinite chain. The main panel shows the
    dependence of the infidelity on the parameters $\nu_0$ and
    $\epsilon$. $\delta F = 10^{-3}$ on the black line across the diagonal. In
    the white region, the optical potential is insufficient to decouple the
    pinned ions from the other ions in the chain. The insets show the gate duration
    $\tau$ and the crosstalk $C$ for fixed infidelity $\delta F = 10^{-3}$.}
  \label{fig:gate_performance_not_optimized}
\end{figure}

We study how infidelity, crosstalk and gate duration depend on the tweezer pinning strength as
measured by the dimensionless optical trapping frequency $\nu_0$, and the
strength of the residual Coulomb coupling $\epsilon$ in
Fig.~\ref{fig:gate_performance_not_optimized}. The main panel of this figure
shows the infidelity $\delta F$ for $p = 6$ and for optimized choices of gate
duration and detuning as discussed above. For large values of $\epsilon$ in the
white region, the gap which separates the COM and stretch bands is more than
half of the gap which separates the stretch band from the bands of the ions
which are not pinned and, therefore, the pinned ions are not sufficiently decoupled
from the ions which are not pinned. In this region, gates cannot be performed
with reasonable infidelity. High-fidelity gates can be performed for values of
$\epsilon$ below the black diagonal line, which corresponds to a threshold value
of $\delta F = 10^{-3}$.

We next analyze how the gate speed and crosstalk are affected by the optical
trapping frequency $\nu_0$ for a fixed value of the infidelity
$\delta F = 10^{-3}$, i.e., for $\epsilon$ on the black diagonal line in
Fig.~\ref{fig:gate_performance_not_optimized}. The inset in the upper left
corner of the figure shows the gate duration for various values of the unit cell
size $p$. The gate duration
is set by the splitting between COM and stretch bands, which according to
Eq.~\eqref{eq:nu-n} is proportional to $\epsilon^2$ for $\epsilon \ll 1$. The resulting analytical prediction of the scaling
$\tau \sim 1/\epsilon^2 \sim 1/\nu_0^2$ on the line $\delta F = 10^{-3}$ is
shown as dashed gray line in the inset and agrees well with the numerical
data. (The vertical offset between the gray line and the numerical data is
introduced to improve the visibility.) For $\nu_0$ in the range from $0.1$ to
$0.4$, we obtain gate durations between $\omega_x \tau = 0.05 \times 10^4$ and
$10^4$. For a typical value of $\omega_x = 2 \pi \times 3 \, \mathrm{MHz}$,
this corresponds to gate durations ranging from $27 \, \mu \mathrm{s}$ to
$531 \, \mu \mathrm{s}$.

The crosstalk for fixed infidelity $\delta F = 10^{-3}$ is shown in the inset in
the lower right corner in Fig.~\ref{fig:gate_performance_not_optimized}. While the crosstalk remains approximately constant as a function of $\nu_0$, it can be suppressed by increasing the unit cell size $p$. As already stated above, we find $C < 10^{-2}$ for $p \geq 9$.

In an experimental implementation, an important contribution to the infidelity is due to spontaneous scattering of photons of the tweezer beams. As discussed in Appendix~\ref{sec:feasibility_study}, promising candidates to implement tweezer gates are $^{24}\mathrm{Mg}^+$ ions, for which the infidelity due to scattering of photons is on the order of $10^{-3}$ for reasonable dimensionless parameters $\epsilon$ and $\nu_0$. Tweezer parameters and scattering infidelities for several ionic species are summarized in Table~\ref{tab:tweezer_parameters}. The infidelity due to scattering can be decreased further by decreasing the tweezer intensity and thus the dimensionless optical trapping frequency $\nu_0$. As illustrated in Fig.~\ref{fig:gate_performance_not_optimized}, this results in an increase of the gate infidelity which, however, can be compensated by employing optimal coherent control as described in Sec.~\ref{sec:entangl-gates-time-mod-Rabi}.

\begingroup
\setlength{\tabcolsep}{10pt}
\begin{table*}[t]
\begin{tabular}{l|cccc}
   & Wavelength (nm) & Optical power (mW) & Scattering infidelity $\times 10^3$ & Inter-ion distance ($\mu$m) \\ \hline
$^{24}\mathrm{Mg}^+$ & 400             & 6.4        & 4.9                                 & 15                          \\
$^{40}\mathrm{Ca}^+$ & 532             & 14.5       & 12.0                                & 12.6                        \\
$^{88}\mathrm{Sr}^+$ & 580             & 40.2       & 30.2                                & 9.7                         \\
$^{171}\mathrm{Yb}^+$ & 532             & 202.2      & 38.2                                & 7.8                         \\
$^{138}\mathrm{Ba}^+$ & 675             & 90.0       & 55.0                                & 8.3                        
\end{tabular}
\caption{Tweezer parameters and scattering-induced infidelity for ground-state qubit encoding in different ionic species. The parameters $\epsilon = 0.07$, $\nu_0 = 0.4$ and $\omega_x = 2\pi \times 3$\,MHz are fixed for all species, and the numerical aperture was assumed to be 0.7 (see Appendix~\ref{sec:feasibility_study} for details). The tweezer wavelength is chosen to result in a close-to-optimal scattering infidelity while being experimentally accessible in terms of both power per tweezer spot in the second column and available laser sources.}
\label{tab:tweezer_parameters}
\end{table*}
\endgroup

\subsection{ Finite chains }
\label{sec:gates-not-optimized-finite}

\begin{figure}
  \centering
  \includegraphics[width=\linewidth]{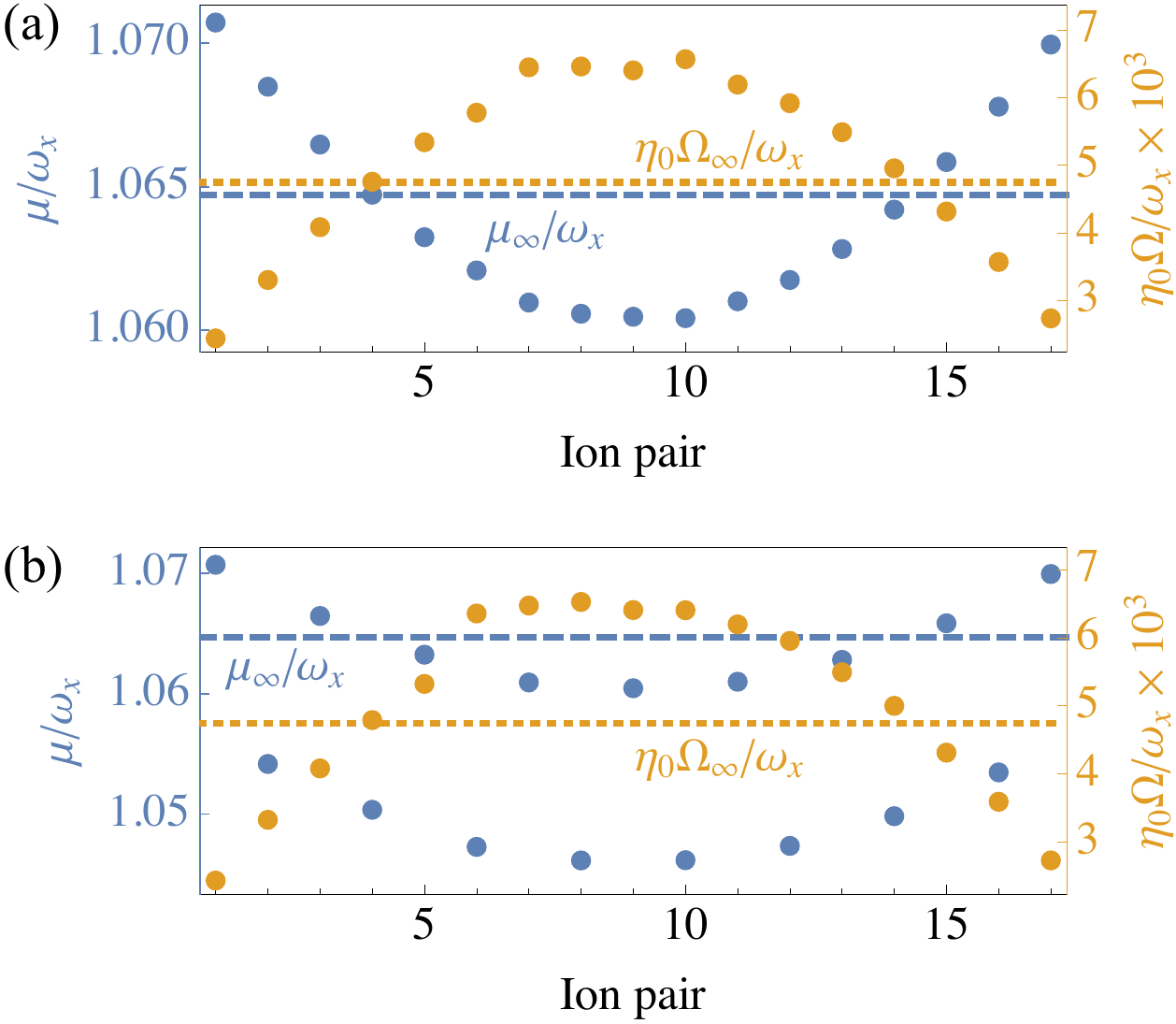}
  \caption{Tweezer gates in finite chains. We show individual detunings and Rabi frequencies for $17$ maximally entangling tweezer gates targeting the stretch mode along a chain of $130$ ions with $15$ buffer ions on each side ($p=6$, $\epsilon \approx 0.07$). Optical tweezers generate a trapping frequency that is uniform $\nu_0 = 0.4$ and alternating between $\nu_{01}=0.4$ and $\nu_{02}=0.36$ for panels (a) and (b), respectively. For comparison, we also show the detuning $\mu_{\infty}$ and Rabi frequency $\Omega_{\infty}$ for an infinite chain with $\epsilon = 0.07$ and $\nu_0 = 0.4$.
  }
  \label{fig:gates_not_optimized_finite}
\end{figure}

Here we show how the methods for implementing parallel tweezer gates can be applied in finite 1D ion strings. For concreteness, we assume harmonic trapping along the trap $z$ axis. In contrast to before the Hamiltonian in question is no longer invariant under discrete translations, which leads to stark changes in the mode spectrum as well as the mode functions themselves (see the discussion around Fig.~\ref{fig:phonon_modes_spectra_finite}). As we point out further below this is not necessarily a negative aspect and indeed further improves the gate performance. For this, the detunings $\mu_i$, the Rabi frequencies $\Omega_i(t)$ in Eq.~\eqref{eq:H-qubit-ph}, and the gate duration have to be controlled for each ion pair. This is due to the fact that for each pair of target ions and in order to fulfill Eq.~\eqref{eq:mu-tau} one has to identify the corresponding COM and stretch mode and choose the detuning and the gate duration accordingly. These deviations are strongest for ions at the ends of the chain where the inter-ion distances are considerably larger than for the rest of the chain. Therefore, we introduce several buffer ions on each end which are not used for quantum computation. We choose the number of buffer ions such that the relative standard deviation for the inter-ion distances of the other ions lies below $10 \, \%$. Furthermore, we choose the axial trapping frequency $\omega_z$ such that the mean inter-ion distance of the non-buffer ions corresponds to $\epsilon=0.07$, i.e., the value that we chose for the infinite system. 

In Fig.~\ref{fig:gates_not_optimized_finite} we show numerical results for a system of $130$ ions, $15$ of which on each end of the chain are used as buffer ions. We choose $p=6$ to implement a total number of $17$ maximally entangling gates in parallel, where the leftmost and the rightmost gates are performed on the ions at positions $i = 16, 17$ and $i = 112, 113$, respectively. This configuration of gates is not symmetric with respect to the center of the trap. However, this asymmetry does not affect the gate performance: Indeed, the performance of the parallel gates for the asymmetric configuration, which we discuss in detail below, is comparable to the gate performance for the symmetric configuration that can be realized by shifting all tweezers by one ion to the right while keeping the total number of gates at $17$. 

The two panels in Fig.~\ref{fig:gates_not_optimized_finite} correspond to different schemes of optical pinning by the tweezers: In panel (a), each pair of target ions is pinned with the same tweezer strength $\nu_0=0.4$, whereas in panel (b), tweezer strengths alternate between the values $\nu_{01}=0.4$ and $\nu_{02}=0.36$. For both cases we show the detuning and the required Rabi frequency for each gate along the chain. As for the infinite chain, the detunings are optimized with regard to optimal infidelity around the estimate given by Eq.~\eqref{eq:mu} for the stretch mode of each pair. The required detunings lie around the value obtained in the infinite case (blue dashed line in Fig.~\ref{fig:gates_not_optimized_finite}) but are lower in the center and higher at the edge of the chain. This is due to the disparity in the mode frequencies: As indicated in Fig.~\ref{fig:phonon_modes_spectra_finite}, modes that are localized at the edge of the ion chain have higher frequencies and smaller gaps compared to those at the center. The maximal gate duration $\omega_x \tau = 2670$ is the same for Fig.~\ref{fig:gates_not_optimized_finite}(a) and Fig.~\ref{fig:gates_not_optimized_finite}(b) and occurs at the edge of the ion chain. This can be attributed to the difference in the inter-ion distances that influence the gap between the COM and the stretch mode of a given ion pair: Larger distances implicate weaker interactions and hence a smaller gap leading to slower gates and \textit{vice versa} (see Eq.~\eqref{eq:mu-tau}).
Furthermore, since the gate duration is larger at the edge, the Rabi frequency, which is also shown in Fig.~\ref{fig:gates_not_optimized_finite}, has to be lower in order to realize a maximally entangling gate.

We note that while the choice of Rabi frequencies shown in Fig.~\ref{fig:gates_not_optimized_finite} leads to maximally entangling gates with $\chi_{i, i + 1} = - \pi/4$ for all pairs of qubits, different and independent values of $\chi_{i, i + 1}$ can be achieved by lowering the individual Rabi frequencies.

As before we quantify the performance of the gate through average infidelity Eq.~\eqref{eq:infidelity} and average crosstalk Eq.~\eqref{eq:crosstalk}. For panels (a) and (b) in Fig.~\ref{fig:gates_not_optimized_finite} we get $\delta F_{a} = 9 \times 10^{-4}$, $C_a = 2.8 \times 10^{-2}$ and $\delta F_{b} = 1.7 \times 10^{-3}$, $C_b = 6.5 \times 10^{-3}$ respectively. If we compare these values with those obtained in the infinite chain, i.e., $\delta F_\infty = 5.7 \times 10^{-4}$ and $C_\infty = 4.1 \times 10^{-2}$, we find that while the infidelity is slightly worse, crosstalk is slightly better in the finite case. The former is due to rare outliers with infidelities of order $10^{-2}$ in the middle of the chain whereas the latter stems from the variation in mode frequency (and hence detuning) and the tighter localization of the COM and stretch modes for the different gates. The alternating tweezer frequencies in Fig.~\ref{fig:gates_not_optimized_finite}(b) amplify the latter effect which leads to further suppression of crosstalk.

\subsection{Infidelity and over-/underrotation errors from tweezer misadjustments}
\label{sec:tweezer-imperfections}

\begin{figure}
  \centering  
  \includegraphics[width=\linewidth]{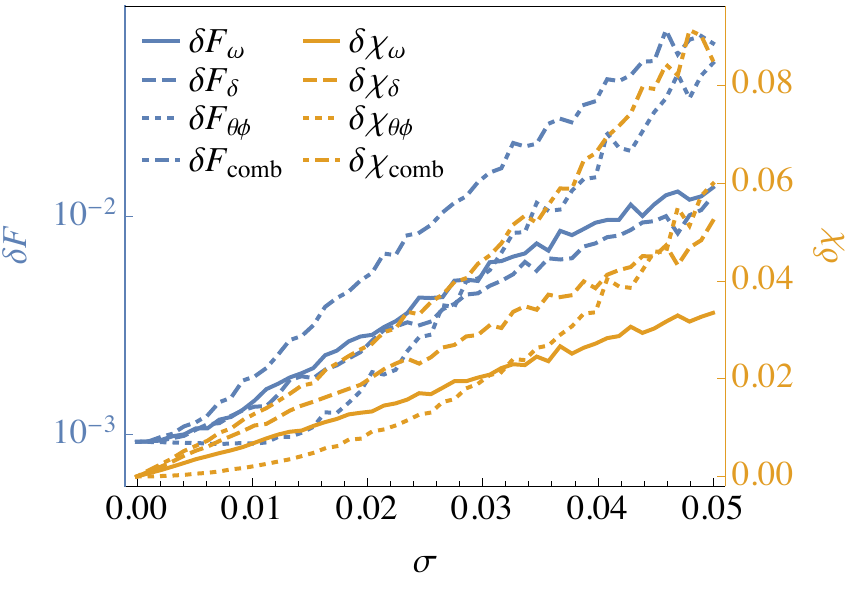}
  \caption{Tweezer misadjustments. We show the infidelity~\eqref{eq:infidelity} (blue) and the over-/underrotation error~\eqref{eq:inaccuracy} (orange) as a function of the strength of fluctuations $\sigma$ for three types of misadjustments and for their combination. For each type of misadjustment, the infidelity and the over-/underrotation error are averaged over $40$ realizations. We consider here a chain of $130$ ions with $15$ buffer ions on each side, and parameters $p=6$, $\epsilon \approx 0.07$, and $\nu_0 = 0.4$.}
  \label{fig:imperfections}
\end{figure}

An important question for the experimental implementation of tweezer gates concerns the sensitivity of the gate performance to misadjustments of the optical tweezer array. To address this question, we consider three distinct types of misadjustments: Deviations of the focuses of the tweezers from the equilibrium positions of the ions, variations in the optical trapping frequencies due to intensity fluctuations of the tweezers, and misalignment of the tweezers with the $y$ direction. 
As figures of merit, we study how these imperfections affect the infidelity as well as the over-/underrotation error. 

Including tweezer misadjustments in the optical potential~\eqref{eq:V-twz} for ion $i \in \{ 1, \dotsc, N \}$ yields
\begin{equation}
  \label{eq:V-twz-imp}
  V^{\mathrm{twz}}_i(\mathbf{r}_i) = \frac{m}{2} \left( \omega_{0, i} + \delta \omega_i \right)^2 
  \left\lvert \Pi(\theta_i,\phi_i)  \left(\delta \mathbf{r}_{i} - \boldsymbol{\delta}_i \right) \right\rvert^2.
\end{equation}
Here, $\delta \omega_i$ is a shift of the optical trapping frequency, $\boldsymbol{\delta}_i$ is the deviation of the focus of the tweezer from the equilibrium position of the ion $\mathbf{r}_{i,0}$ in the absence of an optical potential, and $\delta \mathbf{r}_i = \mathbf{r}_i - \mathbf{r}_{i,0}$ is the displacement of the ion from $\mathbf{r}_{i,0}$. In Appendix~\ref{sec:appendix-gate-imperfections}, we explain how the shift of the equilibrium position of the ion due to deviations of the focus of the tweezer from $\mathbf{r}_{i,0}$ can be calculated perturbatively. Finally, the angles $\theta_i$ and $\phi_i$ describe the misalignment of the tweezer beam with the $y$ axis, and $\Pi (\theta_i,\phi_i)$ is the projector onto the plane orthogonal to the tweezer beam axis. We assume that fluctuations of the parameters $\delta \omega_i$, $\boldsymbol{\delta}_i$, $\theta_i$, and $\phi_i$ are normally distributed around zero, independent for each ion, and constant on the timescale of gate operations.

Figure~\ref{fig:imperfections} shows the infidelity and over-/underrotation error as a function of the strength of misadjustments. In particular, to generate the data shown in the figure, 40 samples of each of the dimensionless parameters $50 \delta \omega_i/\omega_{0,i}$, $\boldsymbol{\delta}_i/l_0$ where $l_0 = (e^2/4 \pi \epsilon_0 m \omega_x^2)^{1/3}$, $\theta_i$, and $\phi_i$ are drawn from a Gaussian distribution with width $\sigma$. The factor of $50$ for shifts of the optical trapping frequency is introduced so that the infidelities and over-/underrotation errors are comparable for all types of misadjustments on the range of values of $\sigma$ shown in Fig.~\ref{fig:imperfections}. The assumption of those misadjustments being constant during the execution of a gate is motivated by the observation that beam pointing instabilities as well as laser intensity noise typically fall off rapidly above tens to hundreds of Hz~\cite{kanai2008beampointing, seifert2006intensitynoise}. If we require $\delta F \lesssim 10^{-2}$ and $\delta \chi \lesssim 4 \times 10^{-2}$, this allows for standard deviations $\sigma$ of approximately 0.04 to 0.05 (the combination of all three types of misadjustments leads to the slightly more stringent requirement $\sigma \lesssim 0.02$ to 0.03), which can be related to misadjustments of $70 \,  \mathrm{nm}$ for the tweezer focuses, $2^\circ$ for the incidence angles, and $10 \, \mathrm{kHz}$ for the pinning frequencies, corresponding to relative intensity errors of $3 \times 10^{-3}$ in a typical experiment with $d = 10 \, \mu \mathrm{m}$ and $\omega_x = 2 \pi \times 3 \, \mathrm{MHz}$. All three conditions can be satisfied in state-of-the-art experiments.

\subsection{Dynamical reconfiguration of tweezer arrays}
\label{sec:dynam-reconf-tweez}

To perform consecutive layers of the quantum circuit shown in
Fig.~\ref{fig:schematic}(d) the tweezer array has to be reconfigured
dynamically. In particular, the second layer in the circuit in
Fig.~\ref{fig:schematic}(d) can be implemented by switching off the tweezers
which are focused on the ions which are affected by the gates $U_1$, $U_2$, and
$U_3$, and by subsequently switching on optical tweezers focused on
the equilibrium positions of the ions which are affected by the gates $U_4$ and
$U_5$.

The dynamical reconfiguration of the optical tweezer array can cause heating by
exciting phonon modes. Crucially, throughout the switching process, the phonon
spectrum remains gapped, i.e., the smallest phonon frequency is larger than zero as in the right panel in
Fig.~\ref{fig:phonon_modes_spectra_infinite}(c), and heating is suppressed if
the switching is performed adiabatically with respect to the phonon gap. Based
on adiabatic perturbation theory, we derive conditions for adiabaticity for the
worst-case scenario of an infinite ion chain in
Appendix~\ref{sec:adiab-switch-tweez}. In this derivation, we assume for
simplicity that all phonon modes are cooled to their ground state. We consider a
switching protocol in which initially the first two ions within each unit cell
of size $p$ are pinned, and at the end of the protocol the second and third ion
are pinned. That is, during a time $\tau_{\mathrm{s}}$, optical tweezers on the
first and third ion are simultaneously switched off and on, respectively.

This process is adiabatic, i.e., the excitation of phonon modes is suppressed,
if $\omega_x \tau_{\mathrm{s}} \gg 8$ for $p = 4$ and
$\omega_x \tau_{\mathrm{s}} \gg 11$ for $p = 6$, for $\epsilon = 0.07$ and
$\nu_0 = 0.4$. Consequently, the switching time can be much shorter than the
gate duration, which is $\omega_x \tau \approx 1400$ for gates with minimal
control for the same values of $\epsilon$ and $\nu_0$. Therefore, the total time
it takes to execute a quantum circuit is dominated by the time for gate
operations.  Even shorter switching times are permissible for smaller values of
$\nu_0$ and larger values of $\epsilon$.

\section{Optimized tweezer gates}
\label{sec:optimized-gates}

\begin{figure}
  \centering  
  \includegraphics[width=\linewidth]{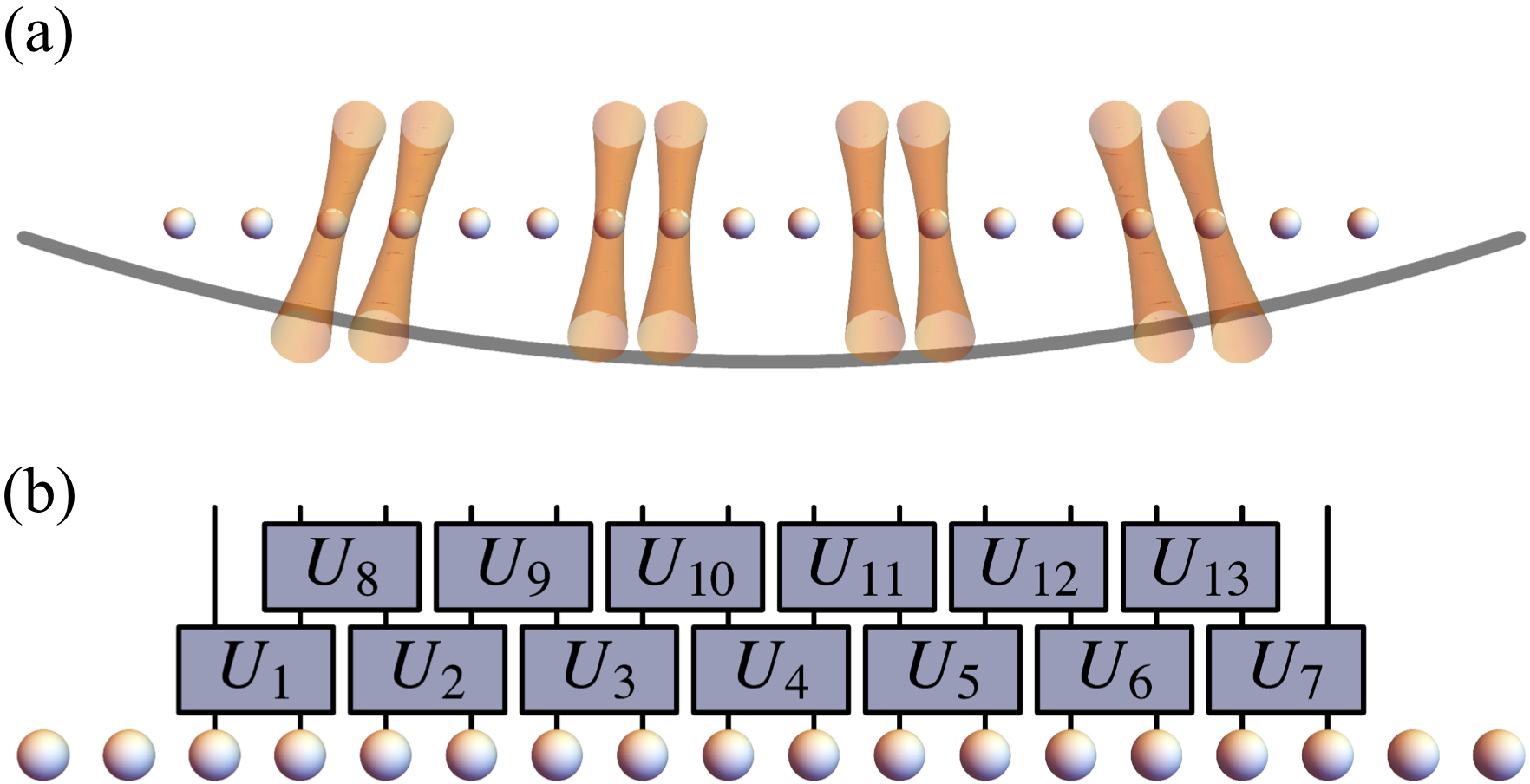}
  \caption{Schematic setup for dense circuits. (a) In a setup in which pairs of ions which
    pinned and not pinned alternate, gates can be performed on all ions in
    parallel. This leads to the realization of dense circuits as illustrated in
    (b). Maintaining high fidelity and low crosstalk in such circuits requires
    optimal coherent control techniques.}
  \label{fig:schematic_dense}
\end{figure}

The optimal choice of detunings and gate durations described above enables the
implementation of parallel entangling tweezer gates in finite and infinite ion chains with a simple laser pulse. However, the
density of the resulting quantum circuits is restricted by crosstalk. This
limitation can be overcome and dense circuits as shown in Fig.~\ref{fig:schematic_dense} can be realized with techniques of optimal coherent control where gate operations are decomposed into multiple laser pulses.

\subsection{Infinite chains}
\label{sec:optimized-gates-infinite}

To optimize the operation of parallel entangling gates, we consider here the
temporal modulation of the amplitudes of the laser pules which drive the
gates~\cite{Zhu2006a, Roos2008, Choi2014, Debnath2016, Wu2018, Landsman2019a, Figgatt2019, Lu2019a}. Originally, optimization of the amplitude shape was devised to
implement fast gates with high fidelity. Our focus is on suppressing crosstalk
in order to implement circuits with high density and fidelity. The control
problem to be solved can be stated in terms of two conditions: The first condition reads $\chi_{i, i'} = \chi_{i, i'}^0$, where we set $\chi_{i, i'}^0 = - \pi/4$ for $\left( i, i' \right) \in I$ and $\chi_{i, i'}^0 = 0$ for $\left( i, i' \right) \in I'$, and where the sets of pairs of ions $I$ and $I'$ are defined as in Eqs.~\eqref{eq:U-0} and~\eqref{eq:U-1-C}. The second condition,
$\alpha^n_i = 0$, ensures that there is no infidelity due to
residual entanglement between qubits and phonon modes. To meet these conditions,
we introduce as independent control parameters a variable number $S$ of pulse
amplitudes $\Omega_i^s$ and detunings $\mu_i$ for ions
$i \in \{ 1, \dotsc, N \}$. Specifically,  
the laser pulse which affects ion $i$ with detuning $\mu_i$ is divided into $S$ segments of equal duration
with constant pulse amplitude
$\Omega_i^s$ within a segment such that
$\Omega_i(t) = \Omega_i^s$ for $\left( s - 1 \right) \tau/S \leq t < s \tau/S$.
If both of the above conditions are satisfied exactly, the implemented gate $U$
in Eq.~\eqref{eq:U-gate} is identical to the ideal gate $U_0$ in
Eq.~\eqref{eq:U-0}. In practice, however, it is not possible to find exact
solutions of this control problem. Instead, we search for approximate solutions by formulating the above
conditions as an unconstrained optimization problem, i.e., we formulate a cost
function $L$ which has to be minimized with respect to $\boldsymbol{\Omega}_i = \left( \Omega_i^1, \dotsc, \Omega_i^S \right)$
and $\mu_i$ for given $\tau$. A minimum with $L = 0$ would correspond to an
exact solution of the control problem.

The definition of a cost function $L$ is not unique and we work with the
 choice $L = L_{\chi} + L_{\alpha}$, where
\begin{equation}
  \label{eq:L-alpha-chi}
  L_{\alpha} = \sum_{i = 1}^N \left( \sum_{n = 1}^N \alpha_i^n \right)^2, \quad
  L_{\chi} = \sum_{\left( i, i' \right) \in J} \left( \chi_{i, i'} - \chi_{i,
      i'}^0 \right)^2,
\end{equation}
and $J$ is a set of ion pairs which is specified
below. $L_{\chi}$ corresponds to the simplest polynomial in Rabi frequencies which has a
minimum with $L_{\chi} = 0$ at $\chi_{i, i'} = \chi_{i, i'}^0$. The square in
the definition of $L_{\alpha}$ ensures that $L_{\alpha}$ is a polynomial of Rabi
frequencies of the same order as $L_{\chi}$. More details can be found in Appendix~\ref{sec:entangl-gates-time-mod-Rabi}. 

\begin{figure*}
  \centering  
  \includegraphics[width=.7\linewidth]{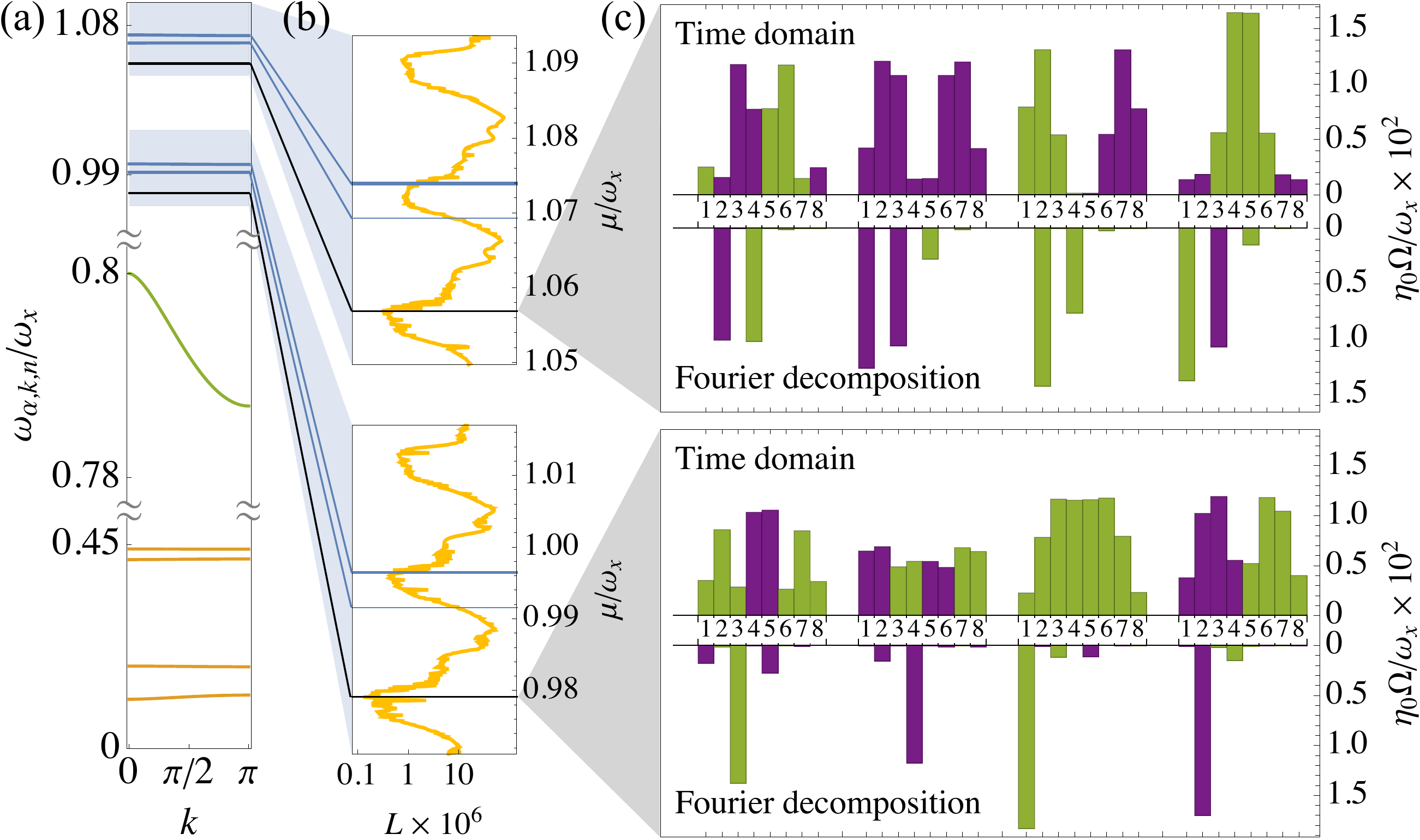}
  \caption{Pulse optimization in an infinite chain. (a) Phonon mode
    spectrum for $p = 4$, $\nu_0 = 0.4$, and $\epsilon = 0.07$. (b) The
    minimization of the cost function $L$ is performed independently for pinned
    and not pinned ions. The upper and lower panels show the cost function for
    optimal sequences of Rabi frequencies for a range of detunings around the
    COM and stretch bands of pinned and not pinned ions,
    respectively. (c) For the optimal detunings we show the corresponding $G = 4$ independent sequences of $S = 8$ Rabi
    frequencies both in the time domain and in discrete Fourier space. Purple
    and green correspond to positive and negative Rabi frequencies
    respectively. The Fourier representations are dominated by few modes which
    belong to subspaces of Fourier space which alternate along the chain: A pulse
    sequence which contains only odd Fourier components is followed by a
    sequence that is composed of even Fourier components and \textit{vice
      versa.}}
  \label{fig:pulse_optimization_infinite_system}
\end{figure*}

We consider now an infinite chain which is subdivided by optical tweezers into unit cells of size $p = 4$. Our goal is to perform entangling gates on both the set of pinned ions as well as the set of ions which are not pinned to realize a maximally dense quantum circuit as illustrated in
Fig.~\ref{fig:pulse_optimization_infinite_system}. Crosstalk between these two sets of ions is suppressed through the small
spatial overlap of the respective phonon modes. Therefore, we minimize the cost function for both sets independently, and we calculate the infidelity and crosstalk which result from performing the independently optimized gates simultaneously \textit{a posteriori}.

To suppress crosstalk within each set of ions, it is sufficient to allow for only a small number $G$ of independent sequences of Rabi frequencies $\boldsymbol{\Omega}_i$: First, we choose the sequences of Rabi frequencies to be the same for two neighboring pinned or not pinned ions, i.e., we set $\boldsymbol{\Omega}_{2 i - 1} = \boldsymbol{\Omega}_{2 i}$. Second, since crosstalk is negligible for sufficiently distant pairs of ions, we limit the ``active'' suppression of crosstalk through the minimazition of $L$ to a group of $G$ neighboring pairs of ions within each subset. The resulting $G$ independent sequences $\boldsymbol{\Omega}_i$ are applied
periodically in space, that is, we set $\boldsymbol{\Omega}_i = \boldsymbol{\Omega}_{i + 4 G}$. Thus, for gates on pinned ions, the set $J$ in Eq.~\eqref{eq:L-alpha-chi} contains
pairs $\left( i, i' \right)$ for which the first is any one of the pinned ions
up to unit cell $G$, $i \in \{ 1, 2, 5, 6, \dotsc, 4 G - 3, 4 G - 2 \}$ and $i'$
runs over all ions to the right of $i$. For gates on ions which are not pinned,
the first ion $i$ in $\left( i, i' \right)$ is in the set
$i \in \{ 3, 4, 7, 8, \dotsc, 4 G - 1, 4 G \}$ and again $i'$ runs over all ions
to the right of $i$.




The gate optimization for infinite systems is illustrated in
Fig.~\ref{fig:pulse_optimization_infinite_system} for $\omega_x \tau = 1500$, $S = 8$, and $G = 4$. Panel~\ref{fig:pulse_optimization_infinite_system}(a) shows the mode spectrum for $p = 4$ with COM and stretch bands both
for the set of pinned and not pinned ions. We minimize $L$ independently for both sets of ions for detunings $\mu$ in ranges indicated by blue
shaded areas. Panel~\ref{fig:pulse_optimization_infinite_system}(b) shows the corresponding values of the cost
function as the detuning is varied. The optimal detuning for each set is determined by the global minimum of $L$.
Panel~\ref{fig:pulse_optimization_infinite_system}(c)
shows the optimal sequences of Rabi frequencies both in the time domain and in
discrete Fourier space. The Fourier representation is defined as
\begin{equation}
  \label{eq:discrete-Fourier-transform}
  \widetilde{\Omega} = \mathsf{F} \Omega, \quad \mathsf{F}_{s, s'} =
  \frac{1}{\sqrt{S}} \sin( \pi s \left( s' - 1/2 \right) \! /S ),
\end{equation}
which corresponds to a discrete sine transform. Interestingly, the Fourier
representations of the optimal pulse sequences are first of all dominated by few
Fourier modes. Moreover, the dominant Fourier modes alternate along the ion chain, with a
pulse sequence which contains only odd Fourier components being followed by a
sequence that is composed of even Fourier components and \textit{vice versa.}
This observation hints at a mechanism to suppress crosstalk which is akin to
refocusing circuits~\cite{Nebendahl2009, Muller2011}. We stress that this mechanism is ``discovered'' here by an unbiased
optimization algorithm.

If the independently optimized sets of gates are performed
simultaneously, we find an average over-/underrotation error and crosstalk per gate of $\delta \chi \approx 2.3 \times 10^{-7}$ and
$C \approx 2.2 \times 10^{-3}$, respectively, and an average infidelity of
$\delta F \approx 3.3 \times 10^{-4}$. While the infidelity is comparable to the results presented in Sec.~\ref{sec:gates-not-optimized-infinite}, crosstalk is significantly lower even though we consider here entangling gates acting on \emph{all} ions in parallel. The values of the infidelity and crosstalk as well as the gate speed can be
improved further by increasing the number of segments $S$ and the number of independent
gates $G$. As noted above, this will lead to a concomitant increase of the
required maximum Rabi frequency, which sets a limit on the achievable gate
performance. At the same time, the complexity of the optimization problem as
determined by the number of independent parameters grows linearly both with $S$
and $G$.

\subsection{Finite chains}
\label{sec:optimized-gates-finite}
\begin{figure}
  \centering  
  \includegraphics[width=.9\linewidth]{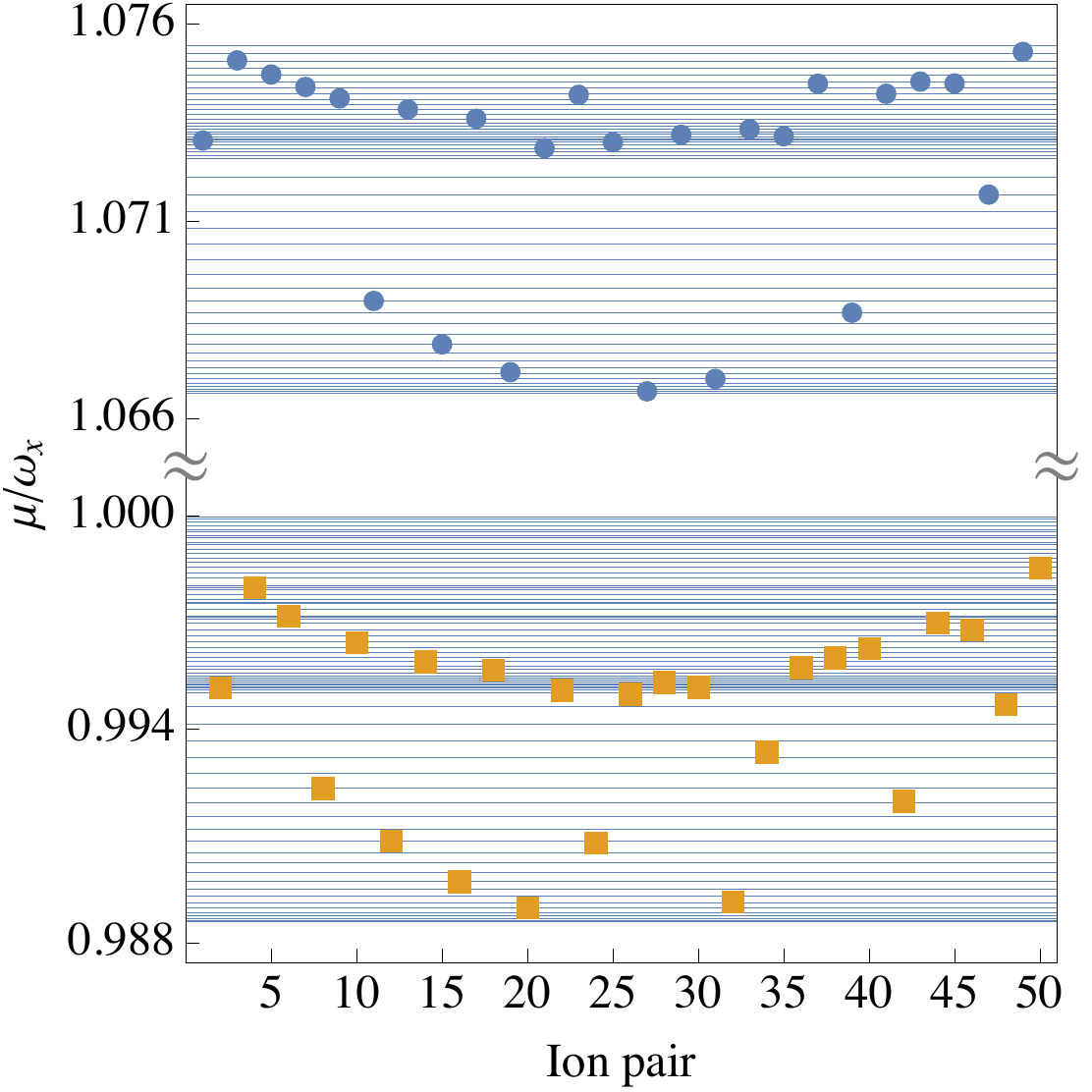}
  \caption{Pulse optimization in a finite chain. We show the chosen detunings (see main text) to implement a dense circuit of $50$ parallel, maximally entangling tweezer gates in a chain of $130$ ions with $15$ buffer ions on each side ($p=4$, $\nu_0=0.4$, $\epsilon \approx 0.07$). The horizontal lines mark the mode frequencies. We allow $S=8$ segments and set the gate duration to $\omega_x \tau = 1500$, which requires a maximal Rabi frequency of $\eta_0 \Omega / \omega_x = 0.007$. We get average infidelity $\delta F = 10^{-5}$ and crosstalk $C = 2.77 \times 10^{-4}$.}
  \label{fig:pulse_optimization_finite_system}
\end{figure}

The results of the previous section show that optimal control can successfully be employed in order to implement dense and parallel tweezer gates in a periodic chain of equidistant ions. For this the periodicity of the system is a crucial requirement since then a small number of optimized pulse sequences can be repeated along the chain. We now introduce a method for finite chains that does not require periodicity. The central idea is to employ optimal control techniques given in Ref.~\cite{Wu2018} in order to independently optimize the infidelities for each tweezer gate and subsequently suppress crosstalk through the choice of laser detunings. 

As in the previous section we divide the gate duration into $S$ segments with different but constant Rabi frequencies $\Omega_i^s$ for $s \in \{ 1 \dots S \}$ and ions $i \in \{1 \dots N \}$. The authors of Ref.~\cite{Wu2018} describe how to choose such a sequence of Rabi frequencies in order to implement a single two-qubit gate for given gate duration and laser detuning and with minimal infidelity. Below we extend their ideas to dense quantum circuits of parallel tweezer gates as shown in Fig.~\ref{fig:schematic_dense}(b). Our method works as follows: In a first step for each target ion pair $(i,i') \in I$ we determine a set of detunings $\mu$ in the vicinity of the corresponding localized modes for which an optimized pulse sequence yields an infidelity below a certain threshold value $\delta F_\text{thresh}$. Secondly we apply an iterative optimization algorithm to choose a detuning $\mu_{(i,i')}$ from each of these sets such that the total crosstalk becomes small for $\boldsymbol{\mu}=(\mu_{(i,i')})_{(i,i') \in I}$. This is done by looping through the target pairs $I$ from the left end of the chain to the right end while applying the following routine: 
For the first pair, as well as in the case that for a given pair there is no detuning for which $\delta F < \delta F_\text{thresh}$, select the detuning that yields the lowest infidelity. Else select the detuning that minimizes crosstalk with all other pairs for which the detuning has already been fixed.
Below we choose to iterate the optimization $5$ times which is sufficient to achieve convergence.

In Fig.~\ref{fig:pulse_optimization_finite_system} we show numerical results for a system of $130$ ions with $15$ buffer ions on each side. As in the infinite case we choose $p=4$ in order to implement a dense circuit consisting of $50$ maximally entangling tweezer gates in parallel, using individual optimal control and the iterative optimization as described above, where we set $\delta F_\text{thresh} = 10^{-3}$. As in the previous section we allow $S=8$ segments for each pulse sequence and we choose a gate duration of $\omega_x \tau = 1500$. This yields an infidelity $\delta F = 10^{-5}$ and a crosstalk $C = 2.77 \times 10^{-4}$ for a maximal Rabi frequency of $\eta_0 \Omega / \omega_x = 0.007$. If higher Rabi frequencies are available one could increase the number of segments for the pulse sequences in order to further improve infidelity or to speed up the gate~\cite{Wu2018}.

\section{Outlook}
\label{sec:outlook}

In this work, we developed the implementation of scalable parallel gate operations using localized transverse phonon modes generated by optical tweezers. To be concrete, we considered quantum circuits with spatially recurring structures of nearest-neighbor two-qubit gates as illustrated in Figs.~\ref{fig:schematic}(d) and~\ref{fig:schematic_dense}(b). The dynamical reconfigurability of programmable tweezer arrays enables reshaping phonon modes on the fly. This is a key feature of \emph{optical} segmentation of ion chains and facilitates the efficient implementation of universal parallelized quantum circuits.

Immediate extensions of the methods developed in this paper are illustrated schematically in Fig.~\ref{fig:schematic}(a). First, \emph{multi-qubit} gates can be performed on subregisters which are separated by ``optical tweezer walls''~\cite{Shen2019}; second, the COM and stretch modes of pairs of \emph{distant} ions can be used to implement entangling gates for qubits which are not nearest neighbors. Combining these capabilities leads to the realization of 1D quantum networks which connect nodes that correspond to subregisters of long 1D chains.

Beyond these opportunities for quantum algorithms and gate-based \emph{digital} quantum simulation, designer phonon modes which are shaped through optical potentials open up new possibilities for \emph{analog} quantum simulation, which can be realized through virtual far off-resonant excitation of phonon modes.


In addition to the applications of optical tweezers in the implementation of quantum gates and the design of Hamiltonians for analog quantum simulation, they also provide new possibilities to tackle challenges on a more fundamental level of quantum hardware design. For example, while we focused here on programming the phonon mode structure for a given configuration of the ion chain, where the equilibrium positions of the ions are fixed by the trapping potential and Coulomb interactions, and tweezers are focused on the equilibrium positions, also the equilibrium positions themselves can be shifted by using optical forces. This enables, e.g., to achieve uniform ion spacings along the chain to facilitate individual control by addressed laser beams for gate operations, and thus provides an alternative to anharmonic potentials~\cite{Lin2009, Wu2018}. Further, laser cooling of phonon modes can be carried out more efficiently in an ion chain which is divided into subregisters~\cite{Shen2019}.

An interesting question concerns the extension of the methods developed in this paper to 2D and 3D structures~\cite{Itano1998, Drewsen1998, Mortensen2006}. Further studies are required to elucidate the interplay between micromotion~\cite{Cirac1994, Wang2015}, which is not negligible in specific spatial directions, and the localization properties of phonon modes in higher dimensions.

\section*{Acknowledgements}

We thank R. Islam and R. Blatt for insightful comments and helpful discussions. Theoretical work at Innsbruck is
supported by the European Union program Horizon 2020 under Grants Agreement
No.~817482 (PASQuanS) and No.~731473 (QuantERA via QTFLAG), the US Air Force
Office of Scientific Research (AFOSR) via IOE Grant No.~FA9550-19-1-7044 LASCEM,
by the Simons Collaboration on Ultra-Quantum Matter, which is a grant from the
Simons Foundation (651440, PZ), and by the Institut f\"ur
Quanteninformation. LP, PS, and TM acknowledge support from the Austrian Science Fund (FWF) through the SFB BeyondC: F7102, and the IQI GmbH. PS acknowledges support from the Austrian Research Promotion Agency (FFG) contract 872766.

\section*{Author contributions}

LS and PZ guided the research based on original ideas proposed by PZ. LS and TO
performed the analytical and numerical studies underlying the manuscript. LP, PS
and TM contributed the experimental feasibility study, and experimental
perspective. The manuscript was written by LS supported by TO and PZ. All
authors contributed to the discussion of results.

\appendix

\section{Phonon modes in 1D ion chains with optical tweezers}
\label{sec:ion-chain-tweezers}

In the following, we derive phonon mode matrices $M^n_{\alpha, i}$ and mode
frequencies $\omega_{\alpha, n}$ for finite and infinite 1D chains of trapped ions
which are subject to programmable arrays of optical tweezers.

\subsection{Phonon modes of finite chains}
\label{sec:phonon-modes-finite-chains}

The Hamiltonian for the classical 3D motion of $N$ ions in a harmonic trap reads
\begin{equation}
  \label{eq:H-0}
  H_0 = \sum_{i = 1}^N \left( \frac{p_i^2}{2 m} +
    V(\mathbf{r}_i) \right) + \frac{e^2}{4 \pi \epsilon_0} \sum_{i < i' = 1}^N
  \frac{1}{\abs{\mathbf{r}_i - \mathbf{r}_{i'}}},
\end{equation}
where
$\mathbf{r}_i = \left( r_{x, i}, r_{y, i}, r_{z, i} \right)^{\transpose} =
\left( x_i, y_i, z_i \right)^{\transpose}$
and $\mathbf{p}_i = \left( p_{x, i}, p_{y, i}, p_{z, i} \right)^{\transpose}$
are, respectively, the position and momentum of ion $i$, and the electronic
trapping potential is given by
$V(\mathbf{r}) = \sum_{\alpha \in \{ x, y, z \}} \frac{1}{2} m \omega_{\alpha}
r_{\alpha}^2$.
We assume tight confinement in the transverse $x$ and $y$ directions, such that
the equilibrium positions $\mathbf{r}_{i, 0}$ of the ions are along the $z$
axis, $\mathbf{r}_{i, 0} = \left( 0, 0, z_{i, 0} \right)^{\transpose}$, and form
a linear 1D chain. If the number of ions $N$ is increased, the axial trapping
frequency $\omega_z$ has to be reduced for the linear configuration of the ion
chain to remain stable~\cite{Fishman2008, Shimshoni2011, Welzel2019}.

Phonon modes of the ion chain correspond to quantized small-amplitude
oscillations of the ions around their equilibrium positions. An expansion of the
Hamiltonian Eq.~\eqref{eq:H-0} to second order in
$\delta \mathbf{r}_i = \mathbf{r}_i - \mathbf{r}_{i, 0}$ leads to
$H_0 = \sum_{\alpha \in \{ x, y, z \}} H_{0, \alpha}$ where
\begin{multline}
  \label{eq:H-0-alpha}
  H_{0, \alpha} = \sum_{i = 1}^N \left( \frac{p_{\alpha, i}^2}{2 m} +
    \frac{1}{2} m \omega_{\alpha, i}^2 \delta r_{\alpha, i}^2 \right) \\ +
  \frac{e^2 s_{\alpha}}{4 \pi \epsilon_0} \sum_{i < i' = 1}^N \frac{\delta
    r_{\alpha, i} \delta r_{\alpha, i'}}{\lvert z_{i, 0} - z_{i', 0} \rvert^3},
\end{multline}
with $s_x = s_y = 1$ and $s_z = -2$. To this order of the expansion,
oscillations of the ions in the transverse $x$ and $y$ directions and the
longitudinal $z$ direction decouple. The ions perform harmonic oscillations
around their equilibrium positions with local trapping frequencies
\begin{equation}
  \label{eq:omega-alpha-i}
  \omega_{\alpha, i}^2 = \omega_\alpha^2 - \frac{e^2 s_\alpha}{4 \pi \epsilon_0
    m} \sum\limits_{\substack{i'=1 \\ i'\neq i}}^N \frac{1}{\lvert z_{i, 0} -
    z_{i', 0} \rvert^3}.
\end{equation}
These oscillations are coupled by the residual Coulomb interaction described by
the last term in Eq.~\eqref{eq:H-0-alpha}. To design phonon modes, we consider
adjusting the local trapping frequencies by focusing optical tweezers on the
equilibrium positions of the ions as detailed in
Sec.~\ref{sec:optical-design-of-modes}. The optical potential which is generated
by the tweezers is described by an additional contribution to the Hamiltonian
which reads $H_1 = \sum_{i = 1}^N V^{\mathrm{twz}}_i(\mathbf{r}_i)$, where
\begin{equation}
  \label{eq:V-twz}
  V^{\mathrm{twz}}_i(\mathbf{r}_i) = \frac{1}{2} m \omega_{0, i}^2 \left( \delta
    r_{x, i}^2 + \delta r_{z, i}^2 \right).
\end{equation}
Normal mode coordinates $\xi_{\alpha, n}$ are introduced via the linear
transformation $\delta r_{\alpha, i} = \sum_{n = 1}^N M^n_{\alpha, i} \xi_i$,
where $M^n_{\alpha, i}$ is the mode matrix which diagonalizes the phonon
Hamiltonian $H_{\mathrm{ph}} = H_0 + H_1$, i.e., which brings the Hamiltonian to
a form that corresponds to decoupled harmonic oscillators with frequencies
$\omega_{\alpha, n}$. The normal-mode oscillations of the ion chain can be
quantized by introducing annihilation and creation operators for phonons,
$a_{\alpha, n}$ and $a_{\alpha, n}^{\dagger}$, respectively. In terms of these
operators, the deviation of ion $i$ from its equilibrium position can be
expressed as
\begin{equation}
  \label{eq:ion-position-quantized}
  \delta r_{\alpha, i} = \sqrt{\frac{\hbar}{2 m}} \sum_{n = 1}^N
  \frac{M_{\alpha, i}^n}{\sqrt{\omega_{\alpha, n}}}
  \left( a_{\alpha, n} + a^{\dagger}_{\alpha, n} \right).
\end{equation}
The qubit-phonon Hamiltonian Eq.~\eqref{eq:H-qubit-ph} couples the qubits which
are encoded in individual ions to the quantized normal-mode oscillations of the
ion chain.

\subsection{Phononic band structure of infinite chains}
\label{sec:phon-band-struct}

We proceed to derive the phononic band structure for an infinitely long 1D ion
chain with uniform spacing $d$. The ions are subject to a spatially periodic
array of optical tweezers which subdivides the ion chain into unit cells of size
$p$. Within each unit cell, the first two ions are pinned by optical tweezers
with trapping frequency $\omega_0$, and the remaining $p - 2$ ions are not
pinned as illustrated in Fig.~\ref{fig:phonon_modes_spectra_infinite}(a).

We label the ions by their unit cell $l \in \Z$ and their position
$i = 1, \dotsc, p$ within the unit cell, where the pinned ions correspond to the
positions $i = 1, 2$. The classical Hamiltonian for small-amplitude oscillations
of the ions around their respective equilibrium positions can be written as
$H_{\mathrm{ph}} = \sum_{\alpha \in \{ x, y, z\}} H_{\alpha}$, where
$H_{\alpha}$ is given by the sum of Eq.~\eqref{eq:H-0-alpha} and the optical
potential in Eq.~\eqref{eq:V-twz} in the simultaneous limit $N \to \infty$ and
$\omega_z \to 0$:
\begin{multline}
  \label{eq:H-ph-alpha}
  H_{\alpha} = \sum_{l \in \Z} \sum_{i = 1}^p \left( \frac{p_{\alpha, l, i}^2}{2
      m} + \frac{1}{2} m \widetilde{\omega}_{\alpha, i}^2 \delta r_{\alpha, l, i}^2 \right) \\ +
  \frac{1}{2} s_{\alpha} \sum_{l, l' \in \Z} \sum_{i, i' = 1}^p \delta r_{\alpha, l, i}
  C^{l - l'}_{i - i'} \delta r_{\alpha, l', i'}.
\end{multline}
The local trapping frequency which the ions at position $i$ within each unit
cell experience is given by
\begin{equation}
  \widetilde{\omega}_{\alpha, i}^2 = \omega_{\alpha}^2 - \frac{s_{\alpha} e^2 \zeta(3)}{2 \pi
    \epsilon_0 d^3 m} + \omega_{\alpha, 0}^2 \left( \delta_{i, 1} + \delta_{i, 2} \right),
\end{equation}
where, according to Eq.~\eqref{eq:V-twz},
$\omega_{x, 0} = \omega_{z, 0} = \omega_0$ and $\omega_{y, 0} = 0$, and
$\zeta(s)$ is Riemann zeta function. The translationally invariant coupling
coefficient $C^l_i$ reads
\begin{equation}
  C^l_i =
  \begin{cases}
    0 & \text{for } l = 0 \text{ and } i = 0, \\
    \frac{e^2}{4 \pi \epsilon_0 d^3} \frac{1}{\abs{p l + i}^3} & \text{else.}
  \end{cases}
\end{equation}

\subsubsection{Phononic band structure for periodic tweezer arrays}
\label{sec:phon-band-struct-periodic}

We seek the normal mode matrix which diagonalizes the potential energy
contribution to Eq.~\eqref{eq:H-ph-alpha}. It is convenient to write the latter
as $H_{\mathrm{pot}, \alpha} = \frac{1}{2} m \omega_x^2 V_{\alpha}$, where
$V_{\alpha}$ is dimensionless. For concreteness and to simplify the notation, we
focus in the following on oscillations in the $x$ direction, and we omit the
subscript $\alpha = x$. The normal modes of oscillations in the $y$ and $z$
directions can be found analogously.

To account for the translational invariance of the phonon
Hamiltonian~\eqref{eq:H-ph-alpha} in the unit-cell index $l$, we interpret the
coordinates $\delta x_{l, i}$ as coefficients of a Fourier series,
$c_{k, i} = \sum_{l \in \Z} \e^{-\imag k l} \delta x_{l, i}$, where $k$ is analogous to the quasimomentum of an electron in a solid. In terms of the
new complex coordinates $c_{k, i}$, the potential energy reads
\begin{equation}
  \label{eq:V-c}
  V = \int_{-\pi}^{\pi} \frac{\mathrm{d} k}{2 \pi} \sum_{i, i' =
    1}^p v^k_{i, i'} c_{k, i}^{*} c_{k, i'},
\end{equation}
where
\begin{equation}
  \label{eq:v-J-i-k}
  v^k_{i, i'} = V_i \delta_{i, i'} + J^k_{i - i'}, \qquad J^k_i = \sum_{l \in \Z} J^l_i \e^{-\imag k l},
\end{equation}
with
$V_i = 1 - 2 \epsilon^2 \zeta(3) + \nu_0^2 \left( \delta_{i, 1} + \delta_{i, 2}
\right)$
and $J^l_i = C^l_i/(m \omega_x^2)$. We next introduce new coordinates
$b_{k, n} = \sum_{i = 1}^p B_i^{k, n *} c_{k, i}$, where $B_i^{k, n}$ is the
unitary matrix which diagonalizes $v^k_{i, i'}$ with eigenvalues
$\nu_{k, n} = \omega_{k, n}/\omega_x$. While $V$ is diagonal in terms of the
coordinates $b_{k, n}$, they cannot be interpreted as proper normal mode
coordinates because they are complex and not independent: Since
$\delta x_{l, i}$ are real, it follows that $b_{k, n}^{*} = b_{-k ,n}$.
Therefore, we restrict the range of values of the quasimomentum to
$k \in [0, \pi]$, and we decompose $b_{k, n}$ and $B_i^{k, n}$ into real and
imaginary parts,
$b_{k,n} = 1/\sqrt{2} \left( \xi_{k, n, 1} + \imag \xi_{k, n, 2} \right)$ and
$B_i^{k, n} = \Xi_i^{k, n, 1} + \imag \Xi_i^{k, n, 2}$, to obtain
\begin{equation}  
  \delta x_{l, i} = \int_0^{\pi} \frac{\mathrm{d} k}{2 \pi} \sum_{n = 1}^p
  \sum_{\lambda = 1}^2 M_{l, i}^{k, n, \lambda} \xi_{k, n, \lambda}.
\end{equation}
$\xi_{k, n, \lambda}$ are the desired real and independent normal mode
coordinates, and the normal mode transformation matrices are given by
\begin{equation}
  \label{eq:mode-matrices-periodic}
  \begin{split}
    M_{l, i}^{k, n, 1} & = \sqrt{2} \left( \cos(k l) \Xi_i^{k, n, 1} - \sin(k l)
      \Xi_i^{k, n, 2} \right), \\ M_{l, i}^{k, n, 2} & = - \sqrt{2} \left(
      \cos(k l) \Xi_i^{k, n, 2} + \sin(k l) \Xi_i^{k, n, 1} \right).
  \end{split}
\end{equation}

\subsubsection{Perturbative expansion for strong pinning}
\label{sec:pert-expans-strong-pinning}

For strong pinning $\nu_0 \gg \epsilon$, the matrix $v^k$ can be diagonalized
perturbatively in $J_i^k \propto \epsilon^2$. To zeroth order, we obtain two
degenerate subspaces which correspond to the ions which are pinned and not
pinned. The respective eigenvalues are $1 - 2 \epsilon^2 \zeta(3) + \nu_0^2$ and
$1 - 2 \epsilon^2 \zeta(3)$, and have degeneracy $2$ and $p - 2$. To obtain the
leading corrections to the eigenvalues in degenerate perturbation theory, we
omit the elements $v^k$ which couple the degenerate subspaces, whereupon $v^k$
becomes block-diagonal. The matrix $B^k$ that diagonalizes the $2 \times 2$
block of $v^k$ which describes the pinned ions reads
\begin{equation}
  \label{eq:B-k-1-large-p}
  B^k \sim \Xi^1 = \frac{1}{\sqrt{2}}
  \begin{pmatrix}
    1 & 1 \\ 1 & - 1
  \end{pmatrix},
\end{equation}
where we omit corrections $\sim 1/p^4$ which are small for $p \gtrsim 4$.
Evidently, for $p \to \infty$, the modes which form the highest two bands are
indeed COM and stretch modes of pairs of pinned ions within each unit cell. The
corresponding $2 \times 2$ blocks of mode matrices in
Eq.~\eqref{eq:mode-matrices-periodic} are given by
$M^{k, 1}_l = \sqrt{2} \cos(k l) \Xi^1$ and
$M^{k, 2}_l = - \sqrt{2} \sin(k l) \Xi^1$. Perturbative corrections to the mode
matrices are of order $O(\epsilon^2/\nu_0^2)$. The mode frequencies of the COM
and stretch bands are
\begin{equation}
  \label{eq:nu-k-1-n}
  \nu_{k, n} = \sqrt{1 - 2 \epsilon^2 \zeta(3) + \nu_0^2 + J_0^k \pm \abs{J_1^k}},
\end{equation}
where $n = 1$ for the COM band and $n = 2$ for the stretch band. The width of
these bands is determined by terms in Eq.~\eqref{eq:nu-k-1-n} which depend on
the quasimomentum $k$, i.e., by $J_0^k \pm \abs{J_1^k}$. For $p \gtrsim 4$, we find
\begin{equation}
  \label{eq:COM-stretch-bandwidths}
  \begin{split}
    J_0^k + \abs{J_1^k} & \sim \epsilon^2 \left( 1 + \frac{4}{p^3} \Re \! \left(
        \Li \! \left( \e^{\imag k} \right) \right) \right), \\ J_0^k -
    \abs{J_1^k} & \sim - \epsilon^2 \left( 1 + \frac{12}{p^5} \Re \! \left(
        \mathop{\mathrm{Li}_5} \!  \left( \e^{\imag k} \right) \right) \right),
  \end{split}
\end{equation}
where $\mathrm{Li}_{\alpha}(z) = \sum_{n = 1}^{\infty} \frac{z^n}{n^{\alpha}}$
denotes the polylogarithm. That is, the widths of the COM and the stretch bands
are suppressed as $1/p^3$ and $1/p^5$, respectively. For $p \to \infty$,
these bands become flat with frequencies
\begin{equation}
  \label{eq:nu-n}
  \nu_n = \sqrt{1 - 2 \epsilon^2 \zeta(3) + \nu_0^2 \pm \epsilon^2}.
\end{equation}
In this limit, pairs of pinned ions are completely decoupled, and the
corresponding modes are strictly local within unit cells.

\section{Experimental feasibility study}
\label{sec:feasibility_study}

Here we study the experimental feasibility of the proposed scheme to perform parallel tweezer gates in a trapped-ion quantum device. We discuss various sources of experimental imperfections, for both optical and ground-state qubit encodings, and for realizations with different ionic species.

\subsection{Spontaneous scattering}

A significant contribution to the infidelity of tweezer gates is due to scattering of photons of the optical tweezers. In the following, we discuss how the choice of qubit encoding and wavelength of the optical tweezer lasers effects the spontaneous scattering rate and therefore the infidelity of the gate.

Our discussion is based on the following model for the interaction of an optical tweezer beam with a trapped ion: The dipole potential induced by the optical tweezer is given by~\cite{grimm2000optical}
\begin{equation}
    U_{\mathrm{dip}}(\mathbf{r}) = - \frac{1}{2\epsilon_0 c}\Re(\alpha)I(\mathbf{r}),
\end{equation}
where $\epsilon_0$ is the vacuum permittivity, $c$ is the speed of light, $I(\mathbf{r})$ is the intensity of the tweezer beam at position $\mathbf{r}$, and $\alpha$ is the polarizability of the internal state of the ion. In general, the polarizability is different for the two states which encode the qubit. Therefore, a question we have to address below is how to ensure that both qubit states experience the same optical potential. 

The scattering rate in the center of the Gaussian tweezer beam with frequency $\omega$, for a transition with resonance frequency $\omega_0$ and linewidth $\Gamma$, is~\cite{grimm2000optical}
\begin{equation}
    \Gamma_{\mathrm{sc}}(\omega) = \frac{3 c^2}{\hbar \omega_0^3} \left(\frac{\omega}{\omega_0}\right)^3\left(\frac{\Gamma}{\omega_0 - \omega} + \frac{\Gamma}{\omega_0 + \omega} \right)^2 \frac{P}{W_0^2},
\end{equation}
with the total light power $P$ and the beam waist $W_0$.
For a tightly focused beam with wavelength $\lambda$, the beam waist is approximately given by~\cite{bass2009handbook}
\begin{equation}
    W_0 \approx 0.41 \times \frac{\lambda}{\textit{NA}},
    \label{eq:diffraction_limit}
\end{equation}
 where $\textit{NA}$ is the numerical aperture of the focusing optics. All following estimations are carried out assuming $\textit{NA}=0.7$, as shown in Ref.~\cite{araneda2020panopticon} in an ion-trapping experiment. Scattering rates from different transitions are summed up, whereas their trap potential partially cancels if the the sign of the detuning $\omega_0 - \omega$ is opposite.

For a given spontaneous scattering rate $\Gamma_{\mathrm{sc}}$ and gate duration $\tau$, the infidelity of the gate due to scattering of light can be estimated as~\cite{ballance2017high}
\begin{equation}
    \delta F_{\mathrm{sc}} = \frac{3}{2}\Gamma_{\mathrm{sc}} \tau .
\end{equation}

\subsubsection{Ground state qubit encoding}
\label{sec:ground_state_qubit}

\begin{figure}
  \centering  
  \includegraphics[width=.9\linewidth]{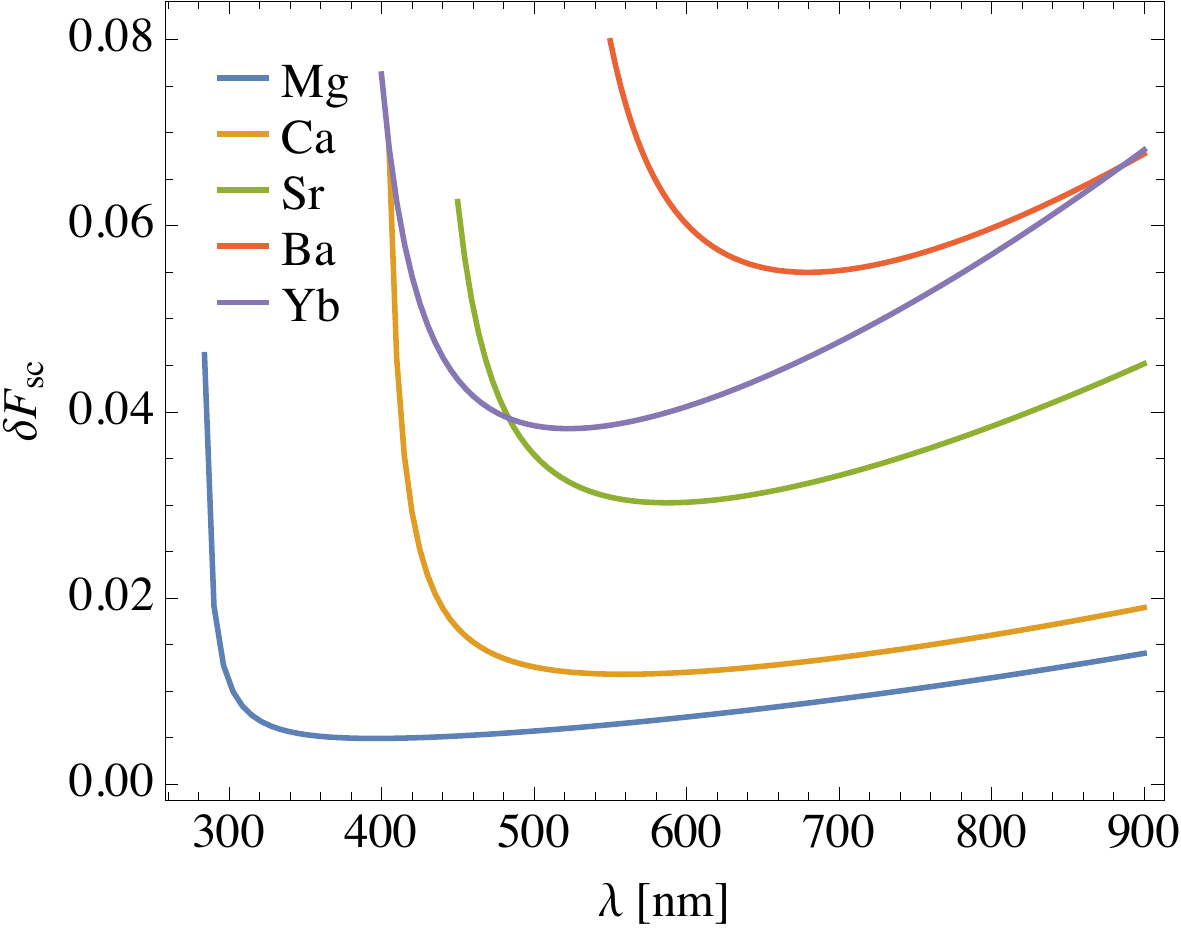}
  \caption{Infidelity due to spontaneous scattering for $\epsilon = 0.07$, $\nu_0 = 0.4$ and $\omega_x = 2\pi \times 3$\,MHz for different atomic species. The steep slope at short wavelengths is caused by the excessive scattering close to the resonance of the $S \Leftrightarrow P$ transition. At long wavelengths, the infidelity increases due to an enlarged spot-size for a diffraction limited spot for the assumed numerical aperture.}
  \label{fig:tweezer_wavelength}
\end{figure}

A first approach to ensure that both qubit states experience the same optical potential is to encode both of them in the ground state of the ion~\cite{Bruzewicz2019}. In this case, the AC-Stark shift for $\pi$-polarized light is not state dependent, and the tweezer wavelength can be varied over a broad range.
We consider the species $^{24}\mathrm{Mg}^+$, $^{40}\mathrm{Ca}^+$, $^{88}\mathrm{Sr}^+$, $^{138}\mathrm{Ba}^+$ and $^{171}\mathrm{Yb}^+$, which are used in experiments targeting quantum information processing~\cite{Bruzewicz2019}.

For all investigated ionic species, we only consider contributions to the optical trapping and to spontaneous scattering from the $S$ to $P$ transitions with linewidths of several MHz~\cite{sansonetti2005handbook}. In Fig.~\ref{fig:tweezer_wavelength}, we show the infidelity $\delta F_{\mathrm{sc}}$ as a function of the tweezer wavelength. The infidelity exhibits a single minimum, which lies in between a regime at short wavelengths near the resonance and a regime at long wavelengths, for which the beam waist~\eqref{eq:diffraction_limit} and, therefore, the required power to reach the strong pinning regime increases. Here we set $\epsilon = 0.07$ and $\nu_0 = 0.4$ as in the examples considered in the main text. This results in $\nu_0^2/\epsilon^2 \approx 32$, deeply in the regime of strong pinning. The gate duration is estimated according to Eqs.~\eqref{eq:nu-n} and~\eqref{eq:mu-tau}.

\begin{figure}
  \centering  
  \includegraphics[width=.9\linewidth]{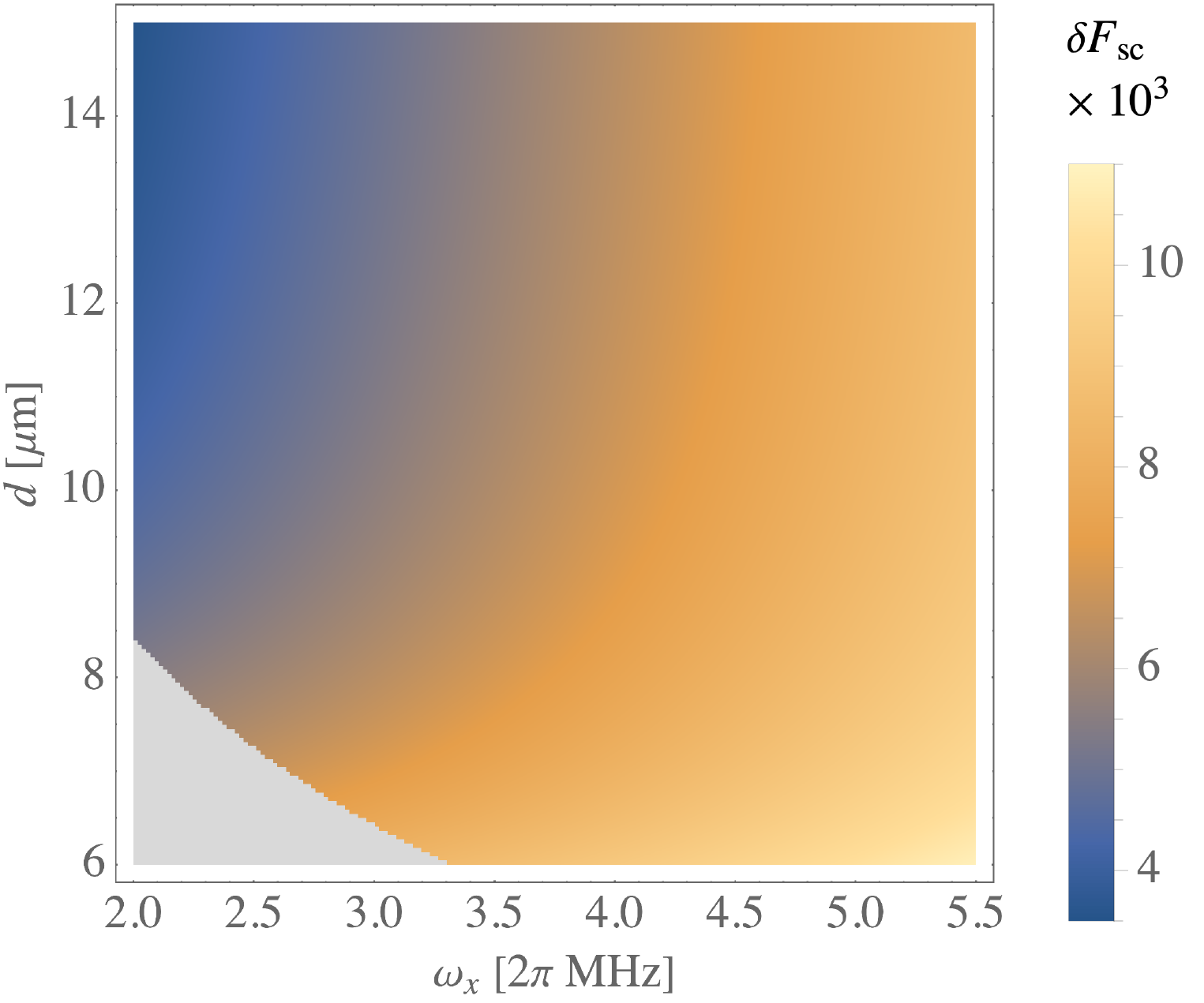}
  \caption{Spontaneous scattering infidelity for parallel gates performed on $^{24}\mathrm{Mg}^+$ ground-state qubits with 400~nm tweezer wavelength for varying ion distance $d$ and radial trap frequency $\omega_x$. By choosing $d$ and $\omega_x$, $\epsilon$ is fixed, and $\nu_0$ is determined in order to reach the strong pinning regime. The gray region marks parameter combinations that lead to a zigzag configuration of the ion string.}
  \label{fig:infidelity_2d}
\end{figure}

The lowest infidelities can be achieved for $^{24}\mathrm{Mg}^+$ followed by $^{40}\mathrm{Ca}^+$. For $^{24}\mathrm{Mg}^+$ ions in a trap with transverse trapping frequency $\omega_x = 2\pi \times 3$\,MHz and a tweezer wavelength of 400\,nm, the contribution of spontaneous scattering to the infidelity is $\delta F_{\mathrm{sc}} \approx 4.9 \times 10^{-3}$. The corresponding inter-ion distance is $d = 15\,\mu \mathrm{m}$. The infidelity caused by spontaneous scattering can be reduced by either increasing $d$ and decreasing $\omega_x$ (see Fig.~\ref{fig:infidelity_2d}), with the downside of increased heating rates~\cite{Brownnutt2015}, or by relaxing the condition on being in the strong trapping regime, which leads to higher infidelities from other sources (see Fig.~\ref{fig:gate_performance_not_optimized}). For $\epsilon = 0.07$ and $\nu_0 = 0.2$, a scattering-induced infidelity of $\delta F_{\mathrm{sc}} \approx 1.2 \times 10^{-3}$ can be achieved.
For the same values of $\omega_x$, $\epsilon$ and $\nu_0$, the infidelity for $^{40}\mathrm{Ca}^+$ ions is $2.8 \times 10^{-3}$ with $d = 12.6\,\mu \mathrm{m}$ and a tweezer wavelength of 532\,nm. The required power of the optical tweezers is on the order of a couple of mW for both species.

\subsubsection{Optical qubit encoding}
\label{sec:optical_qubit}

For an optical qubit, the qubit levels are encoded in two different electronic states. An apparent approach to ensure equal optical trapping potentials for both qubit states is to choose the tweezer wavelength as a magic wavelength with respect to the transition between the two qubit states. However, the polarizabities of $^{40}\mathrm{Ca}^+$, $^{88}\mathrm{Sr}^+$ and $^{138}\mathrm{Ba}^+$, which are commonly used species for optical qubits~\cite{Bruzewicz2019}, are low at the respective magic wavelengths~\cite{kaur2015magic} and, therefore, high light intensities are required to achieve sizeable optical trapping potentials. Unfortunately, the concomitant increase of spontaneous scattering rates leads to relatively high infidelities. 

In particular, magic wavelengths in between the $S_{1/2} \Leftrightarrow P_{1/2}$ and the $S_{1/2} \Leftrightarrow P_{3/2}$ transitions have small detunings from broad transitions, which leads to scattering rates in the regime of hundreds of kHz. Performing gates with typical gate durations $\tau$ on the order of tens to hundreds of $\mathrm{\mu}$s is therefore rendered impossible. 

For the abovementioned species, there also exists a set of magic wavelength that is red detuned from all relevant transitions~\cite{kaur2015magic}. For $^{40}\mathrm{Ca}^+$, this wavelength is at 1271\,nm and therefore has a detuning of over 400\,nm to the nearest relevant transition. Using tweezers at this magic wavelength yields lower scattering rates of hundreds of Hz. However, the resulting infidelities are still in the $10^{-1}$ regime.

An alternative way to reduce the infidelity due to scattering is to use two superimposed tweezer beams at different wavelength to trap each of the two qubit states individually. On the one hand, the gained flexibility with regard to the choice of wavelength could allow for low scattering rates while keeping the required power reasonable, but on the other hand additional experimental challenges would come into play. In particular, to avoid gate errors due to decoherence of the qubit induced by differential trapping potential fluctuations of the two qubit states, precise control over both the power in the tweezer beams as well as the positioning of the tweezers is required.

\subsection{Other experimental imperfections}

\subsubsection{Motional heating and cooling}

Another source of errors in entangling gates is a nonzero occupation of the phonon modes which mediate the entangling interaction, as can be seen in Eq.~\eqref{eq:infidelity}. The occupation of phonon modes increases over time due to electric-field noise. For $N$ ions in a single ion trap, the largest heating rate is proportional to $N$ for the COM mode where all ions oscillate in phase~\cite{Brownnutt2015, joshi2020pgc}. Ions which are pinned with optical tweezers have localized phonon modes and, therefore, we expect that the heating rates for the two modes of each pinned pair are independent of the crystal size. We further expect that this holds also for the case where multiple gates are performed in parallel, since, due to the harmonic trapping potential, the localized modes have different mode frequencies as illustrated in Fig.~\ref{fig:phonon_modes_spectra_finite}, and thus the motion of distant pinned ions is not in phase.

For the proposed tweezer gates in long ion chains, the axial trapping frequency is kept very low to allow for inter-ion distances of 10--15\,$\mu$m. Without additional optical trapping, the spectrum of axial phonon modes reaches down to the axial trapping frequency. Since the motional heating rate is inversely proportional to the frequency, this low axial trapping frequency would result in a high phonon occupation in the axial modes of motion, leading to errors in entangling- and also single-qubit gates~\cite{cetina2020quantum}. However, this effect is suppressed if several ions in the chain are pinned with optical tweezers, and, consequently, the lowest axial phonon frequency is shifted up as illustrated in Figs.~\ref{fig:phonon_modes_spectra_finite} and~\ref{fig:phonon_modes_spectra_infinite}. Moreover, since the oscillations of optically pinned and spectator ions decouple, gates on the pinned ions are only weakly affected by motional heating of phonon modes which are localized on spectator ions.
Finally, the ability to split the ion chain into multiple subsets using the tweezer array can facilitate efficient cooling~\cite{Shen2019}.

\subsubsection{Excess micromotion}

A trapped ion which is not located in the radio frequency (RF) null gives rise to excess micromotion~\cite{mohammadi2019minimizing}. If the amplitude of this micromotion is comparable to the dimension of the optical tweezer, additional gate errors due to RF heating~\cite{Blumel1989Chaos} and modulation of the optical potential are induced. The position of the ion and the RF null~\cite{narayanan2011micromotion} can be overlapped much more precisely than the wavelength of the tweezer light and thus micromotion should not have an influence on the tweezer potential.

\subsubsection{Differential AC Stark shift}

In Appendix~\ref{sec:ground_state_qubit}, we mentioned that for $\pi$-polarized light, the two qubit states encoded in ground states of an ion experience the same optical potential. If there is a contribution of circular polarization, a differential AC Stark shift between the two qubit states occurs. A temporal variation of this differential shift leads to dephasing of the qubit.
Careful alignment and usage of high-quality polarization optics allow for a ratio of $10^{-5}$ of $\sigma$- to $\pi$-polarized light. For a typical optical trap depth of 10\,mK and relative intensity fluctuation of the optical tweezer of $3 \times 10^{-3}$, as assumed in Sec.~\ref{sec:tweezer-imperfections}, this results in differential AC Stark shift fluctuations of a couple of Hz. This should allow a gate fidelity of better than 99.5\,\%~\cite{erhard2019characterizing}.

\section{Tweezer misadjustments}
\label{sec:appendix-gate-imperfections}
In section \ref{sec:tweezer-imperfections} in Eq.~\eqref{eq:V-twz-imp} we state the optical potential that is generated by a misadjusted optical tweezer. Here we describe how we calculate the effect of the shifts of the tweezer focuses. First note that since the optical potential is at most quadratic in the shifts of the tweezers, these shifts affect the modes of the ions only through the corresponding shifts in the equilibrium positions of the ions. We determine these shifts numerically up to second order in the shifts of the tweezers before averaging over several Gaussian realizations as before. To this end we expand the gradient of the total potential for a total number of $N$ ions
\begin{multline}
\nabla_{\boldsymbol{\xi}} V = \underbrace{\sum_m \delta_m \left( \partial_{\delta_m} \nabla_{\boldsymbol{\xi}} V \right)\vert_{\boldsymbol{\xi}_0}}_D \\
+ \underbrace{\frac{1}{2} \sum_{m,n} \delta_m \delta_n  \left( \partial_{\delta_m} \partial_{\delta_n} \nabla_{\boldsymbol{\xi}} V \right) \vert_{\boldsymbol{\xi_0}}}_W
\end{multline}
where $\boldsymbol{\xi} =  \boldsymbol{\xi}_0 + X_1 \boldsymbol{\delta} + \begin{pmatrix} \boldsymbol{\delta}^{\transpose} X_{21} \boldsymbol{\delta} \\ \boldsymbol{\delta}^{\transpose} X_{22} \boldsymbol{\delta} \\ \vdots \end{pmatrix}$ with $\boldsymbol{\xi}^{(0)} = (x_1^{(0)}, x_2^{(0)}, \dots y_1^{(0)}, \dots z_N^{(0)})$ the equilibrium positions of the ions without tweezers, $\boldsymbol{\delta} = (\delta_{x,1}, \delta_{x,2}, \dots \delta_{z,N})$ the shifts of the tweezer focuses and $3N \times 3N$-matrices $X_1$, $\{X_{2m}\}_{m=1}^{3N}$ that determine the new equilibrium positions from the shifts. From the condition $\nabla_{\boldsymbol{\xi}} V =0$ to all orders we obtain 
\begin{equation}
D = 0 \Leftrightarrow  X_1 = H^{-1} P
\end{equation}
where $H_{ij} = ( \partial_{\xi_i} \partial_{\xi_j} V) \vert_{\boldsymbol{\xi}_0}$ and $P_{ij} = (\partial_{\xi_i}  \partial_{\delta_j} V) \vert_{\boldsymbol{\xi}_0}$.
The second order yields
\begin{equation}
W=0 \Leftrightarrow \mathbf{X}_2 = H^{-1} \mathbf{b}
\end{equation}
where $\boldsymbol{X}_2 = (X_{2n})_n$ and $\boldsymbol{b} = (- \frac{1}{2} X_1^T \mathcal{H}_n X_1 )_n $ are vectors whose elements are matrices with $\mathcal{H}_r = \partial_{\xi_r} H$. Hence the multiplication with $H^{-1}$ gives a linear combination of matrices for each element of $\boldsymbol{X}_2$.

\section{Adiabatic switching of tweezer arrays}
\label{sec:adiab-switch-tweez}

Here we derive an estimate, based on adiabatic perturbation theory, for the
excitation of phonons due to the switching of optical tweezers. We consider the
switching protocol described by the phonon Hamiltonian~\eqref{eq:H-ph-alpha}
with time-dependent local oscillation frequencies,
\begin{equation}
  \label{eq:switching-protocol}
  \widetilde{\omega}_{\alpha, i}^2 = \omega_{\alpha}^2 - \frac{s_{\alpha} e^2 \zeta(3)}{2 \pi
    \epsilon_0 d^3 m} + \omega_{\alpha, 0}^2 \left[ \left( 1 - s \right)
    \delta_{i, 1} + \delta_{i, 2} + s \delta_{i, 3} \right],
\end{equation}
which depend linearly on the parameter $s = t/\tau_{\mathrm{s}}$ that varies
between zero and one, and where $\tau_{\mathrm{s}}$ denotes the switching
time. At the beginning of the protocol at $s = 0$, the ions at positions
$i = 1, 2$ are pinned. At the end of the protocol, the optical potential on the
ion at site $i = 1$ is switched off, and the ions at positions $i = 2, 3$ are
pinned. 

The instantaneous eigenstates and energies of the phonon Hamiltonian obey
$H_{\mathrm{ph}}(s) \ket{n(s)} = E_n(s) \ket{n(s)}$, where for simplicity we
label the states with a single index $n \in \N_0$. To obtain an estimate of the
excitation of phonons, we assume that all phonon modes are initially cooled to
their ground state. The transition probability to an excited state $n > 0$ is
given by~\cite{DeGrandi2010}
\begin{equation}
  \label{eq:p-0-n}
  P_{0 \to n} \sim \frac{1}{\tau_{\mathrm{s}}^2} \left( \left. \frac{\braket{n
          | \partial_s H_{\mathrm{ph}} | 0}^2}{\left( E_n - E_0 \right)^4} \right\rvert_{s =
      0} + \left. \frac{\braket{n | \partial_s H_{\mathrm{ph}} | 0}^2}{\left( E_n - E_0
        \right)^4} \right\rvert_{s = 1} \right),
\end{equation}
and the total probability to excite phonon modes is
$P = \sum_{n \neq 0} P_{0 \to n}$. By evaluating the matrix elements of the
phonon Hamiltonian explicitly, we find
\begin{multline}
  P = \frac{1}{ 2 \left( \omega_x \tau_{\mathrm{s}} \right)^2} \\ \times
  \sum_{\alpha \in \{x, z\}} \int_0^{\pi} \frac{\mathrm{d} k}{2 \pi} \sum_{n, n'
    = 1}^p \left( \left. A_{\alpha, k}^{n, n'} \right\rvert_{s = 0} +
    \left. A_{\alpha, k}^{n, n'} \right\rvert_{s = 1} \right),
\end{multline}
where, for $\alpha \in \{x, z\}$,
\begin{equation}  
  A_{\alpha, k}^{n, n'} = \frac{\omega_x^2}{\left( \omega_{\alpha, k, n} +
      \omega_{\alpha, k, n'} \right)^4} \sum_{\lambda, \lambda' = 1}^2 \left(
    c_{\alpha, \lambda, \lambda'}^{k, n, n'} \right)^2,
\end{equation}
and
\begin{equation}  
  c_{\alpha, \lambda, \lambda'}^{k, n, n'} =
  \frac{\omega_0^2}{2 \sqrt{\omega_{\alpha, k, n} \omega_{\alpha, k, n'}}}
  \left( W_{\alpha, 3}^{\lambda, \lambda', k, n, n'} - W_{\alpha, 1}^{\lambda, \lambda', k, n, n'}
  \right),
\end{equation}
with
\begin{equation}
  \label{eq:W}
  \begin{split}
    W_{\alpha, i}^{\lambda, \lambda, k, n, n'} & = \Xi_{\alpha, i}^{k, n, 1} \Xi_{\alpha,
      i}^{k, n', 1} + \Xi_{\alpha, i}^{k, n, 2} \Xi_{\alpha, i}^{k, n', 2}, \\
    W_{\alpha, i}^{1, 2, k, n, n'} & = - W_{\alpha, i}^{2, 1, k, n, n'} \\ & = -
    \Xi_{\alpha, i}^{k, n, 1} \Xi_{\alpha, i}^{k, n', 2} + \Xi_{\alpha, i}^{k,
      n, 2} \Xi_{\alpha, i}^{k, n', 1}.
  \end{split}
\end{equation}
These expressions yield the estimates stated in the main text in
Sec.~\ref{sec:dynam-reconf-tweez}.

\section{Entangling gates with time-modulated Rabi frequencies}
\label{sec:entangl-gates-time-mod-Rabi}

Here we specialize Eqs.~\eqref{eq:alpha} and~\eqref{eq:chi} which determine the
entangling gate Eq.~\eqref{eq:U-gate} to time-modulated Rabi
frequencies~\cite{Wu2018}.

\subsection{Finite chains}
\label{sec:finite-chains}

The qubit-phonon coupling~\eqref{eq:alpha} can be written as
$\alpha^n_i = \mathbf{A}^n_i \cdot \mathbf{R}_i$, where the dimensionless Rabi
frequency is given by
$\mathbf{R}_i = \eta_0 \boldsymbol{\Omega}_i/\omega_x = \left( R_i^1, \dotsc,
  R_i^S \right)$
with the parameter
$\eta_0$ defined in Eq.~\eqref{eq:eta-0}, and where
$A^n_{i, s} = - \imag \sqrt{\omega_x/\omega_n} M^n_i g_i^{n, s}$ with
$g_i^{n, s} = \omega_x \int_{\left( s - 1 \right) \tau/S}^{s \tau/S} \mathrm{d}
t \sin(\mu_i t) \e^{\imag \omega_n t}$.
Further, the qubit-qubit coupling~\eqref{eq:chi} can be written as
$\chi_{i, i'} = \mathbf{R}_i^{\transpose} X_{i, i'} \mathbf{R}_{i'}$, where
$X_{i, i'} = \sum_{n = 1}^N \left( \eta^n_i \eta^n_{i'}/\eta_0^2 \right) f_{i,
  i'}^n$.
$f_{i, i'}^n$ is an $S \times S$ matrix with elements
\begin{multline}  
  f_{i, i'}^{n, s, s'} = \omega_x^2 \int_{\left( s - 1 \right) \tau/S}^{s
    \tau/S} \mathrm{d} t \int_{\left( s' - 1 \right) \tau/S}^{s' \tau/S}
  \mathrm{d} t' \sin(\mu_i t)
  \sin(\mu_{i'} t') \\
  \times \sin(\omega_n \left( t - t' \right)) \quad \text{for } s > s',
\end{multline}
\begin{multline}
      f_{i, i'}^{n, s, s'} = \omega_x^2 \int_{\left( s - 1 \right) \tau/S}^{s \tau/S} \mathrm{d} t
    \int_{\left( s - 1 \right) \tau/S}^t \mathrm{d} t' \left( \sin(\mu_i t)
      \sin(\mu_{i'} t') \right. \\ \left. + \sin(\mu_i t') \sin(\mu_{i'} t)
    \right) \sin(\omega_n \left( t - t' \right)) \quad \text{for } s = s',
\end{multline}
\begin{multline}
    f_{i, i'}^{n, s, s'} = \omega_x^2 \int_{\left( s' - 1 \right) \tau/S}^{s' \tau/S} \mathrm{d} t
    \int_{\left( s - 1 \right) \tau/S}^{s \tau/S} \mathrm{d} t' \sin(\mu_i t')
    \sin(\mu_{i'} t) \\ \times \sin(\omega_n \left( t - t' \right)) \quad
    \text{for } s < s'.
\end{multline}

In terms of the vectors $\mathbf{A}_i^n$, the infidelity per gate
Eq.~\eqref{eq:infidelity} can be written as
$\delta F = \frac{1}{G} \sum_{i = 1}^N \mathbf{R}_i^{\transpose} \Phi_i
\mathbf{R}_i$
where
$\Phi_i^{s, s'} = \frac{8}{5} \sum_{n = 1}^N \Re \! \left( \left( A^n_{i, s}
  \right)^{*} A^n_{i, s'} \right).$

\subsection{Infinite chains}
\label{sec:infinite-chains}

In infinite ion chains, the qubit-qubit coupling can be written as $\chi_{i,
  i'}^{l, l'} = \mathbf{R}_{l, i}^{\transpose} X^{l, l'}_{i, i'} \mathbf{R}_{l',
i'}$, where
\begin{multline}
  \label{eq:X-infinite}
  X^{l, l'}_{i, i'} = 2 \int_0^{\pi} \frac{\mathrm{d} k}{2 \pi} \sum_{n = 1}^p
  \sqrt{\frac{\omega_x}{\omega_{k, n}}} \\ \times \left[ \cos(k \left( l - l'
    \right)) \left( \Xi_i^{k, n, 1} \Xi_{i'}^{k, n, 1} + \Xi_i^{k, n, 2}
      \Xi_{i'}^{k, n, 2} \right) \right. \\ \left. + \sin(k \left( l - l'
    \right)) \left( \Xi_i^{k, n, 1} \Xi_{i'}^{k, n, 2} - \Xi_i^{k, n, 2}
      \Xi_{i'}^{k, n, 1} \right) \right] f_{l, i, l', i'}^{k, n}.
\end{multline}
To obtain $X^{l, l'}_{i, i'}$, we approximate the integral over $k \in [0, \pi]$
by a discrete Riemann sum, and we calculate the matrices $\Xi^{k, n, \lambda}_i$
and the phonon mode frequencies $\omega_{k, n}$ for each discrete value of $k$
numerically as described in Appendix~\ref{sec:ion-chain-tweezers}.

Similarly, for the matrix $\Phi_{l, i}$ which determines the infidelity per gate
we find
\begin{multline}  
  \Phi_{l, i}^{s, s'} = \frac{32}{5} \int_0^{\pi} \frac{\mathrm{d} k}{2 \pi}
  \sum_{n = 1}^p \sqrt{\frac{\omega_x}{\omega_{k, n}}} \sum_{\lambda = 1}^2
  \left( \Xi_i^{k, n, \lambda} \right)^2 \\ \times \Re \! \left( \left( g_{l,
        i}^{k, n, s} \right)^{*} g_{l, i}^{k, n, s'} \right).
\end{multline}


\begin{thebibliography}{91}%
\makeatletter
\providecommand \@ifxundefined [1]{%
 \@ifx{#1\undefined}
}%
\providecommand \@ifnum [1]{%
 \ifnum #1\expandafter \@firstoftwo
 \else \expandafter \@secondoftwo
 \fi
}%
\providecommand \@ifx [1]{%
 \ifx #1\expandafter \@firstoftwo
 \else \expandafter \@secondoftwo
 \fi
}%
\providecommand \natexlab [1]{#1}%
\providecommand \enquote  [1]{``#1''}%
\providecommand \bibnamefont  [1]{#1}%
\providecommand \bibfnamefont [1]{#1}%
\providecommand \citenamefont [1]{#1}%
\providecommand \href@noop [0]{\@secondoftwo}%
\providecommand \href [0]{\begingroup \@sanitize@url \@href}%
\providecommand \@href[1]{\@@startlink{#1}\@@href}%
\providecommand \@@href[1]{\endgroup#1\@@endlink}%
\providecommand \@sanitize@url [0]{\catcode `\\12\catcode `\$12\catcode
  `\&12\catcode `\#12\catcode `\^12\catcode `\_12\catcode `\%12\relax}%
\providecommand \@@startlink[1]{}%
\providecommand \@@endlink[0]{}%
\providecommand \url  [0]{\begingroup\@sanitize@url \@url }%
\providecommand \@url [1]{\endgroup\@href {#1}{\urlprefix }}%
\providecommand \urlprefix  [0]{URL }%
\providecommand \Eprint [0]{\href }%
\providecommand \doibase [0]{http://dx.doi.org/}%
\providecommand \selectlanguage [0]{\@gobble}%
\providecommand \bibinfo  [0]{\@secondoftwo}%
\providecommand \bibfield  [0]{\@secondoftwo}%
\providecommand \translation [1]{[#1]}%
\providecommand \BibitemOpen [0]{}%
\providecommand \bibitemStop [0]{}%
\providecommand \bibitemNoStop [0]{.\EOS\space}%
\providecommand \EOS [0]{\spacefactor3000\relax}%
\providecommand \BibitemShut  [1]{\csname bibitem#1\endcsname}%
\let\auto@bib@innerbib\@empty
\bibitem [{\citenamefont {Cirac}\ and\ \citenamefont
  {Zoller}(1995)}]{Cirac1995}%
  \BibitemOpen
  \bibfield  {author} {\bibinfo {author} {\bibfnamefont {J.~I.}\ \bibnamefont
  {Cirac}}\ and\ \bibinfo {author} {\bibfnamefont {P.}~\bibnamefont {Zoller}},\
  }\bibfield  {title} {\enquote {\bibinfo {title} {{Quantum Computations with
  Cold Trapped Ions}},}\ }\href {\doibase 10.1103/PhysRevLett.74.4091}
  {\bibfield  {journal} {\bibinfo  {journal} {Phys. Rev. Lett.}\ }\textbf
  {\bibinfo {volume} {74}},\ \bibinfo {pages} {4091--4094} (\bibinfo {year}
  {1995})}\BibitemShut {NoStop}%
\bibitem [{\citenamefont {Gardiner}\ and\ \citenamefont
  {Zoller}(2015)}]{gardiner2015quantum}%
  \BibitemOpen
  \bibfield  {author} {\bibinfo {author} {\bibfnamefont {C.}~\bibnamefont
  {Gardiner}}\ and\ \bibinfo {author} {\bibfnamefont {P.}~\bibnamefont
  {Zoller}},\ }\bibfield  {title} {\enquote {\bibinfo {title} {The quantum
  world of ultra-cold atoms and light book ii: The physics of quantum-optical
  devices},}\ }in\ \href@noop {} {\emph {\bibinfo {booktitle} {The Quantum
  World of Ultra-Cold Atoms and Light Book II: The Physics of Quantum-Optical
  Devices}}}\ (\bibinfo  {publisher} {World Scientific},\ \bibinfo {year}
  {2015})\ pp.\ \bibinfo {pages} {1--524}\BibitemShut {NoStop}%
\bibitem [{\citenamefont {Schindler}\ \emph {et~al.}(2013)\citenamefont
  {Schindler}, \citenamefont {Nigg}, \citenamefont {Monz}, \citenamefont
  {Barreiro}, \citenamefont {Martinez}, \citenamefont {Wang}, \citenamefont
  {Quint}, \citenamefont {Brandl}, \citenamefont {Nebendahl}, \citenamefont
  {Roos}, \citenamefont {Chwalla}, \citenamefont {Hennrich},\ and\
  \citenamefont {Blatt}}]{Schindler2013}%
  \BibitemOpen
  \bibfield  {author} {\bibinfo {author} {\bibfnamefont {P.}~\bibnamefont
  {Schindler}}, \bibinfo {author} {\bibfnamefont {D.}~\bibnamefont {Nigg}},
  \bibinfo {author} {\bibfnamefont {T.}~\bibnamefont {Monz}}, \bibinfo {author}
  {\bibfnamefont {J.~T.}\ \bibnamefont {Barreiro}}, \bibinfo {author}
  {\bibfnamefont {E.}~\bibnamefont {Martinez}}, \bibinfo {author}
  {\bibfnamefont {S.~X.}\ \bibnamefont {Wang}}, \bibinfo {author}
  {\bibfnamefont {S.}~\bibnamefont {Quint}}, \bibinfo {author} {\bibfnamefont
  {M.~F.}\ \bibnamefont {Brandl}}, \bibinfo {author} {\bibfnamefont
  {V.}~\bibnamefont {Nebendahl}}, \bibinfo {author} {\bibfnamefont {C.~F.}\
  \bibnamefont {Roos}}, \bibinfo {author} {\bibfnamefont {M.}~\bibnamefont
  {Chwalla}}, \bibinfo {author} {\bibfnamefont {M.}~\bibnamefont {Hennrich}}, \
  and\ \bibinfo {author} {\bibfnamefont {R.}~\bibnamefont {Blatt}},\ }\bibfield
   {title} {\enquote {\bibinfo {title} {{A quantum information processor with
  trapped ions}},}\ }\href {\doibase 10.1088/1367-2630/15/12/123012} {\bibfield
   {journal} {\bibinfo  {journal} {New J. Phys.}\ }\textbf {\bibinfo {volume}
  {15}},\ \bibinfo {pages} {123012} (\bibinfo {year} {2013})}\BibitemShut
  {NoStop}%
\bibitem [{\citenamefont {Bruzewicz}\ \emph {et~al.}(2019)\citenamefont
  {Bruzewicz}, \citenamefont {Chiaverini}, \citenamefont {McConnell},\ and\
  \citenamefont {Sage}}]{Bruzewicz2019}%
  \BibitemOpen
  \bibfield  {author} {\bibinfo {author} {\bibfnamefont {C.~D.}\ \bibnamefont
  {Bruzewicz}}, \bibinfo {author} {\bibfnamefont {J.}~\bibnamefont
  {Chiaverini}}, \bibinfo {author} {\bibfnamefont {R.}~\bibnamefont
  {McConnell}}, \ and\ \bibinfo {author} {\bibfnamefont {J.~M.}\ \bibnamefont
  {Sage}},\ }\bibfield  {title} {\enquote {\bibinfo {title} {{Trapped-ion
  quantum computing: Progress and challenges}},}\ }\href {\doibase
  10.1063/1.5088164} {\bibfield  {journal} {\bibinfo  {journal} {Applied
  Physics Reviews}\ }\textbf {\bibinfo {volume} {6}},\ \bibinfo {pages}
  {021314} (\bibinfo {year} {2019})}\BibitemShut {NoStop}%
\bibitem [{\citenamefont {Ballance}\ \emph {et~al.}(2016)\citenamefont
  {Ballance}, \citenamefont {Harty}, \citenamefont {Linke}, \citenamefont
  {Sepiol},\ and\ \citenamefont {Lucas}}]{Ballance2016}%
  \BibitemOpen
  \bibfield  {author} {\bibinfo {author} {\bibfnamefont {C.~J.}\ \bibnamefont
  {Ballance}}, \bibinfo {author} {\bibfnamefont {T.~P.}\ \bibnamefont {Harty}},
  \bibinfo {author} {\bibfnamefont {N.~M.}\ \bibnamefont {Linke}}, \bibinfo
  {author} {\bibfnamefont {M.~A.}\ \bibnamefont {Sepiol}}, \ and\ \bibinfo
  {author} {\bibfnamefont {D.~M.}\ \bibnamefont {Lucas}},\ }\bibfield  {title}
  {\enquote {\bibinfo {title} {{High-Fidelity Quantum Logic Gates Using
  Trapped-Ion Hyperfine Qubits}},}\ }\href {\doibase
  10.1103/PhysRevLett.117.060504} {\bibfield  {journal} {\bibinfo  {journal}
  {Phys. Rev. Lett.}\ }\textbf {\bibinfo {volume} {117}},\ \bibinfo {pages}
  {060504} (\bibinfo {year} {2016})}\BibitemShut {NoStop}%
\bibitem [{\citenamefont {Shapira}\ \emph {et~al.}(2018)\citenamefont
  {Shapira}, \citenamefont {Shaniv}, \citenamefont {Manovitz}, \citenamefont
  {Akerman},\ and\ \citenamefont {Ozeri}}]{Shapira2018}%
  \BibitemOpen
  \bibfield  {author} {\bibinfo {author} {\bibfnamefont {Y.}~\bibnamefont
  {Shapira}}, \bibinfo {author} {\bibfnamefont {R.}~\bibnamefont {Shaniv}},
  \bibinfo {author} {\bibfnamefont {T.}~\bibnamefont {Manovitz}}, \bibinfo
  {author} {\bibfnamefont {N.}~\bibnamefont {Akerman}}, \ and\ \bibinfo
  {author} {\bibfnamefont {R.}~\bibnamefont {Ozeri}},\ }\bibfield  {title}
  {\enquote {\bibinfo {title} {{Robust Entanglement Gates for Trapped-Ion
  Qubits}},}\ }\href {\doibase 10.1103/PhysRevLett.121.180502} {\bibfield
  {journal} {\bibinfo  {journal} {Phys. Rev. Lett.}\ }\textbf {\bibinfo
  {volume} {121}},\ \bibinfo {pages} {180502} (\bibinfo {year}
  {2018})}\BibitemShut {NoStop}%
\bibitem [{\citenamefont {Negnevitsky}\ \emph {et~al.}(2018)\citenamefont
  {Negnevitsky}, \citenamefont {Marinelli}, \citenamefont {Mehta},
  \citenamefont {Lo}, \citenamefont {Fl{\"{u}}hmann},\ and\ \citenamefont
  {Home}}]{Negnevitsky2018}%
  \BibitemOpen
  \bibfield  {author} {\bibinfo {author} {\bibfnamefont {V.}~\bibnamefont
  {Negnevitsky}}, \bibinfo {author} {\bibfnamefont {M.}~\bibnamefont
  {Marinelli}}, \bibinfo {author} {\bibfnamefont {K.~K.}\ \bibnamefont
  {Mehta}}, \bibinfo {author} {\bibfnamefont {H.~Y.}\ \bibnamefont {Lo}},
  \bibinfo {author} {\bibfnamefont {C.}~\bibnamefont {Fl{\"{u}}hmann}}, \ and\
  \bibinfo {author} {\bibfnamefont {J.~P.}\ \bibnamefont {Home}},\ }\bibfield
  {title} {\enquote {\bibinfo {title} {{Repeated multi-qubit readout and
  feedback with a mixed-species trapped-ion register}},}\ }\href {\doibase
  10.1038/s41586-018-0668-z} {\bibfield  {journal} {\bibinfo  {journal}
  {Nature}\ }\textbf {\bibinfo {volume} {563}},\ \bibinfo {pages} {527--531}
  (\bibinfo {year} {2018})}\BibitemShut {NoStop}%
\bibitem [{\citenamefont {Harty}\ \emph {et~al.}(2014)\citenamefont {Harty},
  \citenamefont {Allcock}, \citenamefont {Ballance}, \citenamefont {Guidoni},
  \citenamefont {Janacek}, \citenamefont {Linke}, \citenamefont {Stacey},\ and\
  \citenamefont {Lucas}}]{Harty2014}%
  \BibitemOpen
  \bibfield  {author} {\bibinfo {author} {\bibfnamefont {T.~P.}\ \bibnamefont
  {Harty}}, \bibinfo {author} {\bibfnamefont {D.~T.~C.}\ \bibnamefont
  {Allcock}}, \bibinfo {author} {\bibfnamefont {C.~J.}\ \bibnamefont
  {Ballance}}, \bibinfo {author} {\bibfnamefont {L.}~\bibnamefont {Guidoni}},
  \bibinfo {author} {\bibfnamefont {H.~A.}\ \bibnamefont {Janacek}}, \bibinfo
  {author} {\bibfnamefont {N.~M.}\ \bibnamefont {Linke}}, \bibinfo {author}
  {\bibfnamefont {D.~N.}\ \bibnamefont {Stacey}}, \ and\ \bibinfo {author}
  {\bibfnamefont {D.~M.}\ \bibnamefont {Lucas}},\ }\bibfield  {title} {\enquote
  {\bibinfo {title} {{High-Fidelity Preparation, Gates, Memory, and Readout of
  a Trapped-Ion Quantum Bit}},}\ }\href {\doibase
  10.1103/PhysRevLett.113.220501} {\bibfield  {journal} {\bibinfo  {journal}
  {Phys. Rev. Lett.}\ }\textbf {\bibinfo {volume} {113}},\ \bibinfo {pages}
  {220501} (\bibinfo {year} {2014})}\BibitemShut {NoStop}%
\bibitem [{\citenamefont {Gaebler}\ \emph {et~al.}(2016)\citenamefont
  {Gaebler}, \citenamefont {Tan}, \citenamefont {Lin}, \citenamefont {Wan},
  \citenamefont {Bowler}, \citenamefont {Keith}, \citenamefont {Glancy},
  \citenamefont {Coakley}, \citenamefont {Knill}, \citenamefont {Leibfried},\
  and\ \citenamefont {Wineland}}]{Gaebler2016}%
  \BibitemOpen
  \bibfield  {author} {\bibinfo {author} {\bibfnamefont {J.~P.}\ \bibnamefont
  {Gaebler}}, \bibinfo {author} {\bibfnamefont {T.~R.}\ \bibnamefont {Tan}},
  \bibinfo {author} {\bibfnamefont {Y.}~\bibnamefont {Lin}}, \bibinfo {author}
  {\bibfnamefont {Y.}~\bibnamefont {Wan}}, \bibinfo {author} {\bibfnamefont
  {R.}~\bibnamefont {Bowler}}, \bibinfo {author} {\bibfnamefont {A.~C.}\
  \bibnamefont {Keith}}, \bibinfo {author} {\bibfnamefont {S.}~\bibnamefont
  {Glancy}}, \bibinfo {author} {\bibfnamefont {K.}~\bibnamefont {Coakley}},
  \bibinfo {author} {\bibfnamefont {E.}~\bibnamefont {Knill}}, \bibinfo
  {author} {\bibfnamefont {D.}~\bibnamefont {Leibfried}}, \ and\ \bibinfo
  {author} {\bibfnamefont {D.~J.}\ \bibnamefont {Wineland}},\ }\bibfield
  {title} {\enquote {\bibinfo {title} {{High-Fidelity Universal Gate Set for
  ${}^9\mathrm{Be}^{+}$ Ion Qubits}},}\ }\href {\doibase
  10.1103/PhysRevLett.117.060505} {\bibfield  {journal} {\bibinfo  {journal}
  {Phys. Rev. Lett.}\ }\textbf {\bibinfo {volume} {117}},\ \bibinfo {pages}
  {060505} (\bibinfo {year} {2016})}\BibitemShut {NoStop}%
\bibitem [{\citenamefont {Wright}\ \emph {et~al.}(2019)\citenamefont {Wright},
  \citenamefont {Beck}, \citenamefont {Debnath}, \citenamefont {Amini},
  \citenamefont {Nam}, \citenamefont {Grzesiak}, \citenamefont {Chen},
  \citenamefont {Pisenti}, \citenamefont {Chmielewski}, \citenamefont
  {Collins}, \citenamefont {Hudek}, \citenamefont {Mizrahi}, \citenamefont
  {Wong-Campos}, \citenamefont {Allen}, \citenamefont {Apisdorf}, \citenamefont
  {Solomon}, \citenamefont {Williams}, \citenamefont {Ducore}, \citenamefont
  {Blinov}, \citenamefont {Kreikemeier}, \citenamefont {Chaplin}, \citenamefont
  {Keesan}, \citenamefont {Monroe},\ and\ \citenamefont {Kim}}]{Wright2019}%
  \BibitemOpen
  \bibfield  {author} {\bibinfo {author} {\bibfnamefont {K.}~\bibnamefont
  {Wright}}, \bibinfo {author} {\bibfnamefont {K.~M.}\ \bibnamefont {Beck}},
  \bibinfo {author} {\bibfnamefont {S.}~\bibnamefont {Debnath}}, \bibinfo
  {author} {\bibfnamefont {J.~M.}\ \bibnamefont {Amini}}, \bibinfo {author}
  {\bibfnamefont {Y.}~\bibnamefont {Nam}}, \bibinfo {author} {\bibfnamefont
  {N.}~\bibnamefont {Grzesiak}}, \bibinfo {author} {\bibfnamefont {J.~S.}\
  \bibnamefont {Chen}}, \bibinfo {author} {\bibfnamefont {N.~C.}\ \bibnamefont
  {Pisenti}}, \bibinfo {author} {\bibfnamefont {M.}~\bibnamefont
  {Chmielewski}}, \bibinfo {author} {\bibfnamefont {C.}~\bibnamefont
  {Collins}}, \bibinfo {author} {\bibfnamefont {K.~M.}\ \bibnamefont {Hudek}},
  \bibinfo {author} {\bibfnamefont {J.}~\bibnamefont {Mizrahi}}, \bibinfo
  {author} {\bibfnamefont {J.~D.}\ \bibnamefont {Wong-Campos}}, \bibinfo
  {author} {\bibfnamefont {S.}~\bibnamefont {Allen}}, \bibinfo {author}
  {\bibfnamefont {J.}~\bibnamefont {Apisdorf}}, \bibinfo {author}
  {\bibfnamefont {P.}~\bibnamefont {Solomon}}, \bibinfo {author} {\bibfnamefont
  {M.}~\bibnamefont {Williams}}, \bibinfo {author} {\bibfnamefont {A.~M.}\
  \bibnamefont {Ducore}}, \bibinfo {author} {\bibfnamefont {A.}~\bibnamefont
  {Blinov}}, \bibinfo {author} {\bibfnamefont {S.~M.}\ \bibnamefont
  {Kreikemeier}}, \bibinfo {author} {\bibfnamefont {V.}~\bibnamefont
  {Chaplin}}, \bibinfo {author} {\bibfnamefont {M.}~\bibnamefont {Keesan}},
  \bibinfo {author} {\bibfnamefont {C.}~\bibnamefont {Monroe}}, \ and\ \bibinfo
  {author} {\bibfnamefont {J.}~\bibnamefont {Kim}},\ }\bibfield  {title}
  {\enquote {\bibinfo {title} {{Benchmarking an 11-qubit quantum computer}},}\
  }\href {\doibase 10.1038/s41467-019-13534-2} {\bibfield  {journal} {\bibinfo
  {journal} {Nature Communications}\ }\textbf {\bibinfo {volume} {10}},\
  \bibinfo {pages} {1--6} (\bibinfo {year} {2019})}\BibitemShut {NoStop}%
\bibitem [{\citenamefont {Erhard}\ \emph {et~al.}(2020)\citenamefont {Erhard},
  \citenamefont {Nautrup}, \citenamefont {Meth}, \citenamefont {Postler},
  \citenamefont {Stricker}, \citenamefont {Ringbauer}, \citenamefont
  {Schindler}, \citenamefont {Briegel}, \citenamefont {Blatt}, \citenamefont
  {Friis},\ and\ \citenamefont {Monz}}]{Erhard2020}%
  \BibitemOpen
  \bibfield  {author} {\bibinfo {author} {\bibfnamefont {A.}~\bibnamefont
  {Erhard}}, \bibinfo {author} {\bibfnamefont {H.~P.}\ \bibnamefont {Nautrup}},
  \bibinfo {author} {\bibfnamefont {M.}~\bibnamefont {Meth}}, \bibinfo {author}
  {\bibfnamefont {L.}~\bibnamefont {Postler}}, \bibinfo {author} {\bibfnamefont
  {R.}~\bibnamefont {Stricker}}, \bibinfo {author} {\bibfnamefont
  {M.}~\bibnamefont {Ringbauer}}, \bibinfo {author} {\bibfnamefont
  {P.}~\bibnamefont {Schindler}}, \bibinfo {author} {\bibfnamefont {H.~J.}\
  \bibnamefont {Briegel}}, \bibinfo {author} {\bibfnamefont {R.}~\bibnamefont
  {Blatt}}, \bibinfo {author} {\bibfnamefont {N.}~\bibnamefont {Friis}}, \ and\
  \bibinfo {author} {\bibfnamefont {T.}~\bibnamefont {Monz}},\ }\bibfield
  {title} {\enquote {\bibinfo {title} {{Entangling logical qubits with lattice
  surgery}},}\ }\href {http://arxiv.org/abs/2006.03071} {\bibfield  {journal}
  {\bibinfo  {journal} {arXiv:2006.03071}\ } (\bibinfo {year}
  {2020})}\BibitemShut {NoStop}%
\bibitem [{\citenamefont {Lu}\ \emph {et~al.}(2019{\natexlab{a}})\citenamefont
  {Lu}, \citenamefont {Zhang}, \citenamefont {Zhang}, \citenamefont {Chen},
  \citenamefont {Shen}, \citenamefont {Zhang}, \citenamefont {Zhang},\ and\
  \citenamefont {Kim}}]{Lu2019}%
  \BibitemOpen
  \bibfield  {author} {\bibinfo {author} {\bibfnamefont {Y.}~\bibnamefont
  {Lu}}, \bibinfo {author} {\bibfnamefont {S.}~\bibnamefont {Zhang}}, \bibinfo
  {author} {\bibfnamefont {K.}~\bibnamefont {Zhang}}, \bibinfo {author}
  {\bibfnamefont {W.}~\bibnamefont {Chen}}, \bibinfo {author} {\bibfnamefont
  {Y.}~\bibnamefont {Shen}}, \bibinfo {author} {\bibfnamefont {J.}~\bibnamefont
  {Zhang}}, \bibinfo {author} {\bibfnamefont {J.~N.}\ \bibnamefont {Zhang}}, \
  and\ \bibinfo {author} {\bibfnamefont {K.}~\bibnamefont {Kim}},\ }\bibfield
  {title} {\enquote {\bibinfo {title} {{Global entangling gates on arbitrary
  ion qubits}},}\ }\href {\doibase 10.1038/s41586-019-1428-4} {\bibfield
  {journal} {\bibinfo  {journal} {Nature}\ }\textbf {\bibinfo {volume} {572}},\
  \bibinfo {pages} {363--367} (\bibinfo {year}
  {2019}{\natexlab{a}})}\BibitemShut {NoStop}%
\bibitem [{\citenamefont {Zhu}\ \emph {et~al.}(2019)\citenamefont {Zhu},
  \citenamefont {Linke}, \citenamefont {Benedetti}, \citenamefont {Landsman},
  \citenamefont {Nguyen}, \citenamefont {Alderete}, \citenamefont
  {Perdomo-Ortiz}, \citenamefont {Korda}, \citenamefont {Garfoot},
  \citenamefont {Brecque}, \citenamefont {Egan}, \citenamefont {Perdomo},\ and\
  \citenamefont {Monroe}}]{Zhu2019}%
  \BibitemOpen
  \bibfield  {author} {\bibinfo {author} {\bibfnamefont {D.}~\bibnamefont
  {Zhu}}, \bibinfo {author} {\bibfnamefont {N.~M.}\ \bibnamefont {Linke}},
  \bibinfo {author} {\bibfnamefont {M.}~\bibnamefont {Benedetti}}, \bibinfo
  {author} {\bibfnamefont {K.~A.}\ \bibnamefont {Landsman}}, \bibinfo {author}
  {\bibfnamefont {N.~H.}\ \bibnamefont {Nguyen}}, \bibinfo {author}
  {\bibfnamefont {C.~H.}\ \bibnamefont {Alderete}}, \bibinfo {author}
  {\bibfnamefont {A.}~\bibnamefont {Perdomo-Ortiz}}, \bibinfo {author}
  {\bibfnamefont {N.}~\bibnamefont {Korda}}, \bibinfo {author} {\bibfnamefont
  {A.}~\bibnamefont {Garfoot}}, \bibinfo {author} {\bibfnamefont
  {C.}~\bibnamefont {Brecque}}, \bibinfo {author} {\bibfnamefont
  {L.}~\bibnamefont {Egan}}, \bibinfo {author} {\bibfnamefont {O.}~\bibnamefont
  {Perdomo}}, \ and\ \bibinfo {author} {\bibfnamefont {C.}~\bibnamefont
  {Monroe}},\ }\bibfield  {title} {\enquote {\bibinfo {title} {{Training of
  quantum circuits on a hybrid quantum computer}},}\ }\href {\doibase
  10.1126/sciadv.aaw9918} {\bibfield  {journal} {\bibinfo  {journal} {Science
  Advances}\ }\textbf {\bibinfo {volume} {5}},\ \bibinfo {pages} {eaaw9918}
  (\bibinfo {year} {2019})}\BibitemShut {NoStop}%
\bibitem [{\citenamefont {Wan}\ \emph {et~al.}(2019)\citenamefont {Wan},
  \citenamefont {Kienzler}, \citenamefont {Erickson}, \citenamefont {Mayer},
  \citenamefont {Tan}, \citenamefont {Wu}, \citenamefont {Vasconcelos},
  \citenamefont {Glancy}, \citenamefont {Knill}, \citenamefont {Wineland},
  \citenamefont {Wilson},\ and\ \citenamefont {Leibfried}}]{Wan2019}%
  \BibitemOpen
  \bibfield  {author} {\bibinfo {author} {\bibfnamefont {Y.}~\bibnamefont
  {Wan}}, \bibinfo {author} {\bibfnamefont {D.}~\bibnamefont {Kienzler}},
  \bibinfo {author} {\bibfnamefont {S.~D.}\ \bibnamefont {Erickson}}, \bibinfo
  {author} {\bibfnamefont {K.~H.}\ \bibnamefont {Mayer}}, \bibinfo {author}
  {\bibfnamefont {T.~R.}\ \bibnamefont {Tan}}, \bibinfo {author} {\bibfnamefont
  {J.~J.}\ \bibnamefont {Wu}}, \bibinfo {author} {\bibfnamefont {H.~M.}\
  \bibnamefont {Vasconcelos}}, \bibinfo {author} {\bibfnamefont
  {S.}~\bibnamefont {Glancy}}, \bibinfo {author} {\bibfnamefont
  {E.}~\bibnamefont {Knill}}, \bibinfo {author} {\bibfnamefont {D.~J.}\
  \bibnamefont {Wineland}}, \bibinfo {author} {\bibfnamefont {A.~C.}\
  \bibnamefont {Wilson}}, \ and\ \bibinfo {author} {\bibfnamefont
  {D.}~\bibnamefont {Leibfried}},\ }\bibfield  {title} {\enquote {\bibinfo
  {title} {{Quantum gate teleportation between separated qubits in a
  trapped-ion processor}},}\ }\href {\doibase 10.1126/science.aaw9415}
  {\bibfield  {journal} {\bibinfo  {journal} {Science}\ }\textbf {\bibinfo
  {volume} {364}},\ \bibinfo {pages} {875--878} (\bibinfo {year}
  {2019})}\BibitemShut {NoStop}%
\bibitem [{\citenamefont {James}(1998)}]{James1998Quantum}%
  \BibitemOpen
  \bibfield  {author} {\bibinfo {author} {\bibfnamefont {D.~F.~V.}\
  \bibnamefont {James}},\ }\bibfield  {title} {\enquote {\bibinfo {title}
  {{Quantum dynamics of cold trapped ions with application to quantum
  computation}},}\ }\href {\doibase 10.1007/s003400050373} {\bibfield
  {journal} {\bibinfo  {journal} {Applied Physics B: Lasers and Optics}\
  }\textbf {\bibinfo {volume} {66}},\ \bibinfo {pages} {181--190} (\bibinfo
  {year} {1998})}\BibitemShut {NoStop}%
\bibitem [{\citenamefont {Schiffer}(1993)}]{Schiffer1993}%
  \BibitemOpen
  \bibfield  {author} {\bibinfo {author} {\bibfnamefont {J.~P.}\ \bibnamefont
  {Schiffer}},\ }\bibfield  {title} {\enquote {\bibinfo {title} {{Phase
  transitions in anisotropically confined ionic crystals}},}\ }\href {\doibase
  10.1103/PhysRevLett.70.818} {\bibfield  {journal} {\bibinfo  {journal} {Phys.
  Rev. Lett.}\ }\textbf {\bibinfo {volume} {70}},\ \bibinfo {pages} {818--821}
  (\bibinfo {year} {1993})}\BibitemShut {NoStop}%
\bibitem [{\citenamefont {Zhu}\ \emph {et~al.}(2006{\natexlab{a}})\citenamefont
  {Zhu}, \citenamefont {Monroe},\ and\ \citenamefont {Duan}}]{Zhu2006}%
  \BibitemOpen
  \bibfield  {author} {\bibinfo {author} {\bibfnamefont {S.~L.}\ \bibnamefont
  {Zhu}}, \bibinfo {author} {\bibfnamefont {C.}~\bibnamefont {Monroe}}, \ and\
  \bibinfo {author} {\bibfnamefont {L.~M.}\ \bibnamefont {Duan}},\ }\bibfield
  {title} {\enquote {\bibinfo {title} {{Trapped ion quantum computation with
  transverse phonon modes}},}\ }\href {\doibase 10.1103/PhysRevLett.97.050505}
  {\bibfield  {journal} {\bibinfo  {journal} {Phys. Rev. Lett.}\ }\textbf
  {\bibinfo {volume} {97}},\ \bibinfo {pages} {050505} (\bibinfo {year}
  {2006}{\natexlab{a}})}\BibitemShut {NoStop}%
\bibitem [{\citenamefont {Zhu}\ \emph {et~al.}(2006{\natexlab{b}})\citenamefont
  {Zhu}, \citenamefont {Monroe},\ and\ \citenamefont {Duan}}]{Zhu2006a}%
  \BibitemOpen
  \bibfield  {author} {\bibinfo {author} {\bibfnamefont {S.-L.}\ \bibnamefont
  {Zhu}}, \bibinfo {author} {\bibfnamefont {C.}~\bibnamefont {Monroe}}, \ and\
  \bibinfo {author} {\bibfnamefont {L.-M.}\ \bibnamefont {Duan}},\ }\bibfield
  {title} {\enquote {\bibinfo {title} {{Arbitrary-speed quantum gates within
  large ion crystals through minimum control of laser beams}},}\ }\href
  {\doibase 10.1209/epl/i2005-10424-4} {\bibfield  {journal} {\bibinfo
  {journal} {Europhys. Lett.}\ }\textbf {\bibinfo {volume} {73}},\ \bibinfo
  {pages} {485--491} (\bibinfo {year} {2006}{\natexlab{b}})}\BibitemShut
  {NoStop}%
\bibitem [{\citenamefont {Roos}(2008)}]{Roos2008}%
  \BibitemOpen
  \bibfield  {author} {\bibinfo {author} {\bibfnamefont {C.~F.}\ \bibnamefont
  {Roos}},\ }\bibfield  {title} {\enquote {\bibinfo {title} {{Ion trap quantum
  gates with amplitude-modulated laser beams}},}\ }\href {\doibase
  10.1088/1367-2630/10/1/013002} {\bibfield  {journal} {\bibinfo  {journal}
  {New J. Phys.}\ }\textbf {\bibinfo {volume} {10}},\ \bibinfo {pages} {013002}
  (\bibinfo {year} {2008})}\BibitemShut {NoStop}%
\bibitem [{\citenamefont {Choi}\ \emph {et~al.}(2014)\citenamefont {Choi},
  \citenamefont {Debnath}, \citenamefont {Manning}, \citenamefont {Figgatt},
  \citenamefont {Gong}, \citenamefont {Duan},\ and\ \citenamefont
  {Monroe}}]{Choi2014}%
  \BibitemOpen
  \bibfield  {author} {\bibinfo {author} {\bibfnamefont {T.}~\bibnamefont
  {Choi}}, \bibinfo {author} {\bibfnamefont {S.}~\bibnamefont {Debnath}},
  \bibinfo {author} {\bibfnamefont {T.~A.}\ \bibnamefont {Manning}}, \bibinfo
  {author} {\bibfnamefont {C.}~\bibnamefont {Figgatt}}, \bibinfo {author}
  {\bibfnamefont {Z.-X.}\ \bibnamefont {Gong}}, \bibinfo {author}
  {\bibfnamefont {L.-M.}\ \bibnamefont {Duan}}, \ and\ \bibinfo {author}
  {\bibfnamefont {C.}~\bibnamefont {Monroe}},\ }\bibfield  {title} {\enquote
  {\bibinfo {title} {{Optimal Quantum Control of Multimode Couplings between
  Trapped Ion Qubits for Scalable Entanglement}},}\ }\href {\doibase
  10.1103/PhysRevLett.112.190502} {\bibfield  {journal} {\bibinfo  {journal}
  {Phys. Rev. Lett.}\ }\textbf {\bibinfo {volume} {112}},\ \bibinfo {pages}
  {190502} (\bibinfo {year} {2014})}\BibitemShut {NoStop}%
\bibitem [{\citenamefont {Debnath}\ \emph {et~al.}(2016)\citenamefont
  {Debnath}, \citenamefont {Linke}, \citenamefont {Figgatt}, \citenamefont
  {Landsman}, \citenamefont {Wright},\ and\ \citenamefont
  {Monroe}}]{Debnath2016}%
  \BibitemOpen
  \bibfield  {author} {\bibinfo {author} {\bibfnamefont {S.}~\bibnamefont
  {Debnath}}, \bibinfo {author} {\bibfnamefont {N.~M.}\ \bibnamefont {Linke}},
  \bibinfo {author} {\bibfnamefont {C.}~\bibnamefont {Figgatt}}, \bibinfo
  {author} {\bibfnamefont {K.~A.}\ \bibnamefont {Landsman}}, \bibinfo {author}
  {\bibfnamefont {K.}~\bibnamefont {Wright}}, \ and\ \bibinfo {author}
  {\bibfnamefont {C.}~\bibnamefont {Monroe}},\ }\bibfield  {title} {\enquote
  {\bibinfo {title} {{Demonstration of a small programmable quantum computer
  with atomic qubits}},}\ }\href {\doibase 10.1038/nature18648} {\bibfield
  {journal} {\bibinfo  {journal} {Nature}\ }\textbf {\bibinfo {volume} {536}},\
  \bibinfo {pages} {63--66} (\bibinfo {year} {2016})}\BibitemShut {NoStop}%
\bibitem [{\citenamefont {Wu}\ \emph {et~al.}(2018)\citenamefont {Wu},
  \citenamefont {Wang},\ and\ \citenamefont {Duan}}]{Wu2018}%
  \BibitemOpen
  \bibfield  {author} {\bibinfo {author} {\bibfnamefont {Y.}~\bibnamefont
  {Wu}}, \bibinfo {author} {\bibfnamefont {S.-T.}\ \bibnamefont {Wang}}, \ and\
  \bibinfo {author} {\bibfnamefont {L.-M.}\ \bibnamefont {Duan}},\ }\bibfield
  {title} {\enquote {\bibinfo {title} {{Noise analysis for high-fidelity
  quantum entangling gates in an anharmonic linear Paul trap}},}\ }\href
  {\doibase 10.1103/PhysRevA.97.062325} {\bibfield  {journal} {\bibinfo
  {journal} {Phys. Rev. A}\ }\textbf {\bibinfo {volume} {97}},\ \bibinfo
  {pages} {062325} (\bibinfo {year} {2018})}\BibitemShut {NoStop}%
\bibitem [{\citenamefont {Landsman}\ \emph {et~al.}(2019)\citenamefont
  {Landsman}, \citenamefont {Wu}, \citenamefont {Leung}, \citenamefont {Zhu},
  \citenamefont {Linke}, \citenamefont {Brown}, \citenamefont {Duan},\ and\
  \citenamefont {Monroe}}]{Landsman2019a}%
  \BibitemOpen
  \bibfield  {author} {\bibinfo {author} {\bibfnamefont {K.~A.}\ \bibnamefont
  {Landsman}}, \bibinfo {author} {\bibfnamefont {Y.}~\bibnamefont {Wu}},
  \bibinfo {author} {\bibfnamefont {P.~H.}\ \bibnamefont {Leung}}, \bibinfo
  {author} {\bibfnamefont {D.}~\bibnamefont {Zhu}}, \bibinfo {author}
  {\bibfnamefont {N.~M.}\ \bibnamefont {Linke}}, \bibinfo {author}
  {\bibfnamefont {K.~R.}\ \bibnamefont {Brown}}, \bibinfo {author}
  {\bibfnamefont {L.}~\bibnamefont {Duan}}, \ and\ \bibinfo {author}
  {\bibfnamefont {C.}~\bibnamefont {Monroe}},\ }\bibfield  {title} {\enquote
  {\bibinfo {title} {{Two-qubit entangling gates within arbitrarily long chains
  of trapped ions}},}\ }\href {\doibase 10.1103/physreva.100.022332} {\bibfield
   {journal} {\bibinfo  {journal} {Phys. Rev. A}\ }\textbf {\bibinfo {volume}
  {100}},\ \bibinfo {pages} {022332} (\bibinfo {year} {2019})}\BibitemShut
  {NoStop}%
\bibitem [{\citenamefont {Figgatt}\ \emph {et~al.}(2019)\citenamefont
  {Figgatt}, \citenamefont {Ostrander}, \citenamefont {Linke}, \citenamefont
  {Landsman}, \citenamefont {Zhu}, \citenamefont {Maslov},\ and\ \citenamefont
  {Monroe}}]{Figgatt2019}%
  \BibitemOpen
  \bibfield  {author} {\bibinfo {author} {\bibfnamefont {C.}~\bibnamefont
  {Figgatt}}, \bibinfo {author} {\bibfnamefont {A.}~\bibnamefont {Ostrander}},
  \bibinfo {author} {\bibfnamefont {N.~M.}\ \bibnamefont {Linke}}, \bibinfo
  {author} {\bibfnamefont {K.~A.}\ \bibnamefont {Landsman}}, \bibinfo {author}
  {\bibfnamefont {D.}~\bibnamefont {Zhu}}, \bibinfo {author} {\bibfnamefont
  {D.}~\bibnamefont {Maslov}}, \ and\ \bibinfo {author} {\bibfnamefont
  {C.}~\bibnamefont {Monroe}},\ }\bibfield  {title} {\enquote {\bibinfo {title}
  {{Parallel entangling operations on a universal ion-trap quantum
  computer}},}\ }\href {\doibase 10.1038/s41586-019-1427-5} {\bibfield
  {journal} {\bibinfo  {journal} {Nature}\ }\textbf {\bibinfo {volume} {572}},\
  \bibinfo {pages} {368--372} (\bibinfo {year} {2019})}\BibitemShut {NoStop}%
\bibitem [{\citenamefont {Lu}\ \emph {et~al.}(2019{\natexlab{b}})\citenamefont
  {Lu}, \citenamefont {Zhang}, \citenamefont {Zhang}, \citenamefont {Chen},
  \citenamefont {Shen}, \citenamefont {Zhang}, \citenamefont {Zhang},\ and\
  \citenamefont {Kim}}]{Lu2019a}%
  \BibitemOpen
  \bibfield  {author} {\bibinfo {author} {\bibfnamefont {Y.}~\bibnamefont
  {Lu}}, \bibinfo {author} {\bibfnamefont {S.}~\bibnamefont {Zhang}}, \bibinfo
  {author} {\bibfnamefont {K.}~\bibnamefont {Zhang}}, \bibinfo {author}
  {\bibfnamefont {W.}~\bibnamefont {Chen}}, \bibinfo {author} {\bibfnamefont
  {Y.}~\bibnamefont {Shen}}, \bibinfo {author} {\bibfnamefont {J.}~\bibnamefont
  {Zhang}}, \bibinfo {author} {\bibfnamefont {J.-N.}\ \bibnamefont {Zhang}}, \
  and\ \bibinfo {author} {\bibfnamefont {K.}~\bibnamefont {Kim}},\ }\bibfield
  {title} {\enquote {\bibinfo {title} {{Global entangling gates on arbitrary
  ion qubits}},}\ }\href {\doibase 10.1038/s41586-019-1428-4} {\bibfield
  {journal} {\bibinfo  {journal} {Nature}\ }\textbf {\bibinfo {volume} {572}},\
  \bibinfo {pages} {363--367} (\bibinfo {year}
  {2019}{\natexlab{b}})}\BibitemShut {NoStop}%
\bibitem [{\citenamefont {Leung}\ \emph {et~al.}(2018)\citenamefont {Leung},
  \citenamefont {Landsman}, \citenamefont {Figgatt}, \citenamefont {Linke},
  \citenamefont {Monroe},\ and\ \citenamefont {Brown}}]{Leung2018}%
  \BibitemOpen
  \bibfield  {author} {\bibinfo {author} {\bibfnamefont {P.~H.}\ \bibnamefont
  {Leung}}, \bibinfo {author} {\bibfnamefont {K.~A.}\ \bibnamefont {Landsman}},
  \bibinfo {author} {\bibfnamefont {C.}~\bibnamefont {Figgatt}}, \bibinfo
  {author} {\bibfnamefont {N.~M.}\ \bibnamefont {Linke}}, \bibinfo {author}
  {\bibfnamefont {C.}~\bibnamefont {Monroe}}, \ and\ \bibinfo {author}
  {\bibfnamefont {K.~R.}\ \bibnamefont {Brown}},\ }\bibfield  {title} {\enquote
  {\bibinfo {title} {{Robust 2-Qubit Gates in a Linear Ion Crystal Using a
  Frequency-Modulated Driving Force}},}\ }\href {\doibase
  10.1103/PhysRevLett.120.020501} {\bibfield  {journal} {\bibinfo  {journal}
  {Phys. Rev. Lett.}\ }\textbf {\bibinfo {volume} {120}},\ \bibinfo {pages}
  {020501} (\bibinfo {year} {2018})}\BibitemShut {NoStop}%
\bibitem [{\citenamefont {Kielpinski}\ \emph {et~al.}(2002)\citenamefont
  {Kielpinski}, \citenamefont {Monroe},\ and\ \citenamefont
  {Wineland}}]{Kielpinski2002}%
  \BibitemOpen
  \bibfield  {author} {\bibinfo {author} {\bibfnamefont {D.}~\bibnamefont
  {Kielpinski}}, \bibinfo {author} {\bibfnamefont {C.}~\bibnamefont {Monroe}},
  \ and\ \bibinfo {author} {\bibfnamefont {D.~J.}\ \bibnamefont {Wineland}},\
  }\bibfield  {title} {\enquote {\bibinfo {title} {{Architecture for a
  large-scale ion-trap quantum computer}},}\ }\href {\doibase
  10.1038/nature00784} {\bibfield  {journal} {\bibinfo  {journal} {Nature}\
  }\textbf {\bibinfo {volume} {417}},\ \bibinfo {pages} {709--711} (\bibinfo
  {year} {2002})}\BibitemShut {NoStop}%
\bibitem [{\citenamefont {Kaufmann}\ \emph {et~al.}(2017)\citenamefont
  {Kaufmann}, \citenamefont {Ruster}, \citenamefont {Schmiegelow},
  \citenamefont {Luda}, \citenamefont {Kaushal}, \citenamefont {Schulz},
  \citenamefont {von Lindenfels}, \citenamefont {Schmidt-Kaler},\ and\
  \citenamefont {Poschinger}}]{Kaufmann2017}%
  \BibitemOpen
  \bibfield  {author} {\bibinfo {author} {\bibfnamefont {H.}~\bibnamefont
  {Kaufmann}}, \bibinfo {author} {\bibfnamefont {T.}~\bibnamefont {Ruster}},
  \bibinfo {author} {\bibfnamefont {C.~T.}\ \bibnamefont {Schmiegelow}},
  \bibinfo {author} {\bibfnamefont {M.~A.}\ \bibnamefont {Luda}}, \bibinfo
  {author} {\bibfnamefont {V.}~\bibnamefont {Kaushal}}, \bibinfo {author}
  {\bibfnamefont {J.}~\bibnamefont {Schulz}}, \bibinfo {author} {\bibfnamefont
  {D.}~\bibnamefont {von Lindenfels}}, \bibinfo {author} {\bibfnamefont
  {F.}~\bibnamefont {Schmidt-Kaler}}, \ and\ \bibinfo {author} {\bibfnamefont
  {U.~G.}\ \bibnamefont {Poschinger}},\ }\bibfield  {title} {\enquote {\bibinfo
  {title} {{Scalable Creation of Long-Lived Multipartite Entanglement}},}\
  }\href {\doibase 10.1103/PhysRevLett.119.150503} {\bibfield  {journal}
  {\bibinfo  {journal} {Phys. Rev. Lett.}\ }\textbf {\bibinfo {volume} {119}},\
  \bibinfo {pages} {150503} (\bibinfo {year} {2017})}\BibitemShut {NoStop}%
\bibitem [{\citenamefont {Blakestad}\ \emph {et~al.}(2009)\citenamefont
  {Blakestad}, \citenamefont {Ospelkaus}, \citenamefont {VanDevender},
  \citenamefont {Amini}, \citenamefont {Britton}, \citenamefont {Leibfried},\
  and\ \citenamefont {Wineland}}]{blakestad-prl-102-153002}%
  \BibitemOpen
  \bibfield  {author} {\bibinfo {author} {\bibfnamefont {R.~B.}\ \bibnamefont
  {Blakestad}}, \bibinfo {author} {\bibfnamefont {C.}~\bibnamefont
  {Ospelkaus}}, \bibinfo {author} {\bibfnamefont {A.~P.}\ \bibnamefont
  {VanDevender}}, \bibinfo {author} {\bibfnamefont {J.~M.}\ \bibnamefont
  {Amini}}, \bibinfo {author} {\bibfnamefont {J.}~\bibnamefont {Britton}},
  \bibinfo {author} {\bibfnamefont {D.}~\bibnamefont {Leibfried}}, \ and\
  \bibinfo {author} {\bibfnamefont {D.~J.}\ \bibnamefont {Wineland}},\
  }\bibfield  {title} {\enquote {\bibinfo {title} {{High-Fidelity Transport of
  Trapped-Ion Qubits through an $\mathbf{X}$-Junction Trap Array}},}\ }\href
  {\doibase 10.1103/PhysRevLett.102.153002} {\bibfield  {journal} {\bibinfo
  {journal} {Phys. Rev. Lett.}\ }\textbf {\bibinfo {volume} {102}},\ \bibinfo
  {pages} {153002} (\bibinfo {year} {2009})}\BibitemShut {NoStop}%
\bibitem [{\citenamefont {Pino}\ \emph {et~al.}(2020)\citenamefont {Pino},
  \citenamefont {Dreiling}, \citenamefont {Figgatt}, \citenamefont {Gaebler},
  \citenamefont {Moses}, \citenamefont {Allman}, \citenamefont {Baldwin},
  \citenamefont {Foss-Feig}, \citenamefont {Hayes}, \citenamefont {Mayer},
  \citenamefont {Ryan-Anderson},\ and\ \citenamefont {Neyenhuis}}]{Pino2020}%
  \BibitemOpen
  \bibfield  {author} {\bibinfo {author} {\bibfnamefont {J.~M.}\ \bibnamefont
  {Pino}}, \bibinfo {author} {\bibfnamefont {J.~M.}\ \bibnamefont {Dreiling}},
  \bibinfo {author} {\bibfnamefont {C.}~\bibnamefont {Figgatt}}, \bibinfo
  {author} {\bibfnamefont {J.~P.}\ \bibnamefont {Gaebler}}, \bibinfo {author}
  {\bibfnamefont {S.~A.}\ \bibnamefont {Moses}}, \bibinfo {author}
  {\bibfnamefont {M.~S.}\ \bibnamefont {Allman}}, \bibinfo {author}
  {\bibfnamefont {C.~H.}\ \bibnamefont {Baldwin}}, \bibinfo {author}
  {\bibfnamefont {M.}~\bibnamefont {Foss-Feig}}, \bibinfo {author}
  {\bibfnamefont {D.}~\bibnamefont {Hayes}}, \bibinfo {author} {\bibfnamefont
  {K.}~\bibnamefont {Mayer}}, \bibinfo {author} {\bibfnamefont
  {C.}~\bibnamefont {Ryan-Anderson}}, \ and\ \bibinfo {author} {\bibfnamefont
  {B.}~\bibnamefont {Neyenhuis}},\ }\bibfield  {title} {\enquote {\bibinfo
  {title} {{Demonstration of the QCCD trapped-ion quantum computer
  architecture}},}\ }\href {http://arxiv.org/abs/2003.01293} {\bibfield
  {journal} {\bibinfo  {journal} {arXiv:2003.01293}\ } (\bibinfo {year}
  {2020})}\BibitemShut {NoStop}%
\bibitem [{\citenamefont {Leibfried}\ \emph {et~al.}(2003)\citenamefont
  {Leibfried}, \citenamefont {Blatt}, \citenamefont {Monroe},\ and\
  \citenamefont {Wineland}}]{Leibfried2003a}%
  \BibitemOpen
  \bibfield  {author} {\bibinfo {author} {\bibfnamefont {D.}~\bibnamefont
  {Leibfried}}, \bibinfo {author} {\bibfnamefont {R.}~\bibnamefont {Blatt}},
  \bibinfo {author} {\bibfnamefont {C.}~\bibnamefont {Monroe}}, \ and\ \bibinfo
  {author} {\bibfnamefont {D.}~\bibnamefont {Wineland}},\ }\bibfield  {title}
  {\enquote {\bibinfo {title} {{Quantum dynamics of single trapped ions}},}\
  }\href {\doibase 10.1103/RevModPhys.75.281} {\bibfield  {journal} {\bibinfo
  {journal} {Rev. Mod. Phys.}\ }\textbf {\bibinfo {volume} {75}},\ \bibinfo
  {pages} {281--324} (\bibinfo {year} {2003})}\BibitemShut {NoStop}%
\bibitem [{\citenamefont {Omran}\ \emph {et~al.}(2019)\citenamefont {Omran},
  \citenamefont {Levine}, \citenamefont {Keesling}, \citenamefont {Semeghini},
  \citenamefont {Wang}, \citenamefont {Ebadi}, \citenamefont {Bernien},
  \citenamefont {Zibrov}, \citenamefont {Pichler}, \citenamefont {Choi},
  \citenamefont {Cui}, \citenamefont {Rossignolo}, \citenamefont {Rembold},
  \citenamefont {Montangero}, \citenamefont {Calarco}, \citenamefont {Endres},
  \citenamefont {Greiner}, \citenamefont {Vuleti{\'c}},\ and\ \citenamefont
  {Lukin}}]{Omran570}%
  \BibitemOpen
  \bibfield  {author} {\bibinfo {author} {\bibfnamefont {A.}~\bibnamefont
  {Omran}}, \bibinfo {author} {\bibfnamefont {H.}~\bibnamefont {Levine}},
  \bibinfo {author} {\bibfnamefont {A.}~\bibnamefont {Keesling}}, \bibinfo
  {author} {\bibfnamefont {G.}~\bibnamefont {Semeghini}}, \bibinfo {author}
  {\bibfnamefont {T.~T.}\ \bibnamefont {Wang}}, \bibinfo {author}
  {\bibfnamefont {S.}~\bibnamefont {Ebadi}}, \bibinfo {author} {\bibfnamefont
  {H.}~\bibnamefont {Bernien}}, \bibinfo {author} {\bibfnamefont {A.~S.}\
  \bibnamefont {Zibrov}}, \bibinfo {author} {\bibfnamefont {H.}~\bibnamefont
  {Pichler}}, \bibinfo {author} {\bibfnamefont {S.}~\bibnamefont {Choi}},
  \bibinfo {author} {\bibfnamefont {J.}~\bibnamefont {Cui}}, \bibinfo {author}
  {\bibfnamefont {M.}~\bibnamefont {Rossignolo}}, \bibinfo {author}
  {\bibfnamefont {P.}~\bibnamefont {Rembold}}, \bibinfo {author} {\bibfnamefont
  {S.}~\bibnamefont {Montangero}}, \bibinfo {author} {\bibfnamefont
  {T.}~\bibnamefont {Calarco}}, \bibinfo {author} {\bibfnamefont
  {M.}~\bibnamefont {Endres}}, \bibinfo {author} {\bibfnamefont
  {M.}~\bibnamefont {Greiner}}, \bibinfo {author} {\bibfnamefont
  {V.}~\bibnamefont {Vuleti{\'c}}}, \ and\ \bibinfo {author} {\bibfnamefont
  {M.~D.}\ \bibnamefont {Lukin}},\ }\bibfield  {title} {\enquote {\bibinfo
  {title} {Generation and manipulation of schr{\"o}dinger cat states in rydberg
  atom arrays},}\ }\href {\doibase 10.1126/science.aax9743} {\bibfield
  {journal} {\bibinfo  {journal} {Science}\ }\textbf {\bibinfo {volume}
  {365}},\ \bibinfo {pages} {570--574} (\bibinfo {year} {2019})}\BibitemShut
  {NoStop}%
\bibitem [{\citenamefont {Barredo}\ \emph {et~al.}(2018)\citenamefont
  {Barredo}, \citenamefont {Lienhard}, \citenamefont {de~L{\'{e}}s{\'{e}}leuc},
  \citenamefont {Lahaye},\ and\ \citenamefont {Browaeys}}]{Barredo2018}%
  \BibitemOpen
  \bibfield  {author} {\bibinfo {author} {\bibfnamefont {D.}~\bibnamefont
  {Barredo}}, \bibinfo {author} {\bibfnamefont {V.}~\bibnamefont {Lienhard}},
  \bibinfo {author} {\bibfnamefont {S.}~\bibnamefont
  {de~L{\'{e}}s{\'{e}}leuc}}, \bibinfo {author} {\bibfnamefont
  {T.}~\bibnamefont {Lahaye}}, \ and\ \bibinfo {author} {\bibfnamefont
  {A.}~\bibnamefont {Browaeys}},\ }\bibfield  {title} {\enquote {\bibinfo
  {title} {{Synthetic three-dimensional atomic structures assembled atom by
  atom}},}\ }\href {\doibase 10.1038/s41586-018-0450-2} {\bibfield  {journal}
  {\bibinfo  {journal} {Nature}\ }\textbf {\bibinfo {volume} {561}},\ \bibinfo
  {pages} {79--82} (\bibinfo {year} {2018})}\BibitemShut {NoStop}%
\bibitem [{\citenamefont {Shen}\ and\ \citenamefont {Lin}(2020)}]{Shen2019}%
  \BibitemOpen
  \bibfield  {author} {\bibinfo {author} {\bibfnamefont {Y.-C.}\ \bibnamefont
  {Shen}}\ and\ \bibinfo {author} {\bibfnamefont {G.-D.}\ \bibnamefont {Lin}},\
  }\bibfield  {title} {\enquote {\bibinfo {title} {{Scalable quantum computing
  stabilised by optical tweezers on an ion crystal}},}\ }\href {\doibase
  10.1088/1367-2630/ab84b6} {\bibfield  {journal} {\bibinfo  {journal} {New J.
  Phys.}\ }\textbf {\bibinfo {volume} {22}},\ \bibinfo {pages} {053032}
  (\bibinfo {year} {2020})}\BibitemShut {NoStop}%
\bibitem [{\citenamefont {Loye}\ \emph {et~al.}(2020)\citenamefont {Loye},
  \citenamefont {Lages},\ and\ \citenamefont {Shepelyansky}}]{Loye2020}%
  \BibitemOpen
  \bibfield  {author} {\bibinfo {author} {\bibfnamefont {J.}~\bibnamefont
  {Loye}}, \bibinfo {author} {\bibfnamefont {J.}~\bibnamefont {Lages}}, \ and\
  \bibinfo {author} {\bibfnamefont {D.~L.}\ \bibnamefont {Shepelyansky}},\
  }\bibfield  {title} {\enquote {\bibinfo {title} {{Properties of phonon modes
  of an ion-trap quantum computer in the Aubry phase}},}\ }\href {\doibase
  10.1103/PhysRevA.101.032349} {\bibfield  {journal} {\bibinfo  {journal}
  {Phys. Rev. A}\ }\textbf {\bibinfo {volume} {101}},\ \bibinfo {pages}
  {032349} (\bibinfo {year} {2020})}\BibitemShut {NoStop}%
\bibitem [{\citenamefont {Ivanov}\ \emph {et~al.}(2009)\citenamefont {Ivanov},
  \citenamefont {Ivanov}, \citenamefont {Vitanov}, \citenamefont {Mering},
  \citenamefont {Fleischhauer},\ and\ \citenamefont {Singer}}]{Ivanov2009}%
  \BibitemOpen
  \bibfield  {author} {\bibinfo {author} {\bibfnamefont {P.~A.}\ \bibnamefont
  {Ivanov}}, \bibinfo {author} {\bibfnamefont {S.~S.}\ \bibnamefont {Ivanov}},
  \bibinfo {author} {\bibfnamefont {N.~V.}\ \bibnamefont {Vitanov}}, \bibinfo
  {author} {\bibfnamefont {A.}~\bibnamefont {Mering}}, \bibinfo {author}
  {\bibfnamefont {M.}~\bibnamefont {Fleischhauer}}, \ and\ \bibinfo {author}
  {\bibfnamefont {K.}~\bibnamefont {Singer}},\ }\bibfield  {title} {\enquote
  {\bibinfo {title} {{Simulation of a quantum phase transition of polaritons
  with trapped ions}},}\ }\href {\doibase 10.1103/PhysRevA.80.060301}
  {\bibfield  {journal} {\bibinfo  {journal} {Phys. Rev. A}\ }\textbf {\bibinfo
  {volume} {80}},\ \bibinfo {pages} {060301} (\bibinfo {year}
  {2009})}\BibitemShut {NoStop}%
\bibitem [{\citenamefont {Abdelrahman}\ \emph {et~al.}(2017)\citenamefont
  {Abdelrahman}, \citenamefont {Khosravani}, \citenamefont {Gessner},
  \citenamefont {Buchleitner}, \citenamefont {Breuer}, \citenamefont {Gorman},
  \citenamefont {Masuda}, \citenamefont {Pruttivarasin}, \citenamefont {Ramm},
  \citenamefont {Schindler},\ and\ \citenamefont
  {H{\"{a}}ffner}}]{Abdelrahman2017}%
  \BibitemOpen
  \bibfield  {author} {\bibinfo {author} {\bibfnamefont {A.}~\bibnamefont
  {Abdelrahman}}, \bibinfo {author} {\bibfnamefont {O.}~\bibnamefont
  {Khosravani}}, \bibinfo {author} {\bibfnamefont {M.}~\bibnamefont {Gessner}},
  \bibinfo {author} {\bibfnamefont {A.}~\bibnamefont {Buchleitner}}, \bibinfo
  {author} {\bibfnamefont {H.~P.}\ \bibnamefont {Breuer}}, \bibinfo {author}
  {\bibfnamefont {D.}~\bibnamefont {Gorman}}, \bibinfo {author} {\bibfnamefont
  {R.}~\bibnamefont {Masuda}}, \bibinfo {author} {\bibfnamefont
  {T.}~\bibnamefont {Pruttivarasin}}, \bibinfo {author} {\bibfnamefont
  {M.}~\bibnamefont {Ramm}}, \bibinfo {author} {\bibfnamefont {P.}~\bibnamefont
  {Schindler}}, \ and\ \bibinfo {author} {\bibfnamefont {H.}~\bibnamefont
  {H{\"{a}}ffner}},\ }\bibfield  {title} {\enquote {\bibinfo {title} {{Local
  probe of single phonon dynamics in warm ion crystals}},}\ }\href {\doibase
  10.1038/ncomms15712} {\bibfield  {journal} {\bibinfo  {journal} {Nature
  Communications}\ }\textbf {\bibinfo {volume} {8}},\ \bibinfo {pages} {1--5}
  (\bibinfo {year} {2017})}\BibitemShut {NoStop}%
\bibitem [{\citenamefont {Barredo}\ \emph {et~al.}(2020)\citenamefont
  {Barredo}, \citenamefont {Lienhard}, \citenamefont {Scholl}, \citenamefont
  {{De L{\'{e}}s{\'{e}}leuc}}, \citenamefont {Boulier}, \citenamefont
  {Browaeys},\ and\ \citenamefont {Lahaye}}]{Barredo2020}%
  \BibitemOpen
  \bibfield  {author} {\bibinfo {author} {\bibfnamefont {D.}~\bibnamefont
  {Barredo}}, \bibinfo {author} {\bibfnamefont {V.}~\bibnamefont {Lienhard}},
  \bibinfo {author} {\bibfnamefont {P.}~\bibnamefont {Scholl}}, \bibinfo
  {author} {\bibfnamefont {S.}~\bibnamefont {{De L{\'{e}}s{\'{e}}leuc}}},
  \bibinfo {author} {\bibfnamefont {T.}~\bibnamefont {Boulier}}, \bibinfo
  {author} {\bibfnamefont {A.}~\bibnamefont {Browaeys}}, \ and\ \bibinfo
  {author} {\bibfnamefont {T.}~\bibnamefont {Lahaye}},\ }\bibfield  {title}
  {\enquote {\bibinfo {title} {{Three-Dimensional Trapping of Individual
  Rydberg Atoms in Ponderomotive Bottle Beam Traps}},}\ }\href {\doibase
  10.1103/PhysRevLett.124.023201} {\bibfield  {journal} {\bibinfo  {journal}
  {Phys. Rev. Lett.}\ }\textbf {\bibinfo {volume} {124}},\ \bibinfo {pages}
  {023201} (\bibinfo {year} {2020})}\BibitemShut {NoStop}%
\bibitem [{\citenamefont {Schneider}\ \emph {et~al.}(2010)\citenamefont
  {Schneider}, \citenamefont {Enderlein}, \citenamefont {Huber},\ and\
  \citenamefont {Schaetz}}]{Schneider2010}%
  \BibitemOpen
  \bibfield  {author} {\bibinfo {author} {\bibfnamefont {C.}~\bibnamefont
  {Schneider}}, \bibinfo {author} {\bibfnamefont {M.}~\bibnamefont
  {Enderlein}}, \bibinfo {author} {\bibfnamefont {T.}~\bibnamefont {Huber}}, \
  and\ \bibinfo {author} {\bibfnamefont {T.}~\bibnamefont {Schaetz}},\
  }\bibfield  {title} {\enquote {\bibinfo {title} {{Optical trapping of an
  ion}},}\ }\href {\doibase 10.1038/nphoton.2010.236} {\bibfield  {journal}
  {\bibinfo  {journal} {Nature Photonics}\ }\textbf {\bibinfo {volume} {4}},\
  \bibinfo {pages} {772--775} (\bibinfo {year} {2010})}\BibitemShut {NoStop}%
\bibitem [{\citenamefont {Martinez}\ \emph {et~al.}(2016)\citenamefont
  {Martinez}, \citenamefont {Muschik}, \citenamefont {Schindler}, \citenamefont
  {Nigg}, \citenamefont {Erhard}, \citenamefont {Heyl}, \citenamefont {Hauke},
  \citenamefont {Dalmonte}, \citenamefont {Monz}, \citenamefont {Zoller},\ and\
  \citenamefont {Blatt}}]{Martinez2016}%
  \BibitemOpen
  \bibfield  {author} {\bibinfo {author} {\bibfnamefont {E.~A.}\ \bibnamefont
  {Martinez}}, \bibinfo {author} {\bibfnamefont {C.~A.}\ \bibnamefont
  {Muschik}}, \bibinfo {author} {\bibfnamefont {P.}~\bibnamefont {Schindler}},
  \bibinfo {author} {\bibfnamefont {D.}~\bibnamefont {Nigg}}, \bibinfo {author}
  {\bibfnamefont {A.}~\bibnamefont {Erhard}}, \bibinfo {author} {\bibfnamefont
  {M.}~\bibnamefont {Heyl}}, \bibinfo {author} {\bibfnamefont {P.}~\bibnamefont
  {Hauke}}, \bibinfo {author} {\bibfnamefont {M.}~\bibnamefont {Dalmonte}},
  \bibinfo {author} {\bibfnamefont {T.}~\bibnamefont {Monz}}, \bibinfo {author}
  {\bibfnamefont {P.}~\bibnamefont {Zoller}}, \ and\ \bibinfo {author}
  {\bibfnamefont {R.}~\bibnamefont {Blatt}},\ }\bibfield  {title} {\enquote
  {\bibinfo {title} {{Real-time dynamics of lattice gauge theories with a
  few-qubit quantum computer}},}\ }\href {\doibase 10.1038/nature18318}
  {\bibfield  {journal} {\bibinfo  {journal} {Nature}\ }\textbf {\bibinfo
  {volume} {534}},\ \bibinfo {pages} {516--519} (\bibinfo {year}
  {2016})}\BibitemShut {NoStop}%
\bibitem [{\citenamefont {Barreiro}\ \emph {et~al.}(2011)\citenamefont
  {Barreiro}, \citenamefont {M{\"{u}}ller}, \citenamefont {Schindler},
  \citenamefont {Nigg}, \citenamefont {Monz}, \citenamefont {Chwalla},
  \citenamefont {Hennrich}, \citenamefont {Roos}, \citenamefont {Zoller},\ and\
  \citenamefont {Blatt}}]{Barreiro2011}%
  \BibitemOpen
  \bibfield  {author} {\bibinfo {author} {\bibfnamefont {J.~T.}\ \bibnamefont
  {Barreiro}}, \bibinfo {author} {\bibfnamefont {M.}~\bibnamefont
  {M{\"{u}}ller}}, \bibinfo {author} {\bibfnamefont {P.}~\bibnamefont
  {Schindler}}, \bibinfo {author} {\bibfnamefont {D.}~\bibnamefont {Nigg}},
  \bibinfo {author} {\bibfnamefont {T.}~\bibnamefont {Monz}}, \bibinfo {author}
  {\bibfnamefont {M.}~\bibnamefont {Chwalla}}, \bibinfo {author} {\bibfnamefont
  {M.}~\bibnamefont {Hennrich}}, \bibinfo {author} {\bibfnamefont {C.~F.}\
  \bibnamefont {Roos}}, \bibinfo {author} {\bibfnamefont {P.}~\bibnamefont
  {Zoller}}, \ and\ \bibinfo {author} {\bibfnamefont {R.}~\bibnamefont
  {Blatt}},\ }\bibfield  {title} {\enquote {\bibinfo {title} {{An open-system
  quantum simulator with trapped ions}},}\ }\href {\doibase
  10.1038/nature09801} {\bibfield  {journal} {\bibinfo  {journal} {Nature}\
  }\textbf {\bibinfo {volume} {470}},\ \bibinfo {pages} {486--491} (\bibinfo
  {year} {2011})}\BibitemShut {NoStop}%
\bibitem [{\citenamefont {Lanyon}\ \emph {et~al.}(2011)\citenamefont {Lanyon},
  \citenamefont {Hempel}, \citenamefont {Nigg}, \citenamefont {M{\"{u}}ller},
  \citenamefont {Gerritsma}, \citenamefont {Z{\"{a}}hringer}, \citenamefont
  {Schindler}, \citenamefont {Barreiro}, \citenamefont {Rambach}, \citenamefont
  {Kirchmair}, \citenamefont {Hennrich}, \citenamefont {Zoller}, \citenamefont
  {Blatt},\ and\ \citenamefont {Roos}}]{Lanyon57}%
  \BibitemOpen
  \bibfield  {author} {\bibinfo {author} {\bibfnamefont {B.~P.}\ \bibnamefont
  {Lanyon}}, \bibinfo {author} {\bibfnamefont {C.}~\bibnamefont {Hempel}},
  \bibinfo {author} {\bibfnamefont {D.}~\bibnamefont {Nigg}}, \bibinfo {author}
  {\bibfnamefont {M.}~\bibnamefont {M{\"{u}}ller}}, \bibinfo {author}
  {\bibfnamefont {R.}~\bibnamefont {Gerritsma}}, \bibinfo {author}
  {\bibfnamefont {F.}~\bibnamefont {Z{\"{a}}hringer}}, \bibinfo {author}
  {\bibfnamefont {P.}~\bibnamefont {Schindler}}, \bibinfo {author}
  {\bibfnamefont {J.~T.}\ \bibnamefont {Barreiro}}, \bibinfo {author}
  {\bibfnamefont {M.}~\bibnamefont {Rambach}}, \bibinfo {author} {\bibfnamefont
  {G.}~\bibnamefont {Kirchmair}}, \bibinfo {author} {\bibfnamefont
  {M.}~\bibnamefont {Hennrich}}, \bibinfo {author} {\bibfnamefont
  {P.}~\bibnamefont {Zoller}}, \bibinfo {author} {\bibfnamefont
  {R.}~\bibnamefont {Blatt}}, \ and\ \bibinfo {author} {\bibfnamefont {C.~F.}\
  \bibnamefont {Roos}},\ }\bibfield  {title} {\enquote {\bibinfo {title}
  {{Universal Digital Quantum Simulation with Trapped Ions}},}\ }\href
  {\doibase 10.1126/science.1208001} {\bibfield  {journal} {\bibinfo  {journal}
  {Science}\ }\textbf {\bibinfo {volume} {334}},\ \bibinfo {pages} {57--61}
  (\bibinfo {year} {2011})}\BibitemShut {NoStop}%
\bibitem [{\citenamefont {Arute}\ \emph {et~al.}(2019)\citenamefont {Arute},
  \citenamefont {Arya}, \citenamefont {Babbush}, \citenamefont {Bacon},
  \citenamefont {Bardin}, \citenamefont {Barends}, \citenamefont {Biswas},
  \citenamefont {Boixo}, \citenamefont {Brandao}, \citenamefont {Buell},
  \citenamefont {Burkett}, \citenamefont {Chen}, \citenamefont {Chen},
  \citenamefont {Chiaro}, \citenamefont {Collins}, \citenamefont {Courtney},
  \citenamefont {Dunsworth}, \citenamefont {Farhi}, \citenamefont {Foxen},
  \citenamefont {Fowler}, \citenamefont {Gidney}, \citenamefont {Giustina},
  \citenamefont {Graff}, \citenamefont {Guerin}, \citenamefont {Habegger},
  \citenamefont {Harrigan}, \citenamefont {Hartmann}, \citenamefont {Ho},
  \citenamefont {Hoffmann}, \citenamefont {Huang}, \citenamefont {Humble},
  \citenamefont {Isakov}, \citenamefont {Jeffrey}, \citenamefont {Jiang},
  \citenamefont {Kafri}, \citenamefont {Kechedzhi}, \citenamefont {Kelly},
  \citenamefont {Klimov}, \citenamefont {Knysh}, \citenamefont {Korotkov},
  \citenamefont {Kostritsa}, \citenamefont {Landhuis}, \citenamefont
  {Lindmark}, \citenamefont {Lucero}, \citenamefont {Lyakh}, \citenamefont
  {Mandr{\`{a}}}, \citenamefont {McClean}, \citenamefont {McEwen},
  \citenamefont {Megrant}, \citenamefont {Mi}, \citenamefont {Michielsen},
  \citenamefont {Mohseni}, \citenamefont {Mutus}, \citenamefont {Naaman},
  \citenamefont {Neeley}, \citenamefont {Neill}, \citenamefont {Niu},
  \citenamefont {Ostby}, \citenamefont {Petukhov}, \citenamefont {Platt},
  \citenamefont {Quintana}, \citenamefont {Rieffel}, \citenamefont {Roushan},
  \citenamefont {Rubin}, \citenamefont {Sank}, \citenamefont {Satzinger},
  \citenamefont {Smelyanskiy}, \citenamefont {Sung}, \citenamefont
  {Trevithick}, \citenamefont {Vainsencher}, \citenamefont {Villalonga},
  \citenamefont {White}, \citenamefont {Yao}, \citenamefont {Yeh},
  \citenamefont {Zalcman}, \citenamefont {Neven},\ and\ \citenamefont
  {Martinis}}]{Arute2019}%
  \BibitemOpen
  \bibfield  {author} {\bibinfo {author} {\bibfnamefont {F.}~\bibnamefont
  {Arute}}, \bibinfo {author} {\bibfnamefont {K.}~\bibnamefont {Arya}},
  \bibinfo {author} {\bibfnamefont {R.}~\bibnamefont {Babbush}}, \bibinfo
  {author} {\bibfnamefont {D.}~\bibnamefont {Bacon}}, \bibinfo {author}
  {\bibfnamefont {J.~C.}\ \bibnamefont {Bardin}}, \bibinfo {author}
  {\bibfnamefont {R.}~\bibnamefont {Barends}}, \bibinfo {author} {\bibfnamefont
  {R.}~\bibnamefont {Biswas}}, \bibinfo {author} {\bibfnamefont
  {S.}~\bibnamefont {Boixo}}, \bibinfo {author} {\bibfnamefont {F.~G. S.~L.}\
  \bibnamefont {Brandao}}, \bibinfo {author} {\bibfnamefont {D.~A.}\
  \bibnamefont {Buell}}, \bibinfo {author} {\bibfnamefont {B.}~\bibnamefont
  {Burkett}}, \bibinfo {author} {\bibfnamefont {Y.}~\bibnamefont {Chen}},
  \bibinfo {author} {\bibfnamefont {Z.}~\bibnamefont {Chen}}, \bibinfo {author}
  {\bibfnamefont {B.}~\bibnamefont {Chiaro}}, \bibinfo {author} {\bibfnamefont
  {R.}~\bibnamefont {Collins}}, \bibinfo {author} {\bibfnamefont
  {W.}~\bibnamefont {Courtney}}, \bibinfo {author} {\bibfnamefont
  {A.}~\bibnamefont {Dunsworth}}, \bibinfo {author} {\bibfnamefont
  {E.}~\bibnamefont {Farhi}}, \bibinfo {author} {\bibfnamefont
  {B.}~\bibnamefont {Foxen}}, \bibinfo {author} {\bibfnamefont
  {A.}~\bibnamefont {Fowler}}, \bibinfo {author} {\bibfnamefont
  {C.}~\bibnamefont {Gidney}}, \bibinfo {author} {\bibfnamefont
  {M.}~\bibnamefont {Giustina}}, \bibinfo {author} {\bibfnamefont
  {R.}~\bibnamefont {Graff}}, \bibinfo {author} {\bibfnamefont
  {K.}~\bibnamefont {Guerin}}, \bibinfo {author} {\bibfnamefont
  {S.}~\bibnamefont {Habegger}}, \bibinfo {author} {\bibfnamefont {M.~P.}\
  \bibnamefont {Harrigan}}, \bibinfo {author} {\bibfnamefont {M.~J.}\
  \bibnamefont {Hartmann}}, \bibinfo {author} {\bibfnamefont {A.}~\bibnamefont
  {Ho}}, \bibinfo {author} {\bibfnamefont {M.}~\bibnamefont {Hoffmann}},
  \bibinfo {author} {\bibfnamefont {T.}~\bibnamefont {Huang}}, \bibinfo
  {author} {\bibfnamefont {T.~S.}\ \bibnamefont {Humble}}, \bibinfo {author}
  {\bibfnamefont {S.~V.}\ \bibnamefont {Isakov}}, \bibinfo {author}
  {\bibfnamefont {E.}~\bibnamefont {Jeffrey}}, \bibinfo {author} {\bibfnamefont
  {Z.}~\bibnamefont {Jiang}}, \bibinfo {author} {\bibfnamefont
  {D.}~\bibnamefont {Kafri}}, \bibinfo {author} {\bibfnamefont
  {K.}~\bibnamefont {Kechedzhi}}, \bibinfo {author} {\bibfnamefont
  {J.}~\bibnamefont {Kelly}}, \bibinfo {author} {\bibfnamefont {P.~V.}\
  \bibnamefont {Klimov}}, \bibinfo {author} {\bibfnamefont {S.}~\bibnamefont
  {Knysh}}, \bibinfo {author} {\bibfnamefont {A.}~\bibnamefont {Korotkov}},
  \bibinfo {author} {\bibfnamefont {F.}~\bibnamefont {Kostritsa}}, \bibinfo
  {author} {\bibfnamefont {D.}~\bibnamefont {Landhuis}}, \bibinfo {author}
  {\bibfnamefont {M.}~\bibnamefont {Lindmark}}, \bibinfo {author}
  {\bibfnamefont {E.}~\bibnamefont {Lucero}}, \bibinfo {author} {\bibfnamefont
  {D.}~\bibnamefont {Lyakh}}, \bibinfo {author} {\bibfnamefont
  {S.}~\bibnamefont {Mandr{\`{a}}}}, \bibinfo {author} {\bibfnamefont {J.~R.}\
  \bibnamefont {McClean}}, \bibinfo {author} {\bibfnamefont {M.}~\bibnamefont
  {McEwen}}, \bibinfo {author} {\bibfnamefont {A.}~\bibnamefont {Megrant}},
  \bibinfo {author} {\bibfnamefont {X.}~\bibnamefont {Mi}}, \bibinfo {author}
  {\bibfnamefont {K.}~\bibnamefont {Michielsen}}, \bibinfo {author}
  {\bibfnamefont {M.}~\bibnamefont {Mohseni}}, \bibinfo {author} {\bibfnamefont
  {J.}~\bibnamefont {Mutus}}, \bibinfo {author} {\bibfnamefont
  {O.}~\bibnamefont {Naaman}}, \bibinfo {author} {\bibfnamefont
  {M.}~\bibnamefont {Neeley}}, \bibinfo {author} {\bibfnamefont
  {C.}~\bibnamefont {Neill}}, \bibinfo {author} {\bibfnamefont {M.~Y.}\
  \bibnamefont {Niu}}, \bibinfo {author} {\bibfnamefont {E.}~\bibnamefont
  {Ostby}}, \bibinfo {author} {\bibfnamefont {A.}~\bibnamefont {Petukhov}},
  \bibinfo {author} {\bibfnamefont {J.~C.}\ \bibnamefont {Platt}}, \bibinfo
  {author} {\bibfnamefont {C.}~\bibnamefont {Quintana}}, \bibinfo {author}
  {\bibfnamefont {E.~G.}\ \bibnamefont {Rieffel}}, \bibinfo {author}
  {\bibfnamefont {P.}~\bibnamefont {Roushan}}, \bibinfo {author} {\bibfnamefont
  {N.~C.}\ \bibnamefont {Rubin}}, \bibinfo {author} {\bibfnamefont
  {D.}~\bibnamefont {Sank}}, \bibinfo {author} {\bibfnamefont {K.~J.}\
  \bibnamefont {Satzinger}}, \bibinfo {author} {\bibfnamefont {V.}~\bibnamefont
  {Smelyanskiy}}, \bibinfo {author} {\bibfnamefont {K.~J.}\ \bibnamefont
  {Sung}}, \bibinfo {author} {\bibfnamefont {M.~D.}\ \bibnamefont
  {Trevithick}}, \bibinfo {author} {\bibfnamefont {A.}~\bibnamefont
  {Vainsencher}}, \bibinfo {author} {\bibfnamefont {B.}~\bibnamefont
  {Villalonga}}, \bibinfo {author} {\bibfnamefont {T.}~\bibnamefont {White}},
  \bibinfo {author} {\bibfnamefont {Z.~J.}\ \bibnamefont {Yao}}, \bibinfo
  {author} {\bibfnamefont {P.}~\bibnamefont {Yeh}}, \bibinfo {author}
  {\bibfnamefont {A.}~\bibnamefont {Zalcman}}, \bibinfo {author} {\bibfnamefont
  {H.}~\bibnamefont {Neven}}, \ and\ \bibinfo {author} {\bibfnamefont {J.~M.}\
  \bibnamefont {Martinis}},\ }\bibfield  {title} {\enquote {\bibinfo {title}
  {{Quantum supremacy using a programmable superconducting processor}},}\
  }\href {\doibase 10.1038/s41586-019-1666-5} {\bibfield  {journal} {\bibinfo
  {journal} {Nature}\ }\textbf {\bibinfo {volume} {574}},\ \bibinfo {pages}
  {505--510} (\bibinfo {year} {2019})}\BibitemShut {NoStop}%
\bibitem [{\citenamefont {Nahum}\ \emph {et~al.}(2017)\citenamefont {Nahum},
  \citenamefont {Ruhman}, \citenamefont {Vijay},\ and\ \citenamefont
  {Haah}}]{Nahum2017}%
  \BibitemOpen
  \bibfield  {author} {\bibinfo {author} {\bibfnamefont {A.}~\bibnamefont
  {Nahum}}, \bibinfo {author} {\bibfnamefont {J.}~\bibnamefont {Ruhman}},
  \bibinfo {author} {\bibfnamefont {S.}~\bibnamefont {Vijay}}, \ and\ \bibinfo
  {author} {\bibfnamefont {J.}~\bibnamefont {Haah}},\ }\bibfield  {title}
  {\enquote {\bibinfo {title} {{Quantum Entanglement Growth under Random
  Unitary Dynamics}},}\ }\href {\doibase 10.1103/PhysRevX.7.031016} {\bibfield
  {journal} {\bibinfo  {journal} {Phys. Rev. X}\ }\textbf {\bibinfo {volume}
  {7}},\ \bibinfo {pages} {031016} (\bibinfo {year} {2017})}\BibitemShut
  {NoStop}%
\bibitem [{\citenamefont {Nahum}\ \emph {et~al.}(2018)\citenamefont {Nahum},
  \citenamefont {Vijay},\ and\ \citenamefont {Haah}}]{Nahum2018}%
  \BibitemOpen
  \bibfield  {author} {\bibinfo {author} {\bibfnamefont {A.}~\bibnamefont
  {Nahum}}, \bibinfo {author} {\bibfnamefont {S.}~\bibnamefont {Vijay}}, \ and\
  \bibinfo {author} {\bibfnamefont {J.}~\bibnamefont {Haah}},\ }\bibfield
  {title} {\enquote {\bibinfo {title} {{Operator Spreading in Random Unitary
  Circuits}},}\ }\href {\doibase 10.1103/PhysRevX.8.021014} {\bibfield
  {journal} {\bibinfo  {journal} {Phys. Rev. X}\ }\textbf {\bibinfo {volume}
  {8}},\ \bibinfo {pages} {21014} (\bibinfo {year} {2018})}\BibitemShut
  {NoStop}%
\bibitem [{\citenamefont {{Von Keyserlingk}}\ \emph {et~al.}(2018)\citenamefont
  {{Von Keyserlingk}}, \citenamefont {Rakovszky}, \citenamefont {Pollmann},\
  and\ \citenamefont {Sondhi}}]{VonKeyserlingk2018}%
  \BibitemOpen
  \bibfield  {author} {\bibinfo {author} {\bibfnamefont {C.~W.}\ \bibnamefont
  {{Von Keyserlingk}}}, \bibinfo {author} {\bibfnamefont {T.}~\bibnamefont
  {Rakovszky}}, \bibinfo {author} {\bibfnamefont {F.}~\bibnamefont {Pollmann}},
  \ and\ \bibinfo {author} {\bibfnamefont {S.~L.}\ \bibnamefont {Sondhi}},\
  }\bibfield  {title} {\enquote {\bibinfo {title} {{Operator Hydrodynamics,
  OTOCs, and Entanglement Growth in Systems without Conservation Laws}},}\
  }\href {\doibase 10.1103/PhysRevX.8.021013} {\bibfield  {journal} {\bibinfo
  {journal} {Phys. Rev. X}\ }\textbf {\bibinfo {volume} {8}},\ \bibinfo {pages}
  {21013} (\bibinfo {year} {2018})}\BibitemShut {NoStop}%
\bibitem [{\citenamefont {Rakovszky}\ \emph {et~al.}(2018)\citenamefont
  {Rakovszky}, \citenamefont {Pollmann},\ and\ \citenamefont {{Von
  Keyserlingk}}}]{Rakovszky2018}%
  \BibitemOpen
  \bibfield  {author} {\bibinfo {author} {\bibfnamefont {T.}~\bibnamefont
  {Rakovszky}}, \bibinfo {author} {\bibfnamefont {F.}~\bibnamefont {Pollmann}},
  \ and\ \bibinfo {author} {\bibfnamefont {C.~W.}\ \bibnamefont {{Von
  Keyserlingk}}},\ }\bibfield  {title} {\enquote {\bibinfo {title} {{Diffusive
  Hydrodynamics of Out-of-Time-Ordered Correlators with Charge
  Conservation}},}\ }\href {\doibase 10.1103/PhysRevX.8.031058} {\bibfield
  {journal} {\bibinfo  {journal} {Phys. Rev. X}\ }\textbf {\bibinfo {volume}
  {8}},\ \bibinfo {pages} {31058} (\bibinfo {year} {2018})}\BibitemShut
  {NoStop}%
\bibitem [{\citenamefont {Khemani}\ \emph {et~al.}(2018)\citenamefont
  {Khemani}, \citenamefont {Vishwanath},\ and\ \citenamefont
  {Huse}}]{Khemani2018}%
  \BibitemOpen
  \bibfield  {author} {\bibinfo {author} {\bibfnamefont {V.}~\bibnamefont
  {Khemani}}, \bibinfo {author} {\bibfnamefont {A.}~\bibnamefont {Vishwanath}},
  \ and\ \bibinfo {author} {\bibfnamefont {D.~A.}\ \bibnamefont {Huse}},\
  }\bibfield  {title} {\enquote {\bibinfo {title} {{Operator Spreading and the
  Emergence of Dissipative Hydrodynamics under Unitary Evolution with
  Conservation Laws}},}\ }\href {\doibase 10.1103/PhysRevX.8.031057} {\bibfield
   {journal} {\bibinfo  {journal} {Phys. Rev. X}\ }\textbf {\bibinfo {volume}
  {8}},\ \bibinfo {pages} {031057} (\bibinfo {year} {2018})}\BibitemShut
  {NoStop}%
\bibitem [{\citenamefont {Milburn}(1999)}]{Milburn1999}%
  \BibitemOpen
  \bibfield  {author} {\bibinfo {author} {\bibfnamefont {G.~J.}\ \bibnamefont
  {Milburn}},\ }\bibfield  {title} {\enquote {\bibinfo {title} {{Simulating
  nonlinear spin models in an ion trap}},}\ }\href
  {http://arxiv.org/abs/quant-ph/9908037} {\bibfield  {journal} {\bibinfo
  {journal} {arXiv:quant-ph/9908037}\ } (\bibinfo {year} {1999})}\BibitemShut
  {NoStop}%
\bibitem [{\citenamefont {M{\o}lmer}\ and\ \citenamefont
  {S{\o}rensen}(1999)}]{Molmer1999}%
  \BibitemOpen
  \bibfield  {author} {\bibinfo {author} {\bibfnamefont {K.}~\bibnamefont
  {M{\o}lmer}}\ and\ \bibinfo {author} {\bibfnamefont {A.}~\bibnamefont
  {S{\o}rensen}},\ }\bibfield  {title} {\enquote {\bibinfo {title}
  {{Multiparticle Entanglement of Hot Trapped Ions}},}\ }\href {\doibase
  10.1103/PhysRevLett.82.1835} {\bibfield  {journal} {\bibinfo  {journal}
  {Phys. Rev. Lett.}\ }\textbf {\bibinfo {volume} {82}},\ \bibinfo {pages}
  {1835--1838} (\bibinfo {year} {1999})}\BibitemShut {NoStop}%
\bibitem [{\citenamefont {S{\o}rensen}\ and\ \citenamefont
  {M{\o}lmer}(1999)}]{Sorensen1999}%
  \BibitemOpen
  \bibfield  {author} {\bibinfo {author} {\bibfnamefont {A.}~\bibnamefont
  {S{\o}rensen}}\ and\ \bibinfo {author} {\bibfnamefont {K.}~\bibnamefont
  {M{\o}lmer}},\ }\bibfield  {title} {\enquote {\bibinfo {title} {{Quantum
  Computation with Ions in Thermal Motion}},}\ }\href {\doibase
  10.1103/PhysRevLett.82.1971} {\bibfield  {journal} {\bibinfo  {journal}
  {Phys. Rev. Lett.}\ }\textbf {\bibinfo {volume} {82}},\ \bibinfo {pages}
  {1971--1974} (\bibinfo {year} {1999})}\BibitemShut {NoStop}%
\bibitem [{\citenamefont {S{\o}rensen}\ and\ \citenamefont
  {M{\o}lmer}(2000)}]{Sorensen2000}%
  \BibitemOpen
  \bibfield  {author} {\bibinfo {author} {\bibfnamefont {A.}~\bibnamefont
  {S{\o}rensen}}\ and\ \bibinfo {author} {\bibfnamefont {K.}~\bibnamefont
  {M{\o}lmer}},\ }\bibfield  {title} {\enquote {\bibinfo {title} {{Entanglement
  and quantum computation with ions in thermal motion}},}\ }\href {\doibase
  10.1103/PhysRevA.62.022311} {\bibfield  {journal} {\bibinfo  {journal} {Phys.
  Rev. A}\ }\textbf {\bibinfo {volume} {62}},\ \bibinfo {pages} {022311}
  (\bibinfo {year} {2000})}\BibitemShut {NoStop}%
\bibitem [{\citenamefont {Zhu}\ and\ \citenamefont {Wang}(2003)}]{Zhu2003}%
  \BibitemOpen
  \bibfield  {author} {\bibinfo {author} {\bibfnamefont {S.-L.}\ \bibnamefont
  {Zhu}}\ and\ \bibinfo {author} {\bibfnamefont {Z.~D.}\ \bibnamefont {Wang}},\
  }\bibfield  {title} {\enquote {\bibinfo {title} {{Unconventional Geometric
  Quantum Computation}},}\ }\href {\doibase 10.1103/PhysRevLett.91.187902}
  {\bibfield  {journal} {\bibinfo  {journal} {Phys. Rev. Lett.}\ }\textbf
  {\bibinfo {volume} {91}},\ \bibinfo {pages} {187902} (\bibinfo {year}
  {2003})}\BibitemShut {NoStop}%
\bibitem [{\citenamefont {Garc{\'i}a-Ripoll}\ \emph {et~al.}(2005)\citenamefont
  {Garc{\'i}a-Ripoll}, \citenamefont {Zoller},\ and\ \citenamefont
  {Cirac}}]{Garcia-Ripoll2005}%
  \BibitemOpen
  \bibfield  {author} {\bibinfo {author} {\bibfnamefont {J.~J.}\ \bibnamefont
  {Garc{\'i}a-Ripoll}}, \bibinfo {author} {\bibfnamefont {P.}~\bibnamefont
  {Zoller}}, \ and\ \bibinfo {author} {\bibfnamefont {J.~I.}\ \bibnamefont
  {Cirac}},\ }\bibfield  {title} {\enquote {\bibinfo {title} {{Coherent control
  of trapped ions using off-resonant lasers}},}\ }\href {\doibase
  10.1103/PhysRevA.71.062309} {\bibfield  {journal} {\bibinfo  {journal} {Phys.
  Rev. A}\ }\textbf {\bibinfo {volume} {71}},\ \bibinfo {pages} {062309}
  (\bibinfo {year} {2005})}\BibitemShut {NoStop}%
\bibitem [{\citenamefont {Monroe}\ \emph {et~al.}(2019)\citenamefont {Monroe},
  \citenamefont {Campbell}, \citenamefont {Duan}, \citenamefont {Gong},
  \citenamefont {Gorshkov}, \citenamefont {Hess}, \citenamefont {Islam},
  \citenamefont {Kim}, \citenamefont {Pagano}, \citenamefont {Richerme},
  \citenamefont {Senko},\ and\ \citenamefont {Yao}}]{Monroe2019}%
  \BibitemOpen
  \bibfield  {author} {\bibinfo {author} {\bibfnamefont {C.}~\bibnamefont
  {Monroe}}, \bibinfo {author} {\bibfnamefont {W.~C.}\ \bibnamefont
  {Campbell}}, \bibinfo {author} {\bibfnamefont {L.~M.}\ \bibnamefont {Duan}},
  \bibinfo {author} {\bibfnamefont {Z.~X.}\ \bibnamefont {Gong}}, \bibinfo
  {author} {\bibfnamefont {A.~V.}\ \bibnamefont {Gorshkov}}, \bibinfo {author}
  {\bibfnamefont {P.}~\bibnamefont {Hess}}, \bibinfo {author} {\bibfnamefont
  {R.}~\bibnamefont {Islam}}, \bibinfo {author} {\bibfnamefont
  {K.}~\bibnamefont {Kim}}, \bibinfo {author} {\bibfnamefont {G.}~\bibnamefont
  {Pagano}}, \bibinfo {author} {\bibfnamefont {P.}~\bibnamefont {Richerme}},
  \bibinfo {author} {\bibfnamefont {C.}~\bibnamefont {Senko}}, \ and\ \bibinfo
  {author} {\bibfnamefont {N.~Y.}\ \bibnamefont {Yao}},\ }\bibfield  {title}
  {\enquote {\bibinfo {title} {{Programmable Quantum Simulations of Spin
  Systems with Trapped Ions}},}\ }\href {http://arxiv.org/abs/1912.07845}
  {\bibfield  {journal} {\bibinfo  {journal} {arXiv:1912.07845}\ } (\bibinfo
  {year} {2019})}\BibitemShut {NoStop}%
\bibitem [{\citenamefont {Kraus}\ and\ \citenamefont
  {Cirac}(2001)}]{Kraus2001}%
  \BibitemOpen
  \bibfield  {author} {\bibinfo {author} {\bibfnamefont {B.}~\bibnamefont
  {Kraus}}\ and\ \bibinfo {author} {\bibfnamefont {J.~I.}\ \bibnamefont
  {Cirac}},\ }\bibfield  {title} {\enquote {\bibinfo {title} {{Optimal creation
  of entanglement using a two-qubit gate}},}\ }\href {\doibase
  10.1103/PhysRevA.63.062309} {\bibfield  {journal} {\bibinfo  {journal} {Phys.
  Rev. A}\ }\textbf {\bibinfo {volume} {63}},\ \bibinfo {pages} {062309}
  (\bibinfo {year} {2001})}\BibitemShut {NoStop}%
\bibitem [{\citenamefont {Nielsen}\ and\ \citenamefont
  {Chuang}(2011)}]{Nielsen2011}%
  \BibitemOpen
  \bibfield  {author} {\bibinfo {author} {\bibfnamefont {M.~A.}\ \bibnamefont
  {Nielsen}}\ and\ \bibinfo {author} {\bibfnamefont {I.~L.}\ \bibnamefont
  {Chuang}},\ }\href@noop {} {\emph {\bibinfo {title} {Quantum Computation and
  Quantum Information: 10th Anniversary Edition}}},\ \bibinfo {edition} {10th}\
  ed.\ (\bibinfo  {publisher} {Cambridge University Press},\ \bibinfo {address}
  {Cambridge},\ \bibinfo {year} {2011})\BibitemShut {NoStop}%
\bibitem [{\citenamefont {Nielsen}(2002)}]{Nielsen2002}%
  \BibitemOpen
  \bibfield  {author} {\bibinfo {author} {\bibfnamefont {M.~A.}\ \bibnamefont
  {Nielsen}},\ }\bibfield  {title} {\enquote {\bibinfo {title} {{A simple
  formula for the average gate fidelity of a quantum dynamical operation}},}\
  }\href {\doibase 10.1016/S0375-9601(02)01272-0} {\bibfield  {journal}
  {\bibinfo  {journal} {Phys. Lett. A}\ }\textbf {\bibinfo {volume} {303}},\
  \bibinfo {pages} {249--252} (\bibinfo {year} {2002})}\BibitemShut {NoStop}%
\bibitem [{\citenamefont {Bengtsson}\ and\ \citenamefont
  {Zyczkowski}(2006)}]{bengtsson_zyczkowski_2006}%
  \BibitemOpen
  \bibfield  {author} {\bibinfo {author} {\bibfnamefont {I.}~\bibnamefont
  {Bengtsson}}\ and\ \bibinfo {author} {\bibfnamefont {K.}~\bibnamefont
  {Zyczkowski}},\ }\href {\doibase 10.1017/CBO9780511535048} {\emph {\bibinfo
  {title} {{Geometry of Quantum States}}}}\ (\bibinfo  {publisher} {Cambridge
  University Press},\ \bibinfo {address} {Cambridge},\ \bibinfo {year}
  {2006})\BibitemShut {NoStop}%
\bibitem [{\citenamefont {Sanders}\ \emph {et~al.}(2015)\citenamefont
  {Sanders}, \citenamefont {Wallman},\ and\ \citenamefont
  {Sanders}}]{Sanders2015}%
  \BibitemOpen
  \bibfield  {author} {\bibinfo {author} {\bibfnamefont {Y.~R.}\ \bibnamefont
  {Sanders}}, \bibinfo {author} {\bibfnamefont {J.~J.}\ \bibnamefont
  {Wallman}}, \ and\ \bibinfo {author} {\bibfnamefont {B.~C.}\ \bibnamefont
  {Sanders}},\ }\bibfield  {title} {\enquote {\bibinfo {title} {{Bounding
  quantum gate error rate based on reported average fidelity}},}\ }\href
  {\doibase 10.1088/1367-2630/18/1/012002} {\bibfield  {journal} {\bibinfo
  {journal} {New J. Phys.}\ }\textbf {\bibinfo {volume} {18}},\ \bibinfo
  {pages} {012002} (\bibinfo {year} {2015})}\BibitemShut {NoStop}%
\bibitem [{\citenamefont {Kueng}\ \emph {et~al.}(2016)\citenamefont {Kueng},
  \citenamefont {Long}, \citenamefont {Doherty},\ and\ \citenamefont
  {Flammia}}]{Kueng2016}%
  \BibitemOpen
  \bibfield  {author} {\bibinfo {author} {\bibfnamefont {R.}~\bibnamefont
  {Kueng}}, \bibinfo {author} {\bibfnamefont {D.~M.}\ \bibnamefont {Long}},
  \bibinfo {author} {\bibfnamefont {A.~C.}\ \bibnamefont {Doherty}}, \ and\
  \bibinfo {author} {\bibfnamefont {S.~T.}\ \bibnamefont {Flammia}},\
  }\bibfield  {title} {\enquote {\bibinfo {title} {{Comparing Experiments to
  the Fault-Tolerance Threshold}},}\ }\href {\doibase
  10.1103/PhysRevLett.117.170502} {\bibfield  {journal} {\bibinfo  {journal}
  {Phys. Rev. Lett.}\ }\textbf {\bibinfo {volume} {117}},\ \bibinfo {pages}
  {170502} (\bibinfo {year} {2016})}\BibitemShut {NoStop}%
\bibitem [{\citenamefont {Aharonov}\ \emph {et~al.}(1998)\citenamefont
  {Aharonov}, \citenamefont {Kitaev},\ and\ \citenamefont
  {Nisan}}]{Aharonov1998}%
  \BibitemOpen
  \bibfield  {author} {\bibinfo {author} {\bibfnamefont {D.}~\bibnamefont
  {Aharonov}}, \bibinfo {author} {\bibfnamefont {A.}~\bibnamefont {Kitaev}}, \
  and\ \bibinfo {author} {\bibfnamefont {N.}~\bibnamefont {Nisan}},\ }\bibfield
   {title} {\enquote {\bibinfo {title} {{Quantum circuits with mixed
  states}},}\ }in\ \href {\doibase 10.1145/276698.276708} {\emph {\bibinfo
  {booktitle} {Proc. thirtieth Annu. ACM Symp. Theory Comput. - STOC '98}}},\
  Vol.~\bibinfo {volume} {1}\ (\bibinfo  {publisher} {ACM Press},\ \bibinfo
  {address} {New York, New York, USA},\ \bibinfo {year} {1998})\ pp.\ \bibinfo
  {pages} {20--30}\BibitemShut {NoStop}%
\bibitem [{\citenamefont {Duan}(2004)}]{Duan2004}%
  \BibitemOpen
  \bibfield  {author} {\bibinfo {author} {\bibfnamefont {L.~M.}\ \bibnamefont
  {Duan}},\ }\bibfield  {title} {\enquote {\bibinfo {title} {{Scaling ion trap
  quantum computation through fast quantum gates}},}\ }\href {\doibase
  10.1103/PhysRevLett.93.100502} {\bibfield  {journal} {\bibinfo  {journal}
  {Phys. Rev. Lett.}\ }\textbf {\bibinfo {volume} {93}},\ \bibinfo {pages}
  {100502} (\bibinfo {year} {2004})}\BibitemShut {NoStop}%
\bibitem [{\citenamefont {Li}\ \emph {et~al.}(2017)\citenamefont {Li},
  \citenamefont {Urban}, \citenamefont {Noel}, \citenamefont {Chuang},
  \citenamefont {Xia}, \citenamefont {Ransford}, \citenamefont {Hemmerling},
  \citenamefont {Wang}, \citenamefont {Li}, \citenamefont {H\"affner},\ and\
  \citenamefont {Zhang}}]{Li2017}%
  \BibitemOpen
  \bibfield  {author} {\bibinfo {author} {\bibfnamefont {H.-K.}\ \bibnamefont
  {Li}}, \bibinfo {author} {\bibfnamefont {E.}~\bibnamefont {Urban}}, \bibinfo
  {author} {\bibfnamefont {C.}~\bibnamefont {Noel}}, \bibinfo {author}
  {\bibfnamefont {A.}~\bibnamefont {Chuang}}, \bibinfo {author} {\bibfnamefont
  {Y.}~\bibnamefont {Xia}}, \bibinfo {author} {\bibfnamefont {A.}~\bibnamefont
  {Ransford}}, \bibinfo {author} {\bibfnamefont {B.}~\bibnamefont
  {Hemmerling}}, \bibinfo {author} {\bibfnamefont {Y.}~\bibnamefont {Wang}},
  \bibinfo {author} {\bibfnamefont {T.}~\bibnamefont {Li}}, \bibinfo {author}
  {\bibfnamefont {H.}~\bibnamefont {H\"affner}}, \ and\ \bibinfo {author}
  {\bibfnamefont {X.}~\bibnamefont {Zhang}},\ }\bibfield  {title} {\enquote
  {\bibinfo {title} {Realization of translational symmetry in trapped cold ion
  rings},}\ }\href {\doibase 10.1103/PhysRevLett.118.053001} {\bibfield
  {journal} {\bibinfo  {journal} {Phys. Rev. Lett.}\ }\textbf {\bibinfo
  {volume} {118}},\ \bibinfo {pages} {053001} (\bibinfo {year}
  {2017})}\BibitemShut {NoStop}%
\bibitem [{\citenamefont {Kanai}\ \emph {et~al.}(2008)\citenamefont {Kanai},
  \citenamefont {Suda}, \citenamefont {Bohman}, \citenamefont {Kaku},
  \citenamefont {Yamaguchi},\ and\ \citenamefont
  {Midorikawa}}]{kanai2008beampointing}%
  \BibitemOpen
  \bibfield  {author} {\bibinfo {author} {\bibfnamefont {T.}~\bibnamefont
  {Kanai}}, \bibinfo {author} {\bibfnamefont {A.}~\bibnamefont {Suda}},
  \bibinfo {author} {\bibfnamefont {S.}~\bibnamefont {Bohman}}, \bibinfo
  {author} {\bibfnamefont {M.}~\bibnamefont {Kaku}}, \bibinfo {author}
  {\bibfnamefont {S.}~\bibnamefont {Yamaguchi}}, \ and\ \bibinfo {author}
  {\bibfnamefont {K.}~\bibnamefont {Midorikawa}},\ }\bibfield  {title}
  {\enquote {\bibinfo {title} {Pointing stabilization of a high-repetition-rate
  high-power femtosecond laser for intense few-cycle pulse generation},}\
  }\href {\doibase 10.1063/1.2842414} {\bibfield  {journal} {\bibinfo
  {journal} {Applied Physics Letters}\ }\textbf {\bibinfo {volume} {92}},\
  \bibinfo {pages} {061106} (\bibinfo {year} {2008})}\BibitemShut {NoStop}%
\bibitem [{\citenamefont {Seifert}\ \emph {et~al.}(2006)\citenamefont
  {Seifert}, \citenamefont {Kwee}, \citenamefont {Heurs}, \citenamefont
  {Willke},\ and\ \citenamefont {Danzmann}}]{seifert2006intensitynoise}%
  \BibitemOpen
  \bibfield  {author} {\bibinfo {author} {\bibfnamefont {F.}~\bibnamefont
  {Seifert}}, \bibinfo {author} {\bibfnamefont {P.}~\bibnamefont {Kwee}},
  \bibinfo {author} {\bibfnamefont {M.}~\bibnamefont {Heurs}}, \bibinfo
  {author} {\bibfnamefont {B.}~\bibnamefont {Willke}}, \ and\ \bibinfo {author}
  {\bibfnamefont {K.}~\bibnamefont {Danzmann}},\ }\bibfield  {title} {\enquote
  {\bibinfo {title} {Laser power stabilization for second-generation
  gravitational wave detectors},}\ }\href {\doibase 10.1364/OL.31.002000}
  {\bibfield  {journal} {\bibinfo  {journal} {Optics letters}\ }\textbf
  {\bibinfo {volume} {31}},\ \bibinfo {pages} {2000--2002} (\bibinfo {year}
  {2006})}\BibitemShut {NoStop}%
\bibitem [{\citenamefont {Nebendahl}\ \emph {et~al.}(2009)\citenamefont
  {Nebendahl}, \citenamefont {H{\"{a}}ffner},\ and\ \citenamefont
  {Roos}}]{Nebendahl2009}%
  \BibitemOpen
  \bibfield  {author} {\bibinfo {author} {\bibfnamefont {V.}~\bibnamefont
  {Nebendahl}}, \bibinfo {author} {\bibfnamefont {H.}~\bibnamefont
  {H{\"{a}}ffner}}, \ and\ \bibinfo {author} {\bibfnamefont {C.~F.}\
  \bibnamefont {Roos}},\ }\bibfield  {title} {\enquote {\bibinfo {title}
  {{Optimal control of entangling operations for trapped-ion quantum
  computing}},}\ }\href {\doibase 10.1103/PhysRevA.79.012312} {\bibfield
  {journal} {\bibinfo  {journal} {Phys. Rev. A}\ }\textbf {\bibinfo {volume}
  {79}},\ \bibinfo {pages} {012312} (\bibinfo {year} {2009})}\BibitemShut
  {NoStop}%
\bibitem [{\citenamefont {M{\"{u}}ller}\ \emph {et~al.}(2011)\citenamefont
  {M{\"{u}}ller}, \citenamefont {Hammerer}, \citenamefont {Zhou}, \citenamefont
  {Roos},\ and\ \citenamefont {Zoller}}]{Muller2011}%
  \BibitemOpen
  \bibfield  {author} {\bibinfo {author} {\bibfnamefont {M.}~\bibnamefont
  {M{\"{u}}ller}}, \bibinfo {author} {\bibfnamefont {K.}~\bibnamefont
  {Hammerer}}, \bibinfo {author} {\bibfnamefont {Y.~L.}\ \bibnamefont {Zhou}},
  \bibinfo {author} {\bibfnamefont {C.~F.}\ \bibnamefont {Roos}}, \ and\
  \bibinfo {author} {\bibfnamefont {P.}~\bibnamefont {Zoller}},\ }\bibfield
  {title} {\enquote {\bibinfo {title} {{Simulating open quantum systems: from
  many-body interactions to stabilizer pumping}},}\ }\href {\doibase
  10.1088/1367-2630/13/8/085007} {\bibfield  {journal} {\bibinfo  {journal}
  {New J. Phys.}\ }\textbf {\bibinfo {volume} {13}},\ \bibinfo {pages} {085007}
  (\bibinfo {year} {2011})}\BibitemShut {NoStop}%
\bibitem [{\citenamefont {Lin}\ \emph {et~al.}(2009)\citenamefont {Lin},
  \citenamefont {Zhu}, \citenamefont {Islam}, \citenamefont {Kim},
  \citenamefont {Chang}, \citenamefont {Korenblit}, \citenamefont {Monroe},\
  and\ \citenamefont {Duan}}]{Lin2009}%
  \BibitemOpen
  \bibfield  {author} {\bibinfo {author} {\bibfnamefont {G.-D.}\ \bibnamefont
  {Lin}}, \bibinfo {author} {\bibfnamefont {S.-L.}\ \bibnamefont {Zhu}},
  \bibinfo {author} {\bibfnamefont {R.}~\bibnamefont {Islam}}, \bibinfo
  {author} {\bibfnamefont {K.}~\bibnamefont {Kim}}, \bibinfo {author}
  {\bibfnamefont {M.-S.}\ \bibnamefont {Chang}}, \bibinfo {author}
  {\bibfnamefont {S.}~\bibnamefont {Korenblit}}, \bibinfo {author}
  {\bibfnamefont {C.}~\bibnamefont {Monroe}}, \ and\ \bibinfo {author}
  {\bibfnamefont {L.-M.}\ \bibnamefont {Duan}},\ }\bibfield  {title} {\enquote
  {\bibinfo {title} {{Large-scale quantum computation in an anharmonic linear
  ion trap}},}\ }\href {\doibase 10.1209/0295-5075/86/60004} {\bibfield
  {journal} {\bibinfo  {journal} {Europhys. Lett.}\ }\textbf {\bibinfo {volume}
  {86}},\ \bibinfo {pages} {60004} (\bibinfo {year} {2009})}\BibitemShut
  {NoStop}%
\bibitem [{\citenamefont {Itano}\ \emph {et~al.}(1998)\citenamefont {Itano},
  \citenamefont {Bollinger}, \citenamefont {Tan}, \citenamefont
  {Jelenkovi{\'{c}}}, \citenamefont {Huang},\ and\ \citenamefont
  {Wineland}}]{Itano1998}%
  \BibitemOpen
  \bibfield  {author} {\bibinfo {author} {\bibfnamefont {W.~M.}\ \bibnamefont
  {Itano}}, \bibinfo {author} {\bibfnamefont {J.~J.}\ \bibnamefont
  {Bollinger}}, \bibinfo {author} {\bibfnamefont {J.~N.}\ \bibnamefont {Tan}},
  \bibinfo {author} {\bibfnamefont {B.}~\bibnamefont {Jelenkovi{\'{c}}}},
  \bibinfo {author} {\bibfnamefont {X.-P.}\ \bibnamefont {Huang}}, \ and\
  \bibinfo {author} {\bibfnamefont {D.~J.}\ \bibnamefont {Wineland}},\
  }\bibfield  {title} {\enquote {\bibinfo {title} {{Bragg Diffraction from
  Crystallized Ion Plasmas}},}\ }\href {\doibase 10.1126/science.279.5351.686}
  {\bibfield  {journal} {\bibinfo  {journal} {Science}\ }\textbf {\bibinfo
  {volume} {279}},\ \bibinfo {pages} {686--689} (\bibinfo {year}
  {1998})}\BibitemShut {NoStop}%
\bibitem [{\citenamefont {Drewsen}\ \emph {et~al.}(1998)\citenamefont
  {Drewsen}, \citenamefont {Brodersen}, \citenamefont {Hornek{\ae}r},
  \citenamefont {Hangst},\ and\ \citenamefont {Schifffer}}]{Drewsen1998}%
  \BibitemOpen
  \bibfield  {author} {\bibinfo {author} {\bibfnamefont {M.}~\bibnamefont
  {Drewsen}}, \bibinfo {author} {\bibfnamefont {C.}~\bibnamefont {Brodersen}},
  \bibinfo {author} {\bibfnamefont {L.}~\bibnamefont {Hornek{\ae}r}}, \bibinfo
  {author} {\bibfnamefont {J.~S.}\ \bibnamefont {Hangst}}, \ and\ \bibinfo
  {author} {\bibfnamefont {J.~P.}\ \bibnamefont {Schifffer}},\ }\bibfield
  {title} {\enquote {\bibinfo {title} {{Large Ion Crystals in a Linear Paul
  Trap}},}\ }\href {\doibase 10.1103/PhysRevLett.81.2878} {\bibfield  {journal}
  {\bibinfo  {journal} {Phys. Rev. Lett.}\ }\textbf {\bibinfo {volume} {81}},\
  \bibinfo {pages} {2878--2881} (\bibinfo {year} {1998})}\BibitemShut {NoStop}%
\bibitem [{\citenamefont {Mortensen}\ \emph {et~al.}(2006)\citenamefont
  {Mortensen}, \citenamefont {Nielsen}, \citenamefont {Matthey},\ and\
  \citenamefont {Drewsen}}]{Mortensen2006}%
  \BibitemOpen
  \bibfield  {author} {\bibinfo {author} {\bibfnamefont {A.}~\bibnamefont
  {Mortensen}}, \bibinfo {author} {\bibfnamefont {E.}~\bibnamefont {Nielsen}},
  \bibinfo {author} {\bibfnamefont {T.}~\bibnamefont {Matthey}}, \ and\
  \bibinfo {author} {\bibfnamefont {M.}~\bibnamefont {Drewsen}},\ }\bibfield
  {title} {\enquote {\bibinfo {title} {{Observation of Three-Dimensional
  Long-Range Order in Small Ion Coulomb Crystals in an rf Trap}},}\ }\href
  {\doibase 10.1103/PhysRevLett.96.103001} {\bibfield  {journal} {\bibinfo
  {journal} {Phys. Rev. Lett.}\ }\textbf {\bibinfo {volume} {96}},\ \bibinfo
  {pages} {103001} (\bibinfo {year} {2006})}\BibitemShut {NoStop}%
\bibitem [{\citenamefont {Cirac}\ \emph {et~al.}(1994)\citenamefont {Cirac},
  \citenamefont {Garay}, \citenamefont {Blatt}, \citenamefont {Parkins},\ and\
  \citenamefont {Zoller}}]{Cirac1994}%
  \BibitemOpen
  \bibfield  {author} {\bibinfo {author} {\bibfnamefont {J.~I.}\ \bibnamefont
  {Cirac}}, \bibinfo {author} {\bibfnamefont {L.~J.}\ \bibnamefont {Garay}},
  \bibinfo {author} {\bibfnamefont {R.}~\bibnamefont {Blatt}}, \bibinfo
  {author} {\bibfnamefont {A.~S.}\ \bibnamefont {Parkins}}, \ and\ \bibinfo
  {author} {\bibfnamefont {P.}~\bibnamefont {Zoller}},\ }\bibfield  {title}
  {\enquote {\bibinfo {title} {{Laser cooling of trapped ions: The influence of
  micromotion}},}\ }\href {\doibase 10.1103/PhysRevA.49.421} {\bibfield
  {journal} {\bibinfo  {journal} {Phys. Rev. A}\ }\textbf {\bibinfo {volume}
  {49}},\ \bibinfo {pages} {421--432} (\bibinfo {year} {1994})}\BibitemShut
  {NoStop}%
\bibitem [{\citenamefont {Wang}\ \emph {et~al.}(2015)\citenamefont {Wang},
  \citenamefont {Shen},\ and\ \citenamefont {Duan}}]{Wang2015}%
  \BibitemOpen
  \bibfield  {author} {\bibinfo {author} {\bibfnamefont {S.~T.}\ \bibnamefont
  {Wang}}, \bibinfo {author} {\bibfnamefont {C.}~\bibnamefont {Shen}}, \ and\
  \bibinfo {author} {\bibfnamefont {L.~M.}\ \bibnamefont {Duan}},\ }\bibfield
  {title} {\enquote {\bibinfo {title} {{Quantum computation under micromotion
  in a planar ion crystal}},}\ }\href {\doibase 10.1038/srep08555} {\bibfield
  {journal} {\bibinfo  {journal} {Sci. Rep.}\ }\textbf {\bibinfo {volume}
  {5}},\ \bibinfo {pages} {8555} (\bibinfo {year} {2015})}\BibitemShut
  {NoStop}%
\bibitem [{\citenamefont {Fishman}\ \emph {et~al.}(2008)\citenamefont
  {Fishman}, \citenamefont {{De Chiara}}, \citenamefont {Calarco},\ and\
  \citenamefont {Morigi}}]{Fishman2008}%
  \BibitemOpen
  \bibfield  {author} {\bibinfo {author} {\bibfnamefont {S.}~\bibnamefont
  {Fishman}}, \bibinfo {author} {\bibfnamefont {G.}~\bibnamefont {{De
  Chiara}}}, \bibinfo {author} {\bibfnamefont {T.}~\bibnamefont {Calarco}}, \
  and\ \bibinfo {author} {\bibfnamefont {G.}~\bibnamefont {Morigi}},\
  }\bibfield  {title} {\enquote {\bibinfo {title} {{Structural phase
  transitions in low-dimensional ion crystals}},}\ }\href {\doibase
  10.1103/PhysRevB.77.064111} {\bibfield  {journal} {\bibinfo  {journal} {Phys.
  Rev. B}\ }\textbf {\bibinfo {volume} {77}},\ \bibinfo {pages} {064111}
  (\bibinfo {year} {2008})}\BibitemShut {NoStop}%
\bibitem [{\citenamefont {Shimshoni}\ \emph {et~al.}(2011)\citenamefont
  {Shimshoni}, \citenamefont {Morigi},\ and\ \citenamefont
  {Fishman}}]{Shimshoni2011}%
  \BibitemOpen
  \bibfield  {author} {\bibinfo {author} {\bibfnamefont {E.}~\bibnamefont
  {Shimshoni}}, \bibinfo {author} {\bibfnamefont {G.}~\bibnamefont {Morigi}}, \
  and\ \bibinfo {author} {\bibfnamefont {S.}~\bibnamefont {Fishman}},\
  }\bibfield  {title} {\enquote {\bibinfo {title} {{Quantum Zigzag Transition
  in Ion Chains}},}\ }\href {\doibase 10.1103/PhysRevLett.106.010401}
  {\bibfield  {journal} {\bibinfo  {journal} {Phys. Rev. Lett.}\ }\textbf
  {\bibinfo {volume} {106}},\ \bibinfo {pages} {010401} (\bibinfo {year}
  {2011})}\BibitemShut {NoStop}%
\bibitem [{\citenamefont {Welzel}\ \emph {et~al.}(2019)\citenamefont {Welzel},
  \citenamefont {Stopp},\ and\ \citenamefont {Schmidt-Kaler}}]{Welzel2019}%
  \BibitemOpen
  \bibfield  {author} {\bibinfo {author} {\bibfnamefont {J.}~\bibnamefont
  {Welzel}}, \bibinfo {author} {\bibfnamefont {F.}~\bibnamefont {Stopp}}, \
  and\ \bibinfo {author} {\bibfnamefont {F.}~\bibnamefont {Schmidt-Kaler}},\
  }\bibfield  {title} {\enquote {\bibinfo {title} {{Spin and motion dynamics
  with zigzag ion crystals in transverse magnetic gradients}},}\ }\href
  {\doibase 10.1088/1361-6455/aaf347} {\bibfield  {journal} {\bibinfo
  {journal} {J. Phys. B At. Mol. Opt. Phys.}\ }\textbf {\bibinfo {volume}
  {52}},\ \bibinfo {pages} {025301} (\bibinfo {year} {2019})}\BibitemShut
  {NoStop}%
\bibitem [{\citenamefont {Grimm}\ \emph {et~al.}(2000)\citenamefont {Grimm},
  \citenamefont {Weidem{\"{u}}ller},\ and\ \citenamefont
  {Ovchinnikov}}]{grimm2000optical}%
  \BibitemOpen
  \bibfield  {author} {\bibinfo {author} {\bibfnamefont {R.}~\bibnamefont
  {Grimm}}, \bibinfo {author} {\bibfnamefont {M.}~\bibnamefont
  {Weidem{\"{u}}ller}}, \ and\ \bibinfo {author} {\bibfnamefont {Y.~B.}\
  \bibnamefont {Ovchinnikov}},\ }\bibfield  {title} {\enquote {\bibinfo {title}
  {{Optical Dipole Traps for Neutral Atoms}},}\ \ }(\bibinfo  {publisher}
  {Academic Press},\ \bibinfo {year} {2000})\ pp.\ \bibinfo {pages}
  {95--170}\BibitemShut {NoStop}%
\bibitem [{\citenamefont {Bass}\ \emph {et~al.}(2009)\citenamefont {Bass},
  \citenamefont {DeCusatis}, \citenamefont {Enoch}, \citenamefont
  {Lakshminarayanan}, \citenamefont {Li}, \citenamefont {Macdonald},
  \citenamefont {Mahajan},\ and\ \citenamefont
  {Van~Stryland}}]{bass2009handbook}%
  \BibitemOpen
  \bibfield  {author} {\bibinfo {author} {\bibfnamefont {M.}~\bibnamefont
  {Bass}}, \bibinfo {author} {\bibfnamefont {C.}~\bibnamefont {DeCusatis}},
  \bibinfo {author} {\bibfnamefont {J.}~\bibnamefont {Enoch}}, \bibinfo
  {author} {\bibfnamefont {V.}~\bibnamefont {Lakshminarayanan}}, \bibinfo
  {author} {\bibfnamefont {G.}~\bibnamefont {Li}}, \bibinfo {author}
  {\bibfnamefont {C.}~\bibnamefont {Macdonald}}, \bibinfo {author}
  {\bibfnamefont {V.}~\bibnamefont {Mahajan}}, \ and\ \bibinfo {author}
  {\bibfnamefont {E.}~\bibnamefont {Van~Stryland}},\ }\href@noop {} {\emph
  {\bibinfo {title} {{Handbook of optics, Volume II: Design, fabrication and
  testing, sources and detectors, radiometry and photometry}}}}\ (\bibinfo
  {publisher} {McGraw-Hill, Inc.},\ \bibinfo {year} {2009})\BibitemShut
  {NoStop}%
\bibitem [{\citenamefont {Araneda}\ \emph {et~al.}(2020)\citenamefont
  {Araneda}, \citenamefont {Cerchiari}, \citenamefont {Higginbottom},
  \citenamefont {Holz}, \citenamefont {Lakhmanskiy}, \citenamefont
  {Ob{\v{s}}il}, \citenamefont {Colombe},\ and\ \citenamefont
  {Blatt}}]{araneda2020panopticon}%
  \BibitemOpen
  \bibfield  {author} {\bibinfo {author} {\bibfnamefont {G.}~\bibnamefont
  {Araneda}}, \bibinfo {author} {\bibfnamefont {G.}~\bibnamefont {Cerchiari}},
  \bibinfo {author} {\bibfnamefont {D.~B.}\ \bibnamefont {Higginbottom}},
  \bibinfo {author} {\bibfnamefont {P.}~\bibnamefont {Holz}}, \bibinfo {author}
  {\bibfnamefont {K.}~\bibnamefont {Lakhmanskiy}}, \bibinfo {author}
  {\bibfnamefont {P.}~\bibnamefont {Ob{\v{s}}il}}, \bibinfo {author}
  {\bibfnamefont {Y.}~\bibnamefont {Colombe}}, \ and\ \bibinfo {author}
  {\bibfnamefont {R.}~\bibnamefont {Blatt}},\ }\bibfield  {title} {\enquote
  {\bibinfo {title} {{The Panopticon device: an integrated
  Paul-trap-hemispherical mirror system for quantum optics}},}\ }\href
  {http://arxiv.org/abs/2006.04828} {\bibfield  {journal} {\bibinfo  {journal}
  {arXiv:2006.04828}\ } (\bibinfo {year} {2020})}\BibitemShut {NoStop}%
\bibitem [{\citenamefont {Ballance}(2017)}]{ballance2017high}%
  \BibitemOpen
  \bibfield  {author} {\bibinfo {author} {\bibfnamefont {C.~J.}\ \bibnamefont
  {Ballance}},\ }\href@noop {} {\emph {\bibinfo {title} {High-fidelity quantum
  logic in Ca+}}}\ (\bibinfo  {publisher} {Springer},\ \bibinfo {year}
  {2017})\BibitemShut {NoStop}%
\bibitem [{\citenamefont {Sansonetti}\ and\ \citenamefont
  {Martin}(2005)}]{sansonetti2005handbook}%
  \BibitemOpen
  \bibfield  {author} {\bibinfo {author} {\bibfnamefont {J.~E.}\ \bibnamefont
  {Sansonetti}}\ and\ \bibinfo {author} {\bibfnamefont {W.~C.}\ \bibnamefont
  {Martin}},\ }\bibfield  {title} {\enquote {\bibinfo {title} {Handbook of
  basic atomic spectroscopic data},}\ }\href {\doibase 10.1063/1.1800011}
  {\bibfield  {journal} {\bibinfo  {journal} {Journal of Physical and Chemical
  Reference Data}\ }\textbf {\bibinfo {volume} {34}},\ \bibinfo {pages}
  {1559--2259} (\bibinfo {year} {2005})}\BibitemShut {NoStop}%
\bibitem [{\citenamefont {Brownnutt}\ \emph {et~al.}(2015)\citenamefont
  {Brownnutt}, \citenamefont {Kumph}, \citenamefont {Rabl},\ and\ \citenamefont
  {Blatt}}]{Brownnutt2015}%
  \BibitemOpen
  \bibfield  {author} {\bibinfo {author} {\bibfnamefont {M.}~\bibnamefont
  {Brownnutt}}, \bibinfo {author} {\bibfnamefont {M.}~\bibnamefont {Kumph}},
  \bibinfo {author} {\bibfnamefont {P.}~\bibnamefont {Rabl}}, \ and\ \bibinfo
  {author} {\bibfnamefont {R.}~\bibnamefont {Blatt}},\ }\bibfield  {title}
  {\enquote {\bibinfo {title} {Ion-trap measurements of electric-field noise
  near surfaces},}\ }\href {\doibase 10.1103/RevModPhys.87.1419} {\bibfield
  {journal} {\bibinfo  {journal} {Rev. Mod. Phys.}\ }\textbf {\bibinfo {volume}
  {87}},\ \bibinfo {pages} {1419--1482} (\bibinfo {year} {2015})}\BibitemShut
  {NoStop}%
\bibitem [{\citenamefont {Kaur}\ \emph {et~al.}(2015)\citenamefont {Kaur},
  \citenamefont {Singh}, \citenamefont {Arora},\ and\ \citenamefont
  {Sahoo}}]{kaur2015magic}%
  \BibitemOpen
  \bibfield  {author} {\bibinfo {author} {\bibfnamefont {J.}~\bibnamefont
  {Kaur}}, \bibinfo {author} {\bibfnamefont {S.}~\bibnamefont {Singh}},
  \bibinfo {author} {\bibfnamefont {B.}~\bibnamefont {Arora}}, \ and\ \bibinfo
  {author} {\bibfnamefont {B.~K.}\ \bibnamefont {Sahoo}},\ }\bibfield  {title}
  {\enquote {\bibinfo {title} {{Magic wavelengths in the alkaline-earth-metal
  ions}},}\ }\href {\doibase 10.1103/PhysRevA.92.031402} {\bibfield  {journal}
  {\bibinfo  {journal} {Phys. Rev. A}\ }\textbf {\bibinfo {volume} {92}},\
  \bibinfo {pages} {031402} (\bibinfo {year} {2015})}\BibitemShut {NoStop}%
\bibitem [{\citenamefont {Joshi}\ \emph {et~al.}(2020)\citenamefont {Joshi},
  \citenamefont {Fabre}, \citenamefont {Maier}, \citenamefont {Brydges},
  \citenamefont {Kiesenhofer}, \citenamefont {Hainzer}, \citenamefont {Blatt},\
  and\ \citenamefont {Roos}}]{joshi2020pgc}%
  \BibitemOpen
  \bibfield  {author} {\bibinfo {author} {\bibfnamefont {M.~K.}\ \bibnamefont
  {Joshi}}, \bibinfo {author} {\bibfnamefont {A.}~\bibnamefont {Fabre}},
  \bibinfo {author} {\bibfnamefont {C.}~\bibnamefont {Maier}}, \bibinfo
  {author} {\bibfnamefont {T.}~\bibnamefont {Brydges}}, \bibinfo {author}
  {\bibfnamefont {D.}~\bibnamefont {Kiesenhofer}}, \bibinfo {author}
  {\bibfnamefont {H.}~\bibnamefont {Hainzer}}, \bibinfo {author} {\bibfnamefont
  {R.}~\bibnamefont {Blatt}}, \ and\ \bibinfo {author} {\bibfnamefont {C.~F.}\
  \bibnamefont {Roos}},\ }\bibfield  {title} {\enquote {\bibinfo {title}
  {{Polarization-gradient cooling of 1D and 2D ion Coulomb crystals}},}\ }\href
  {\doibase 10.1088/1367-2630/abb912} {\bibfield  {journal} {\bibinfo
  {journal} {New J. Phys.}\ }\textbf {\bibinfo {volume} {22}},\ \bibinfo
  {pages} {103013} (\bibinfo {year} {2020})}\BibitemShut {NoStop}%
\bibitem [{\citenamefont {Cetina}\ \emph {et~al.}(2020)\citenamefont {Cetina},
  \citenamefont {Egan}, \citenamefont {Noel}, \citenamefont {Goldman},
  \citenamefont {Risinger}, \citenamefont {Zhu}, \citenamefont {Biswas},\ and\
  \citenamefont {Monroe}}]{cetina2020quantum}%
  \BibitemOpen
  \bibfield  {author} {\bibinfo {author} {\bibfnamefont {M.}~\bibnamefont
  {Cetina}}, \bibinfo {author} {\bibfnamefont {L.~N.}\ \bibnamefont {Egan}},
  \bibinfo {author} {\bibfnamefont {C.~A.}\ \bibnamefont {Noel}}, \bibinfo
  {author} {\bibfnamefont {M.~L.}\ \bibnamefont {Goldman}}, \bibinfo {author}
  {\bibfnamefont {A.~R.}\ \bibnamefont {Risinger}}, \bibinfo {author}
  {\bibfnamefont {D.}~\bibnamefont {Zhu}}, \bibinfo {author} {\bibfnamefont
  {D.}~\bibnamefont {Biswas}}, \ and\ \bibinfo {author} {\bibfnamefont
  {C.}~\bibnamefont {Monroe}},\ }\bibfield  {title} {\enquote {\bibinfo {title}
  {{Quantum Gates on Individually-Addressed Atomic Qubits Subject to Noisy
  Transverse Motion}},}\ }\href {http://arxiv.org/abs/2007.06768} {\bibfield
  {journal} {\bibinfo  {journal} {arXiv:2007.06768}\ } (\bibinfo {year}
  {2020})}\BibitemShut {NoStop}%
\bibitem [{\citenamefont {Mohammadi}\ \emph {et~al.}(2019)\citenamefont
  {Mohammadi}, \citenamefont {Wolf}, \citenamefont {Kr{\"{u}}kow},
  \citenamefont {Dei{\ss}},\ and\ \citenamefont {{Hecker
  Denschlag}}}]{mohammadi2019minimizing}%
  \BibitemOpen
  \bibfield  {author} {\bibinfo {author} {\bibfnamefont {A.}~\bibnamefont
  {Mohammadi}}, \bibinfo {author} {\bibfnamefont {J.}~\bibnamefont {Wolf}},
  \bibinfo {author} {\bibfnamefont {A.}~\bibnamefont {Kr{\"{u}}kow}}, \bibinfo
  {author} {\bibfnamefont {M.}~\bibnamefont {Dei{\ss}}}, \ and\ \bibinfo
  {author} {\bibfnamefont {J.}~\bibnamefont {{Hecker Denschlag}}},\ }\bibfield
  {title} {\enquote {\bibinfo {title} {{Minimizing rf-induced excess
  micromotion of a trapped ion with the help of ultracold atoms}},}\ }\href
  {\doibase 10.1007/s00340-019-7223-y} {\bibfield  {journal} {\bibinfo
  {journal} {Appl. Phys. B}\ }\textbf {\bibinfo {volume} {125}},\ \bibinfo
  {pages} {122} (\bibinfo {year} {2019})}\BibitemShut {NoStop}%
\bibitem [{\citenamefont {Bl\"umel}\ \emph {et~al.}(1989)\citenamefont
  {Bl\"umel}, \citenamefont {Kappler}, \citenamefont {Quint},\ and\
  \citenamefont {Walther}}]{Blumel1989Chaos}%
  \BibitemOpen
  \bibfield  {author} {\bibinfo {author} {\bibfnamefont {R.}~\bibnamefont
  {Bl\"umel}}, \bibinfo {author} {\bibfnamefont {C.}~\bibnamefont {Kappler}},
  \bibinfo {author} {\bibfnamefont {W.}~\bibnamefont {Quint}}, \ and\ \bibinfo
  {author} {\bibfnamefont {H.}~\bibnamefont {Walther}},\ }\bibfield  {title}
  {\enquote {\bibinfo {title} {{Chaos and order of laser-cooled ions in a Paul
  trap}},}\ }\href {\doibase 10.1103/PhysRevA.40.808} {\bibfield  {journal}
  {\bibinfo  {journal} {Phys. Rev. A}\ }\textbf {\bibinfo {volume} {40}},\
  \bibinfo {pages} {808--823} (\bibinfo {year} {1989})}\BibitemShut {NoStop}%
\bibitem [{\citenamefont {Narayanan}\ \emph {et~al.}(2011)\citenamefont
  {Narayanan}, \citenamefont {Daniilidis}, \citenamefont {Möller},
  \citenamefont {Clark}, \citenamefont {Ziesel}, \citenamefont {Singer},
  \citenamefont {Schmidt-Kaler},\ and\ \citenamefont
  {Häffner}}]{narayanan2011micromotion}%
  \BibitemOpen
  \bibfield  {author} {\bibinfo {author} {\bibfnamefont {S.}~\bibnamefont
  {Narayanan}}, \bibinfo {author} {\bibfnamefont {N.}~\bibnamefont
  {Daniilidis}}, \bibinfo {author} {\bibfnamefont {S.~A.}\ \bibnamefont
  {Möller}}, \bibinfo {author} {\bibfnamefont {R.}~\bibnamefont {Clark}},
  \bibinfo {author} {\bibfnamefont {F.}~\bibnamefont {Ziesel}}, \bibinfo
  {author} {\bibfnamefont {K.}~\bibnamefont {Singer}}, \bibinfo {author}
  {\bibfnamefont {F.}~\bibnamefont {Schmidt-Kaler}}, \ and\ \bibinfo {author}
  {\bibfnamefont {H.}~\bibnamefont {Häffner}},\ }\bibfield  {title} {\enquote
  {\bibinfo {title} {Electric field compensation and sensing with a single ion
  in a planar trap},}\ }\href {\doibase 10.1063/1.3665647} {\bibfield
  {journal} {\bibinfo  {journal} {Journal of Applied Physics}\ }\textbf
  {\bibinfo {volume} {110}},\ \bibinfo {pages} {114909} (\bibinfo {year}
  {2011})}\BibitemShut {NoStop}%
\bibitem [{\citenamefont {Erhard}\ \emph {et~al.}(2019)\citenamefont {Erhard},
  \citenamefont {Wallman}, \citenamefont {Postler}, \citenamefont {Meth},
  \citenamefont {Stricker}, \citenamefont {Martinez}, \citenamefont
  {Schindler}, \citenamefont {Monz}, \citenamefont {Emerson},\ and\
  \citenamefont {Blatt}}]{erhard2019characterizing}%
  \BibitemOpen
  \bibfield  {author} {\bibinfo {author} {\bibfnamefont {A.}~\bibnamefont
  {Erhard}}, \bibinfo {author} {\bibfnamefont {J.~J.}\ \bibnamefont {Wallman}},
  \bibinfo {author} {\bibfnamefont {L.}~\bibnamefont {Postler}}, \bibinfo
  {author} {\bibfnamefont {M.}~\bibnamefont {Meth}}, \bibinfo {author}
  {\bibfnamefont {R.}~\bibnamefont {Stricker}}, \bibinfo {author}
  {\bibfnamefont {E.~A.}\ \bibnamefont {Martinez}}, \bibinfo {author}
  {\bibfnamefont {P.}~\bibnamefont {Schindler}}, \bibinfo {author}
  {\bibfnamefont {T.}~\bibnamefont {Monz}}, \bibinfo {author} {\bibfnamefont
  {J.}~\bibnamefont {Emerson}}, \ and\ \bibinfo {author} {\bibfnamefont
  {R.}~\bibnamefont {Blatt}},\ }\bibfield  {title} {\enquote {\bibinfo {title}
  {Characterizing large-scale quantum computers via cycle benchmarking},}\
  }\href {\doibase 10.1038/s41467-019-13068-7} {\bibfield  {journal} {\bibinfo
  {journal} {Nature communications}\ }\textbf {\bibinfo {volume} {10}},\
  \bibinfo {pages} {1--7} (\bibinfo {year} {2019})}\BibitemShut {NoStop}%
\bibitem [{\citenamefont {{De Grandi}}\ and\ \citenamefont
  {Polkovnikov}(2010)}]{DeGrandi2010}%
  \BibitemOpen
  \bibfield  {author} {\bibinfo {author} {\bibfnamefont {C.}~\bibnamefont {{De
  Grandi}}}\ and\ \bibinfo {author} {\bibfnamefont {A.}~\bibnamefont
  {Polkovnikov}},\ }\enquote {\bibinfo {title} {{Adiabatic Perturbation Theory:
  From Landau--Zener Problem to Quenching Through a Quantum Critical Point}},}\
  in\ \href {\doibase 10.1007/978-3-642-11470-0_4} {\emph {\bibinfo {booktitle}
  {Quantum Quenching, Annealing and Computation}}},\ \bibinfo {editor} {edited
  by\ \bibinfo {editor} {\bibfnamefont {A.~K.}\ \bibnamefont {Chandra}},
  \bibinfo {editor} {\bibfnamefont {A.}~\bibnamefont {Das}}, \ and\ \bibinfo
  {editor} {\bibfnamefont {B.~K.}\ \bibnamefont {Chakrabarti}}}\ (\bibinfo
  {publisher} {Springer},\ \bibinfo {address} {Berlin, Heidelberg},\ \bibinfo
  {year} {2010})\ pp.\ \bibinfo {pages} {75--114}\BibitemShut {NoStop}%
\end{thebibliography}

%

\end{document}